\documentclass[notitlepage,prd,twocolumn,showpacs,preprintnumbers,nofootinbib,superscriptaddress,floatfix,tightenlines,longbibliography]{revtex4-1}
\usepackage{mathrsfs}
\usepackage{amssymb}
\usepackage{amsmath}
\usepackage{graphicx}
\usepackage[utf8]{inputenc}
\usepackage{placeins}
\usepackage{amsfonts,amsbsy}
\usepackage{nicefrac}
\usepackage{xcolor}
\definecolor{lcolor}{rgb}{0.5,0,0}
\definecolor{citcolor}{rgb}{0,0.3,0.0}
\usepackage[breaklinks,colorlinks,urlcolor=blue,citecolor=citcolor,linkcolor=lcolor]{hyperref}
\usepackage{comment}
\usepackage{bbm}
\usepackage[normalem]{ulem} 
\usepackage{enumitem}

\newcommand{\qs}{Q_\mathrm{s}}

\newcommand{\Q}{Q_s}

\newcommand{\fig}{Fig.~}
\newcommand{\figs}{Figs.~}
\newcommand{\eq}{Eq.~}

\newcommand{\se}{Sec.~}

\newcommand{\re}{Ref.~}

\newcommand{\app}{Appendix~}
\newcommand{\nr}[1]{(\ref{#1})}
\newcommand{\bs}[1]{\boldsymbol{#1}}

\newcommand{\ud}{\mathrm{d}}

\newcommand{\md}{m_D}
\newcommand{\pmin}{p_{\mathrm{min}}}
\newcommand{\pmax}{p_\mathrm{max}}
\newcommand{\tstar}{T_*}
\newcommand{\pt}{p_T}

\newcommand{\nn}{\nonumber}

\newcommand{\nc}{N_c}

\newcommand{\der}{\mathrm{d}}

\newcommand{\be}{\begin{equation}}
\newcommand{\ee}{\end{equation}}
\newcommand{\bea}{\begin{eqnarray}}
\newcommand{\eea}{\end{eqnarray}}

\newcommand{\derthree}{\frac{\der^3 p }{\left(2 \pi \right)^3}}

\newcommand{\eps}{\varepsilon}

\newcommand{\nbar}{\bar{n}}
\newcommand{\pz}{p_z}
\newcommand{\Pmin}{p_{\text{min}}}
\newcommand{\Pmax}{p_{\text{max}}}
\newcommand{\tauT}{\tau_{\mathrm{BMSS}}}

\newcommand{\distvec}[1]{f_{\bs{#1}}}

\begin{document}

\title{Heavy quark diffusion coefficient in heavy-ion collisions via kinetic theory}

\author{K.~Boguslavski} 
\affiliation{Institute for Theoretical Physics, Technische Universit\"{a}t Wien, 1040 Vienna, Austria }

\author{A.~Kurkela} 
\affiliation{Faculty of Science and Technology, University of Stavanger, 4036 Stavanger, Norway}

\author{T.~Lappi} 
\affiliation{Department of Physics, P.O.~Box 35, 40014 University of Jyv\"{a}skyl\"{a}, Finland}
\affiliation{Helsinki Institute of Physics, P.O.~Box 64, 00014 University of Helsinki, Finland}

\author{F.~Lindenbauer} 
\affiliation{Institute for Theoretical Physics, Technische Universit\"{a}t Wien, 1040 Vienna, Austria }

\author{J.~Peuron} 
\email{jarkko.t.peuron@jyu.fi}
\affiliation{Department of Physics, P.O.~Box 35, 40014 University of Jyv\"{a}skyl\"{a}, Finland}
\affiliation{Helsinki Institute of Physics, P.O.~Box 64, 00014 University of Helsinki, Finland}
\affiliation{Dept. of Physics, Lund University,  S\"{o}lvegatan  14A, Lund,SE-223 62, Sweden}

\begin{abstract}
We compute the heavy quark momentum diffusion coefficient $\kappa$ using QCD kinetic theory for a system going through bottom-up isotropization in the  initial stages of a heavy ion collision. We find that the values of $\kappa$ 
are within 30\% from a thermal system at the same energy density. When matching for other quantities we observe considerably larger deviations. We also observe that the diffusion coefficient in the transverse direction is larger at high occupation numbers, whereas for an underoccupied system the longitudinal diffusion coefficient dominates. 
The behavior of the diffusion coefficient 
can be understood on a qualitative level based on the Debye mass $m_D$ and the effective temperature of soft modes $\tstar$. 
Our findings for the kinetic evolution of $\kappa$ in different directions can be used in phenomenological descriptions of heavy quark diffusion and quarkonium dynamics to include the impact of pre-equilibrium stages.

\end{abstract}

\maketitle

\section{Introduction}

Several interesting signatures of the quark-gluon plasma, such as quarkonium suppression, are sensitive to the dynamics of heavy quarks in the plasma. 
Due to their large mass $m_Q \gg T$, heavy quarks are solely produced in initial hard scattering processes. Furthermore 
they interact strongly with the medium during the entire evolution. The typical momentum exchange of a heavy quark with the QGP is small ($\mathcal{O}(gT)$) and thus a significant change in momentum requires several scatterings with the medium. This is characteristic of diffusive motion, and hence diffusion properties of heavy quarks within the QGP have been subject to extensive research~\cite{He:2022ywp,Rothkopf:2019ipj,Brambilla:2022fqa}.
The interaction of the heavy quark with the medium is conventionally quantified by 
the momentum diffusion coefficient $\kappa$. 

The diffusion coefficient is not only interesting in its own right as a property of the medium---it also plays an instrumental role in understanding and describing the evolution of quarkonia using various phenomenological frameworks \cite{Rothkopf:2019ipj,Brambilla:2022fqa}, including e.g., the open quantum systems approach
\cite{Young:2010jq, Brambilla:2016wgg, Brambilla:2020qwo, Brambilla:2021wkt, Brambilla:2023hkw}.

The behavior of $\kappa$ in equilibrium is relatively well understood \cite{Moore:2004tg,Caron-Huot:2007rwy,Caron-Huot:2008dyw,Brambilla:2020siz,Casalderrey-Solana:2006fio,vanHees:2005wb,Svetitsky:1987gq,vanHees:2007me,Banerjee:2011ra,Capellino:2022nvf,Yao:2020xzw,Banerjee:2022gen}, but out of equilibrium it remains elusive. The existing literature on transport coefficients out of equilibrium \cite{Ipp:2020nfu, Ipp:2020mjc, Avramescu:2023qvv,Carrington:2022bnv,Carrington:2021dvw,Carrington:2020sww,Khowal:2021zoo,Ruggieri:2022kxv,Sun:2019fud,Boguslavski:2020tqz} indicates that the overoccupied glasma stage can have considerable impact on transport phenomena, such as momentum broadening,  jet quenching and heavy quark transport coefficients.
The effect of anisotropic initial stages has been studied using a simplified description of the underlying anisotropic momentum distribution \cite{Romatschke:2006bb, Romatschke:2004au, Hauksson:2021okc}. 
However, to our knowledge, the time evolution of transport coefficients has not yet been extracted during the kinetic evolution between the glasma stage and the hydrodynamical phase.

The aim of this paper is to bridge 
this gap and to study the heavy quark momentum diffusion coefficient $\kappa$ during the bottom-up isotropization process \cite{Baier:2000sb}. With this aim in mind, we use QCD effective kinetic theory with the same initial conditions and numerical setup as in \cite{Kurkela:2015qoa} for the limit of an infinitely massive stationary quark. Furthermore, our aim is to understand how $\kappa$ out of equilibrium compares to similar computations in thermal equilibrium. To this end, we compare values of $\kappa$ to thermal systems with 
 the same screening mass, energy density or effective infrared (IR) temperature. 
 We establish that out of these options $\kappa$ in the pre-equilibrium system is closest to the thermal value at the same energy density. We will also study the transverse (to the direction of the expansion), and longitudinal (along the beam direction) diffusion coefficients separately. We observe that there is a dynamical hierarchy between the longitudinal and transverse directions, which has a natural explanation in terms of the different stages of the bottom-up thermalization process. We will also derive a set of parametric estimates that explain the evolution of the matched quantities and $\kappa$ relative to their equilibrium value 
in terms of the dynamical occupation number and anisotropy of the system.

This paper is structured as follows. In \se \ref{sec:theory} we discuss the kinetic theory framework, how we extract the heavy quark momentum diffusion coefficient and introduce our initial conditions and the relevant physical scales $\md$ and $\tstar$. We continue in \se \ref{sec:results} by presenting our results for the diffusion coefficient during the initial stages in heavy-ion collisions and
discuss the diffusion coefficient in the transverse and longitudinal directions separately. In \se \ref{sec:comparisons} we compare our results to those obtained in lattice and glasma simulations. Finally we conclude in \se \ref{sec:conc}.

\section{Theoretical background}
\label{sec:theory}
Let us start by introducing the
effective kinetic theory, heavy quark diffusion coefficient and other observables relevant for this work. 
Some details of our discretization procedure and discretization effects are discussed in  Appendix \ref{app:numerics}.

\subsection{Effective kinetic theory}
In this paper we use the same numerical setup as in Ref.~\cite{Kurkela:2015qoa}, 
with the aim of reproducing the bottom-up isotropization process \cite{Baier:2000sb} using the numerical framework of QCD effective kinetic theory (EKT) \cite{Arnold:2002zm}. 
Kinetic theory is a quasiparticle description of the quark-gluon plasma in terms of its individual constituents, quarks and gluons. 
Here we consider only gluons for which the central dynamical quantity
is the quasiparticle distribution function 
\begin{equation}
\label{eq:fdef}
f(\bs{p}) = \dfrac{1}{\nu_g}\dfrac{\der N}{\der^3 x\, \der^3 \bs{p}}.
\end{equation}
 This is the phase space density of gluons averaged over their 
 degrees of freedom 
 $\nu_g = 2 d_A,$ including $d_A = N_c^2 -1$ colors and two spins. 
 We assume that the distributions are independent of transverse coordinate, boost invariant  and do not depend on spin.
The time-evolution of the distribution is given by the Boltzmann equation
\begin{equation}
\label{eq:EKTEOMS}
-\dfrac{\partial f(\bs{p})}{\partial \tau } = \mathcal{C}_{1 \leftrightarrow 2  }[f(\bs{p})] + \mathcal{C}_{2 \leftrightarrow 2  }[f(\bs{p})] + \mathcal{C}_{\mathrm{exp} }[f(\bs{p})].
\end{equation}
Here the $\mathcal{C}$ terms describe the number of scatterings/splittings per unit time for momentum state $\bs{p}$. These processes involve the possibility to scatter/split into or out of a state. 
Effective particle 1 to 2 splittings are encoded in $\mathcal{C}_{1 \leftrightarrow 2  }$ and 2 to 2 scatterings are described by $\mathcal{C}_{2 \leftrightarrow 2  }$. 

The boost invariant  expansion is treated as an effective scattering term, taking the simple form \cite{Mueller:1999pi}
\begin{equation}
\mathcal{C}_{\mathrm{exp} }[f(\bs{p})] = - \dfrac{p_z}{\tau} \dfrac{\partial}{\partial p_z} f(\bs{p}).
\end{equation}
These assumptions leave
our distributions $f$ dependent only on the magnitude of the momentum $p$ and on the polar angle $\cos \theta = \hat{\bs{z}} \cdot \hat{\bs{p}}$, i.e $f(\bs{p}) = f(p, \cos \theta_p)$.
Our spherical coordinate system is defined such that 
\begin{align}
\pz &= p \cos \theta \\
\pt &= p \sin \theta.
\end{align}
Momenta in the $x$ and $y$ directions are given in terms of the azimuthal angle $\phi$ 
\begin{align}
p_x &=  \pt \cos \phi  \\
p_y &= \pt \sin \phi.
\end{align}
The 2 to 2 scattering term reads  \cite{Arnold:2002zm}
\begin{align}
\label{eq:C22}
\mathcal{C}_{2 \leftrightarrow 2  }&[f(\tilde{\bs{p}})] = \dfrac{\left( 2 \pi \right)^3}{4 \pi \tilde{p}^2} \dfrac{1}{8 \nu_g} \int \der \Gamma_{\mathrm{PS}} \left|M \right|_{2 \leftrightarrow 2  }^2 \nn \\
& \times  \left( f_{\bs{p}} f_{\bs{k}} (1 + f_{\bs{p^\prime}}) ( 1 + f_{\bs{k^\prime}}) - \distvec{p^\prime} \distvec{k^\prime} (1 + \distvec{p}) (1 + \distvec{k}) \right) \nn \\
& \times \left( \delta(\bs{\tilde{p}} - \bs{p})  + \delta(\bs{\tilde{p}} - \bs{k}) - \delta (\bs{\tilde{p}} - \bs{p^\prime}) - \delta(\bs{\tilde{p}} - \bs{k^\prime})\right),
\end{align}
where $\der \Gamma_{\mathrm{PS}}$ is a phase space integration over the incoming and outgoing momenta $\bs{p},\bs{p^\prime},\bs{k}$ and $\bs{k^\prime}$. The matrix element squared
 $|M|_{2 \leftrightarrow 2  }^2$  at leading order $gg \to gg$ is
 \footnote{Note that there is a typo in \cite{Kurkela:2015qoa} where some factors of $1/4$ were missing.}
\begin{align}
\label{eq:ggMatrixElem}
\dfrac{\left|M\right|_{2 \leftrightarrow 2  }^2}{4 \lambda^2 d_A} = \left( 9 + \dfrac{(s-t)^2}{u^2} + \dfrac{(u-s)^2}{t^2} + \dfrac{(t-u)^2}{s^2} \right).
\end{align}

Here $\lambda = g^2 N_c$ is the 't Hooft coupling. In \eqref{eq:ggMatrixElem} the division by the Mandelstam variables $t$ and $u$ leads to infrared divergences. These can be regulated by including contributions arising from medium modifications. In the employed numerical framework, these are approximated by substituting the denominators in \eqref{eq:ggMatrixElem} with 
\begin{equation}  \label{eq:regt}
 t \to t\, \frac{q^2  + 2\xi_0^2 m^2 }{q^2},
\end{equation}
with  $q = |\bs{q}|$ and $\bs{q}=\bs p' - \bs p$, and symmetrically for $u$.
Note that the limit $t\to 0$ with $s$ fixed is reached only when $q^2 \to 0$. 
We follow the prescription 
of \cite{AbraaoYork:2014hbk} and use $\xi_0 = \nicefrac{e^{\nicefrac{5}{6}}}{\sqrt{8}}$. Here $m$ is  the asymptotic mass in the gluon dispersion relation at high momentum that is perturbatively related to the Debye mass $\md$ as
\begin{align}
    \label{eq:mDdefinition}
    \md^2 = 2 m^2 = 4\lambda \int \frac{\ud^3 p}{(2\pi)^3}\,\frac{f(\bs p)}{p}\,.
\end{align}
The final state Bose enhancement in \eqref{eq:C22} is governed by the  factors $(1 + f)$. The first term  in \eqref{eq:C22} is a loss term for particles scattering out of the state and the second term is the term associated with scattering into the state.  The phase space integration measure is given by
\begin{align}
 \int \der \Gamma_{\mathrm{PS}}  &=  \int_{\bs{p k p' k'}}
 \left(2 \pi \right)^4 \delta\left( P + K - P^\prime - K^\prime \right) \nn \\
 & = \dfrac{1}{2^{11} \pi^7} \int_0^\infty \der q \int_{-q}^q \der \omega \int_{\frac{q-\omega}{2}}^\infty \der p  \int_{\frac{q+\omega}{2}}^\infty \der k \nn \\ 
 & \times \int_{-1}^1 \der x_q \int_0^{2 \pi} \der \phi_{pq} \int_0^{2 \pi} \phi_{kq},
 \end{align}
 with $\omega = q^0,$ $x_q = \cos \theta_q$, 
$\int_{\bs{p}} =  \int \frac{\der p^3}{2 p^0\left(2 \pi \right)^3}$ and $p^0 = p$. 
The simplification above is achieved by changing variables from $p^\prime$ to $q$ and using delta functions to eliminate the $k^\prime$ integral. The residual three-dimensional integrals can be written as integrals over four-momenta, introducing delta functions that constrain the energy component. The integration limits are found by inspecting the energy delta functions. One also makes use of the fact that the system is azimuthally symmetric. 

The collinear 1 to 2 splitting rate  $\mathcal{C}_{1 \leftrightarrow 2}$  is given by 
 \begin{align}
 &\mathcal{C}_{1 \leftrightarrow 2  }\left[f\right] (\bs{\tilde{p}})= \dfrac{\left( 2 \pi \right)^2}{4 \pi \tilde{p}^2} \dfrac{1}{\nu_g}  \int_0^\infty \der p \int_0^{p/2} \der k^\prime \left( 4 \pi \gamma(p ; p^\prime , k^\prime) \right) \nn \\
 &\times \Big[\distvec{p}  \left( 1 + f(x_p , p^\prime) \right) (1+f(x_p , k^\prime)) \nn \\ &  - (1 + \distvec{p}) f(x_p, p^\prime) f(x_p , k^\prime) \Big] \nn \\
 &\times \left[ \delta(\tilde{p} - p) - \delta(\tilde{p} - p^\prime)  - \delta(\tilde{p} - k^\prime) \right].
\end{align}
 Due to collinearity all momenta are pointing in the same direction. We take the particle with momentum $p$ to be the particle that splits, and hence $p^\prime = p - k^\prime$.
Here $\gamma$ parametrizes the differential rate for the splitting processes and is given by 
\begin{equation}
\gamma^g_{gg}(p , p^\prime , k^\prime) = \dfrac{p^4 + p^{\prime 4} + k^{\prime4}}{p^3 p^{\prime^3} k^{\prime^3}} \mathcal{F}_g(p; p^\prime , k^\prime).
\end{equation}
The function $\mathcal{F}$ is computed from 
\begin{equation}
\mathcal{F}(p , p^\prime , k^\prime) = \dfrac{d_A C_A \alpha_s}{2 (2 \pi)^3} \int \dfrac{\der^2 h }{\left( 2 \pi \right)^2} 2 \bs{h} \cdot \mathfrak{Re} \bs{F}_g (\bs{h} , p , p^\prime , k^\prime),
\end{equation}
where $C_A = N_c$ and $\alpha_s = \nicefrac{g^2}{4 \pi}.$

The function $\bs{F}_g$ is a solution of an integral equation 
\begin{align}
\label{eq:integralEqu}
&2 \bs{h} = i \delta E( \bs{h} ; p , p^\prime, k^\prime) \bs{F}_g(\bs{h}; p , p^\prime, k^\prime) +  \nn \\ 
& \times \dfrac{g^2 C_A}{2} T_* \int \dfrac{\der^2 q_\perp}{\left( 2 \pi \right)^2} \left[ \dfrac{1}{q_\perp^2} - \dfrac{1}{q_\perp^2 + m_D^2 } \right] \nn \\ 
& \times \Big[\left(\bs{F}_g(\bs{h}; p , p^\prime, k^\prime) - \bs{F}_g(\bs{h} - k^\prime \bs{q}_\perp; p , p^\prime, k^\prime) \right)  \nn \\
& + \left(\bs{F}_g(\bs{h}; p , p^\prime, k^\prime) - \bs{F}_g(\bs{h} - p^\prime \bs{q}_\perp; p , p^\prime, k^\prime) \right) \nn \\
& + \left(\bs{F}_g(\bs{h}; p , p^\prime, k^\prime) - \bs{F}_g(\bs{h} + p \bs{q}_\perp; p , p^\prime, k^\prime) \right)\Big].
\end{align}
Here $\bs{\hat{n}}$ is a unit vector in the direction of the splitting or merging hard particle(s). The vector $\bs{h}$ is perpendicular to $\bs{\hat{n}}$, and can be related to the transverse momentum \cite{Arnold:2002ja}. In deriving \eqref{eq:integralEqu} the Wightman correlation function
 is evaluated by making use of the 
 sum rule derived in \cite{Aurenche:2002pd} using an isotropic screening approximation.
The energy difference in \eqref{eq:integralEqu} is given by  
\begin{align}
\delta E( \bs{h} ; p , p^\prime, k^\prime) = \dfrac{m^2}{2 k^\prime} + \dfrac{m^2}{2 p^\prime} - \dfrac{m^2}{2 p} + \dfrac{\bs{h}^2}{2 p k^\prime p^\prime }.
\end{align}
The integral equation \eqref{eq:integralEqu} is solved by using the numerical method described in \cite{Ghiglieri:2014kma}.

For a more comprehensive description of EKT we refer the reader to \cite{Arnold:2002zm} and we discuss our discretization procedure in \ref{app:numerics}. 

\subsection{Numerical implementation and discretization effects}
Our numerical framework has several discretization parameters, such as minimum and maximum momenta stored for the distribution function or the number of bins in the angular discretization.
In our simulations, it turns out that the dominant source of discretization effects is the infrared cutoff of the particle distribution function $\pmin$. The reason for this is that due to the longitudinal expansion, the typical momenta become smaller over time.
For instance, the dimensionless ratio of the infrared cutoff and the temperature $\pmin/T(\tau)$ grows with time.
In practice, this means that the thermal equilibrium that the system approaches will also have 
discretization effects. We will take these into account by replacing the continuum thermal quantities
with ones calculated from a thermal distribution with the same cutoff as the EKT simulation.
This correction enables us to compare our nonequilibrium results to the  thermal state that the system is actually approaching.
We will apply this correction to all observables when comparing with equilibrium quantities unless stated otherwise. This effect is discussed in more detail in Appendix~\ref{app:pminDependence}. Our discretization framework and parameters are discussed in more detail in Appendix~\ref{sec:discdetails}.

\subsection{Initial conditions: bottom-up thermalization}

Our initial conditions are chosen to correspond to the bottom-up isotropization scenario, as implemented in Ref.~\cite{Kurkela:2015qoa}. 
The initial distribution function at the time $\Q \tau = 1$ 
is taken to be
\begin{align}
f(p_\perp, p_z) & = \dfrac{2}{\lambda} A \dfrac{\langle p_T \rangle}{\sqrt{p_\perp^2 + (\xi p_z)^2}} \nonumber \\
& \times \exp{\left( \dfrac{-2}{3 \langle p_T \rangle^2 } \left(p_\perp^2 + (\xi p_z)^2  \right) \right)}.
\label{eq:IC}
\end{align}
In this paper we consider the same two sets of initial parameters with different anisotropies $\xi$ as in Ref.
~\cite{Kurkela:2015qoa}, given by
\begin{eqnarray}
    \label{eq:GoodParamsXi10}
    A = 5.24171\,, \quad \langle p_T \rangle = 1.8  \Q\,, \quad \xi = 10 \\
    A = 2.05335\,, \quad \langle p_T \rangle = 1.8 \Q\,, \quad \xi = 4\,.
    \label{eq:GoodParamsXi4}
\end{eqnarray}
In the following we will refer to these only by the value of the anisotropy parameter $\xi$. In our figures the initial condition with $\xi=10$ is always represented by full lines. When $\xi=4$ results are shown for comparison, we use dash-dotted transparent lines.

\subsection{Heavy quark diffusion coefficient}

The expression for the heavy quark momentum diffusion coefficient $\kappa$ for pure glue QCD has been originally derived in \cite{Moore:2004tg}. Keeping only the leading order contributions in $\nicefrac{1}{M}$, where $M$ is the mass of the heavy quark, we obtain the diffusion coefficient as
\begin{align}
\label{eq:kappa_master_formula}
3 \kappa &= \frac{1}{2M}\int_{\bs{k} \bs{k^\prime} \bs{p^\prime}}\left(2 \pi \right)^3 \delta^3\left( \bs{p} +\bs{k} - \bs{p^\prime} - \bs{k^\prime} \right) \nn \\
& \times  2 \pi \delta \left(k^\prime - k \right) \bs{q}^2 
\left[ \left| \mathcal{M}_\kappa \right|^2 f(\bs{k}) (1+f(\bs{k^\prime})) \right].
\end{align}
The heavy quark mass is taken to be larger than any other scale in the system (e.g. $M \gg T, \Q, m_D$). The in- and outgoing heavy quark momenta are given by  $\bs{p}, \bs{p^\prime}$, the momenta 
of gluons in the plasma are labeled by $\bs{k}, \bs{k^\prime}$, and $\bs{q}=\bs{p}-\bs{p^\prime}$ is the momentum transfer to the heavy quark. The gluon distribution function $f$ is obtained from solving the Boltzmann equation \eqref{eq:EKTEOMS}.
A final state Bose-enhancement factor appears in \eqref{eq:kappa_master_formula} similarly as in the Boltzmann equation. 
We set $p^0 = p'{}^0 = M$ 
for the heavy quark. For the gluons and $k^0 = \left| \bs{k} \right|$, $k'{}^0 = \left| \bs{k'} \right|$. 

To leading order in the coupling, the dominant contribution to $\kappa$ is given by $t$-channel gluon exchange \cite{Moore:2004tg}, while other diagrams are suppressed by inverse powers of the heavy quark mass. The corresponding matrix element is then given by
\begin{align}
\label{eq:MatrixElement}
\left| \mathcal{M}_\kappa \right|^2 = N_c C_H g^4\,  \frac{16 M^2 k_0^2 (1+ \cos^2 \theta_{\bs{k} \bs{k^\prime} } )}{\left(q^2 + m_D^2 \right)^2},
\end{align}
where $C_H = \nicefrac{(N_c^2-1)}{2 N_c}$.
Note that due to the heavy quark mass, the HTL propagator reduces in this case to just Debye screening~\cite{Moore:2004tg}, unlike for massless particles in  Eq.~\nr{eq:regt}.

The resulting expression $\left| \mathcal{M}_\kappa \right|^2/M^2$ entering the Boltzmann equation is independent of the heavy quark mass. In our simulations, $\kappa$ is computed using \eq \eqref{eq:kappa_master_formula}. The discretization of \eqref{eq:kappa_master_formula} and a detailed description of how transverse and longitudinal diffusion coefficients are extracted, are explained in more detail in the \app \ref{se:DiscreteKappa}. For more details on the diffusion coefficient we refer the reader to \cite{Moore:2004tg} and to our previous work \cite{Boguslavski:2020tqz}.

Perturbatively, in thermal equilibrium the screening mass \eqref{eq:mDdefinition} becomes
\begin{equation}
\label{eq:mDvsT}
\md^2 = \dfrac{T^2 \lambda}{3}.
\end{equation}
In our kinetic simulations, we compute $\md$ according to Eq.~\eqref{eq:mDdefinition} using the nonequilibrium distribution $f(\bs p)$. The screening mass is also affected by the IR regulator $\pmin$. Thus, when comparing $\md$ to its thermal value, we will apply a discretization correction to it. This is discussed in more detail in \app \ref{sec:mdpmin}. 

For an isotropic distribution in the continuum,  the diffusion coefficient \eqref{eq:kappa_master_formula} becomes 
\begin{align}
\label{eq:kappaLO}
\kappa &= \dfrac{\lambda^2 C_H}{12 N_c \pi^3 } \int_0^\infty \der k\,k^2  f(k) (1 + f(k)) \nn \\
& \quad \times \int_0^{2k}\der q\,\dfrac{q^3}{\left(q^2 + m_D^2 \right)^2}\left( 2 - \dfrac{q^2}{k^2} + \dfrac{q^4}{4 k^4} \right).
\end{align}
This expression is very easy to evaluate in thermal equilibrium using a Bose-Einstein distribution for $f(p)$ and the thermal result \eqref{eq:mDvsT} for $m_D$. 
We take into account the impact of a finite $\pmin$ on the value of $\kappa$ in a thermal system by calculating it using \eqref{eq:kappaLO}, where the screening mass now depends on the infrared cutoff $\pmin$ according to Eq.~\eqref{eq:mDpmin}. 
We will not regulate the momentum transfers by this parameter, since the main $\pmin$ dependence seems to enter $\kappa$ through $\md$. This procedure is described in more detail in Appendix~\ref{sec:kappapmin}, and our final expression for thermal $\kappa$ in the presence of the IR regulator $\pmin$ is given by \eq \eqref{eq:EpicKappa}.

\subsection{Other observables}

Here we describe the observables that we are extracting during the nonequilibrium evolution in addition to the heavy quark diffusion coefficient. 

\subsubsection{Expectation values}
In general, we use the following definition for the expectation value of observable $X$
\begin{equation}
\label{eq:ExpectationValueDef}
\langle X \rangle = \frac{\int \derthree\,  X\, f(\bs p)}{\int \derthree  f(\bs p)}\,.
\end{equation}

\subsubsection{Energy density}
The energy density corresponds to the first moment of the distribution 
\begin{equation}
\label{eq:epsdef}
\eps= \nu_g \int \dfrac{\der^3 p}{\left( 2 \pi \right)^3}\, p f(\bs p),
\end{equation}
where $\nu_g = 2 \left(\nc^2 -1\right)$ for pure glue QCD.

\subsubsection{Temperature}
There is no unambiguous way to define a temperature out of equilibrium. One option is to formulate it effectively  in terms of the (time-dependent) energy density
\begin{equation}
\label{eq:sensibleTemperature}
 T_\eps = \left(\dfrac{30\,\eps}{\pi^2\nu_g} \right)^{1/4} ,
\end{equation}
which agrees with the temperature in thermal equilibrium.
For ideal Bjorken hydrodynamics, the temperature scales as 
\begin{equation}
\label{eq:Tscaling}
T_{id} \sim \left( \Q \tau \right)^{-1/3}.
\end{equation}
This scaling is also expected to hold for $T_\eps$ at sufficiently late times 
due to the approximate $\tau^{-4/3}$ scaling of the energy density in the expanding system.

The effective infrared temperature $\tstar$ is given by 
\begin{equation}
\label{eq:Tstardef}
\tstar = \dfrac{I}{J},
\end{equation}
where 
\begin{align}
I &= \dfrac{1}{2} \int \dfrac{\der^3 p}{\left(2 \pi \right)^3}\, f(\bs p) (1+f(\bs p)) \\
J &= \int \dfrac{\der^3 p }{\left(2 \pi \right)^3} \dfrac{f(\bs p)}{p} = \dfrac{m_D^2}{4 \lambda }.
\end{align}
Similarly to $\kappa$ and $\md$, $\tstar$ is also affected by the IR regulator $\pmin$. Thus we compute the corresponding equilibrium value by taking into account these effects both in $\md$ and in the momentum integral. This is discussed in more detail in Appendix~\ref{sec:tstarpmin}. In thermal equilibrium in the presence of the IR regulator, $\tstar$ is given by \eq \eqref{eq:tstarpminexpression}. 

Slightly different definitions of $\tstar$, based on other integral moments,  have been also introduced in the literature~\cite{Kurkela:2018oqw}, but we do not consider them here.

\subsubsection{Energy momentum tensor}

The components of the energy momentum tensor are obtained as moments of the distribution function by
\begin{equation}
T^{\mu \nu} = \nu_g \int \derthree \dfrac{p^\mu p^\nu }{p} f(\bs p).
\end{equation}
The components relevant for this paper are 
\begin{align}
T_{xx} &= \nu_g \int \derthree f(\bs p) \dfrac{\pt^2}{p} \cos^2\phi \\ 
T_{yy} &= \nu_g  \int \derthree f(\bs p) \dfrac{\pt^2}{p} \sin^2\phi \\
T_{zz} &= \nu_g \int \derthree  f(\bs p) \dfrac{\pz^2}{p}.
\end{align}
 They are connected to the longitudinal and transverse pressure by
\begin{align}
P_T &= \dfrac{T_{xx}+T_{yy}}{2} \\
P_z &= T_{zz}.
\end{align}
The temporal component of the energy-momentum tensor corresponds to the energy density $T_{00} \equiv \eps$.

\begin{figure}
  \centering
  \includegraphics[width=0.48\textwidth]{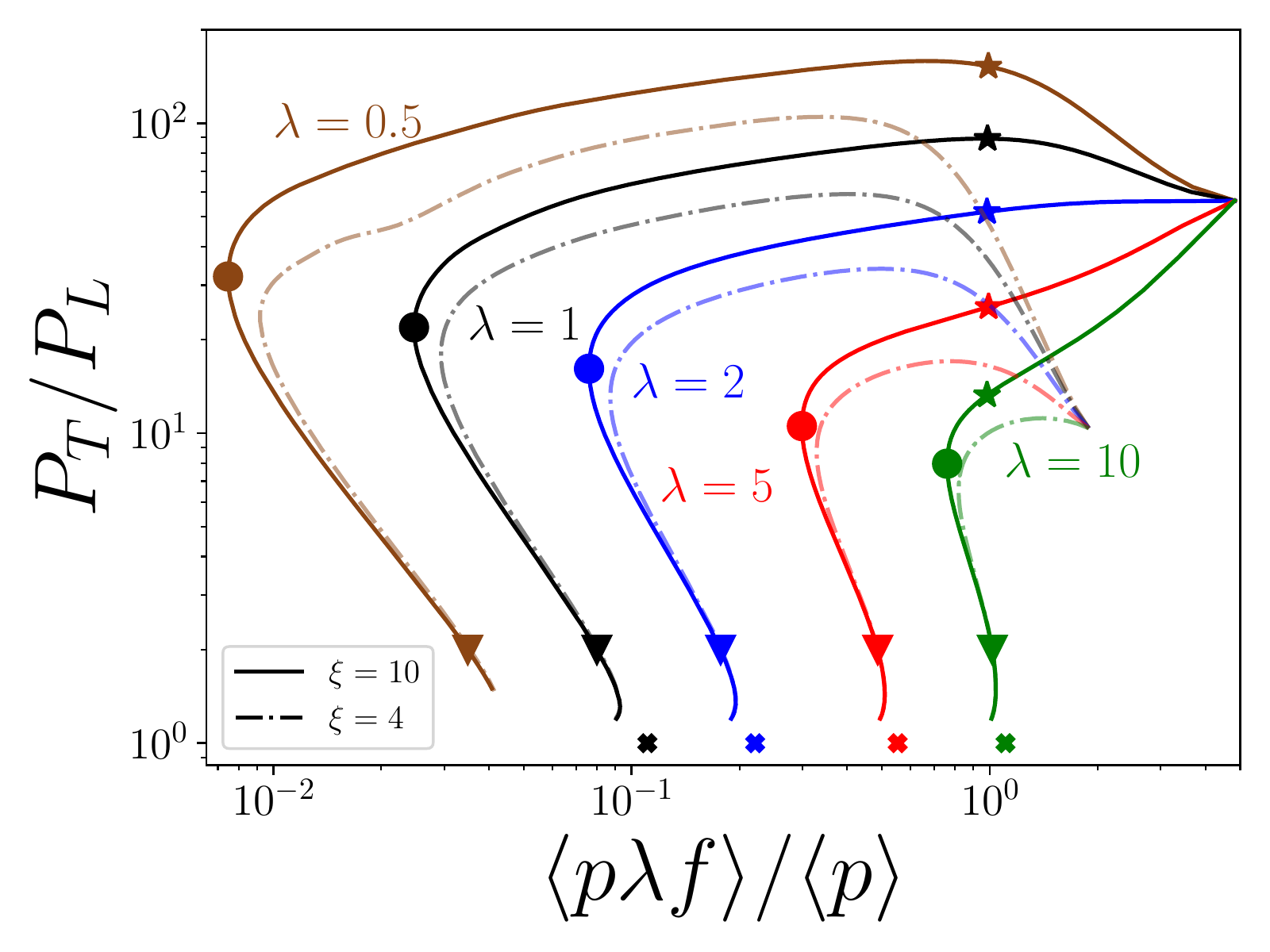}
\caption{Occupation number as a function of anisotropy as in \cite{Kurkela:2015qoa}. The markers indicate different stages of the bottom up thermalization as described in the text. The color coding of the lines indicates the coupling used in the simulation, and all figures use the same color coding. Full lines correspond to the $\xi=10$ initial condition \eqref{eq:GoodParamsXi10}, dash-dotted lines to the $\xi=4$ initial condition \eqref{eq:GoodParamsXi4}. }
\label{fig:occupVsAnisotropy}
\end{figure}

\section{Results }
\label{sec:results}

We start by discussing the bottom-up isotropization process and how it is reproduced in our simulations in \se \ref{sec:results-bottom-up}. We will also discuss how we highlight different stages of the isotropization process. In \se \ref{sec:matching} we will explain how we compare our $\kappa$ results out of equilibrium to thermal values, and especially how we choose the corresponding matching scales. The comparison is then carried out in \se\ref{se:kappaInOutEqu}, first without distinguishing between directions. Then in \se\ref{se:kappaTransLong} we discuss the transverse and longitudinal diffusion coefficients separately. In order to better understand the observed time-evolution of $\kappa$, we will consider how the matching scales evolve during the bottom-up isotropization compared to equilibrium in \se \ref{se:MatchingEvolution}. Finally, we will derive simple parametric estimates that can be used to explain our results in \se \ref{se:paramest}.

\subsection{Different stages of the bottom up thermalization scenario}
\label{sec:results-bottom-up}

Our kinetic theory simulations follow the different phases of the bottom-up thermalization scenario \cite{Baier:2000sb}. It consists of the following stages: during the first stage the overoccupied gluons become more dilute as the system expands, and consequently the occupation number of the hard gluons becomes of the order of unity $f_h \sim 1$ at time scale $\Q \tau \sim \alpha_s^{-3/2}$. At this point the system is no longer describable with classical fields. In the second stage hard gluons radiate softer gluons, creating a soft thermal bath. Remarkably, at the end of this process the hard gluons become underoccupied $f_h \sim \alpha_s$. 
This occurs at the timescales $\Q \tau \sim \alpha_s^{\nicefrac{-5}{2}}$. In the final stage,
the hard particles lose their energy to the soft thermal bath.
For a review on the thermalization processes see e.g.~\cite{Schlichting:2019abc,Berges:2020fwq}.

In this scenario, the system is expected to thermalize parametrically on a timescale of the order of \cite{Baier:2000sb}
\begin{equation}
\label{eq:tauT}
\tauT =  \nicefrac{\alpha_s^{\nicefrac{-13}{5}}}{\Q},
\end{equation}
where $\alpha_s = \frac{\lambda}{4 \pi N_c}$. We will therefore use this quantity to rescale the time in our figures. Note that an alternative time scale, the hydrodynamical relaxation time $\tau_R = \frac{4 \pi \eta/s(\lambda)}{ T}$ with shear viscosity $\eta$ and entropy density $s$, is also often used to rescale the time variable. We will investigate the universality of these time scales for different observables and couplings in a separate paper, while employing only $\tauT$ in the present work.

The bottom-up thermalization process is shown in \fig \ref{fig:occupVsAnisotropy} in terms of the anisotropy $\nicefrac{P_T}{P_L}$ and the mean occupation number $\nicefrac{\langle p\lambda f \rangle}{\langle p \rangle}$ for different couplings as in \re \cite{Kurkela:2015qoa}. In order to illustrate how observables behave during different stages of the thermalization process, we have placed three time markers on the curves in \fig \ref{fig:occupVsAnisotropy}. The first marker (star) is placed during the highly occupied regime, when $f \sim \nicefrac{1}{\lambda}$. For smaller values of the coupling this corresponds to maximal anisotropy. However, for large couplings the first stage of the evolution proceeds differently, and the anisotropy does not increase initially. Hence, we have chosen the occupancy as the criterion for the time marker instead of maximum anisotropy. The second marker (circle) is inserted at the minimum occupancy, which in the bottom-up thermalization scenario is expected to be $f \sim \alpha_s$. The third marker (triangle) is placed at $\nicefrac{P_T}{P_L} = 2.$ The purpose of this is to illustrate when the system is approximately close to equilibrium. We will use the same markers in other figures throughout this paper to allow the reader to connect the time-evolution of observables to the stages of the bottom-up scenario.

The curves for different values of the coupling $\lambda$ also use the same color coding in all figures throughout this paper.  The initial conditions with $\xi=10$ from  \eqref{eq:GoodParamsXi10} are shown as full lines. For comparison, we will often add curves for the initial conditions with $\xi=4$ in \eqref{eq:GoodParamsXi4} as more transparent dash-dotted lines.

We see from \fig \ref{fig:occupVsAnisotropy} that for the extremely anisotropic initial condition ($\xi=10$) the bottom-up picture is better realized at smaller values of the coupling. For intermediate couplings $\lambda = 2,5,10$ the system does not experience an initial growth in anisotropy. Instead, the system takes a more straightforward path to thermal equilibrium, without resolving of the different stages of the bottom-up picture in detail.
However, the third stage of the scenario is still visible and emerges after the circle marker.

\begin{figure}[tbhp!]
  \centering
  \includegraphics[width=0.45\textwidth]{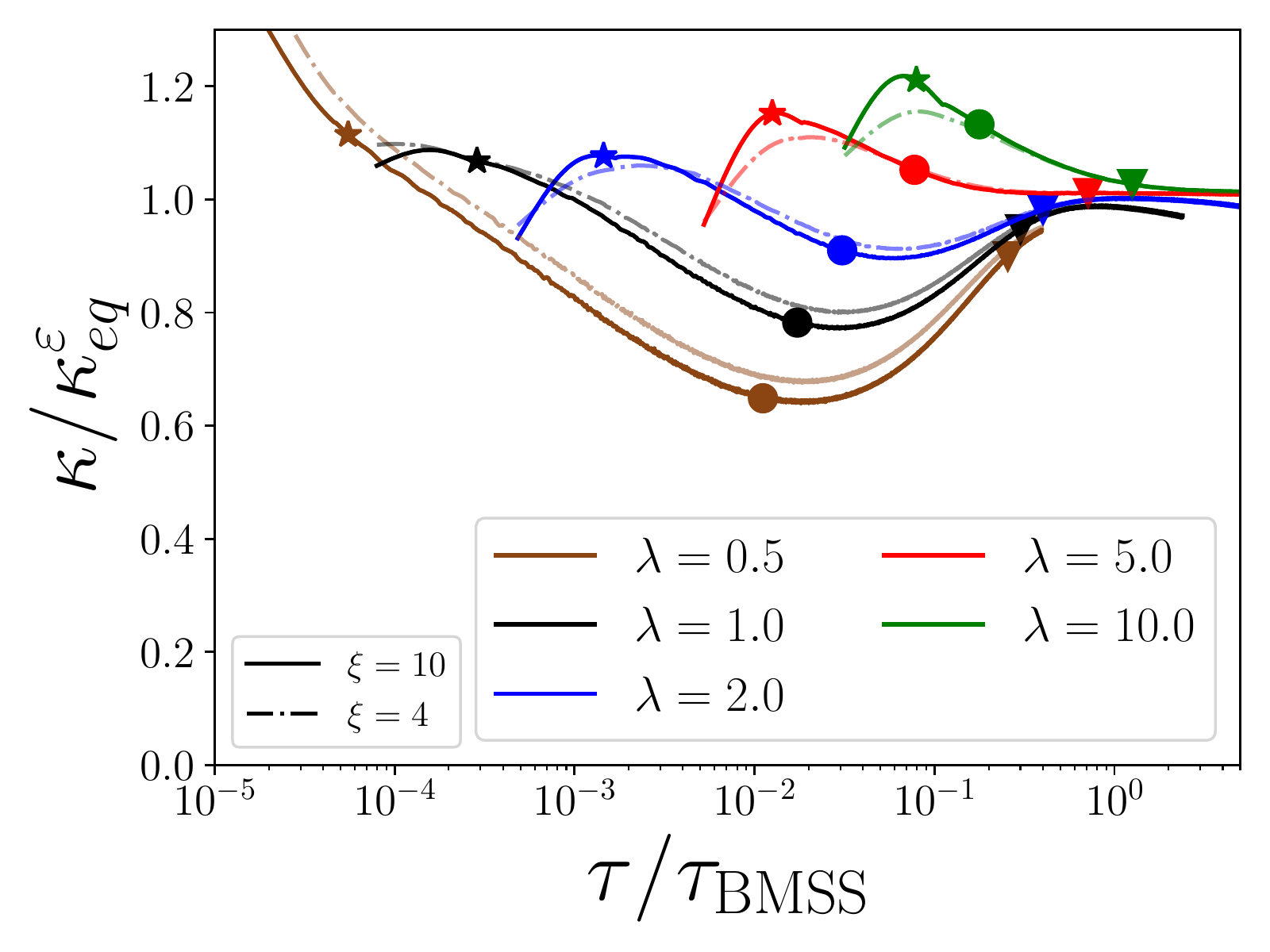}

  \includegraphics[width=0.45\textwidth]{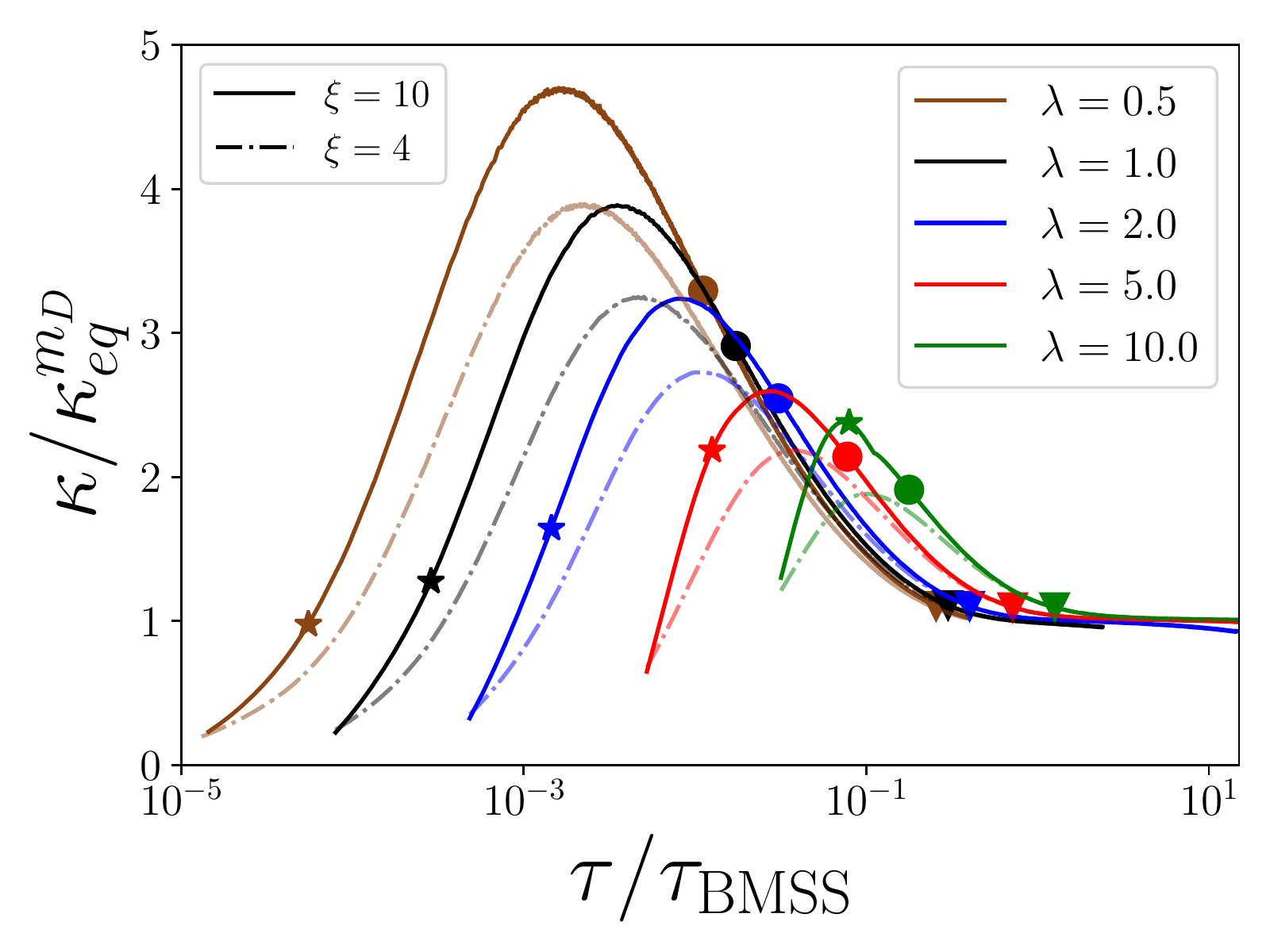}

  \includegraphics[width=0.45\textwidth]{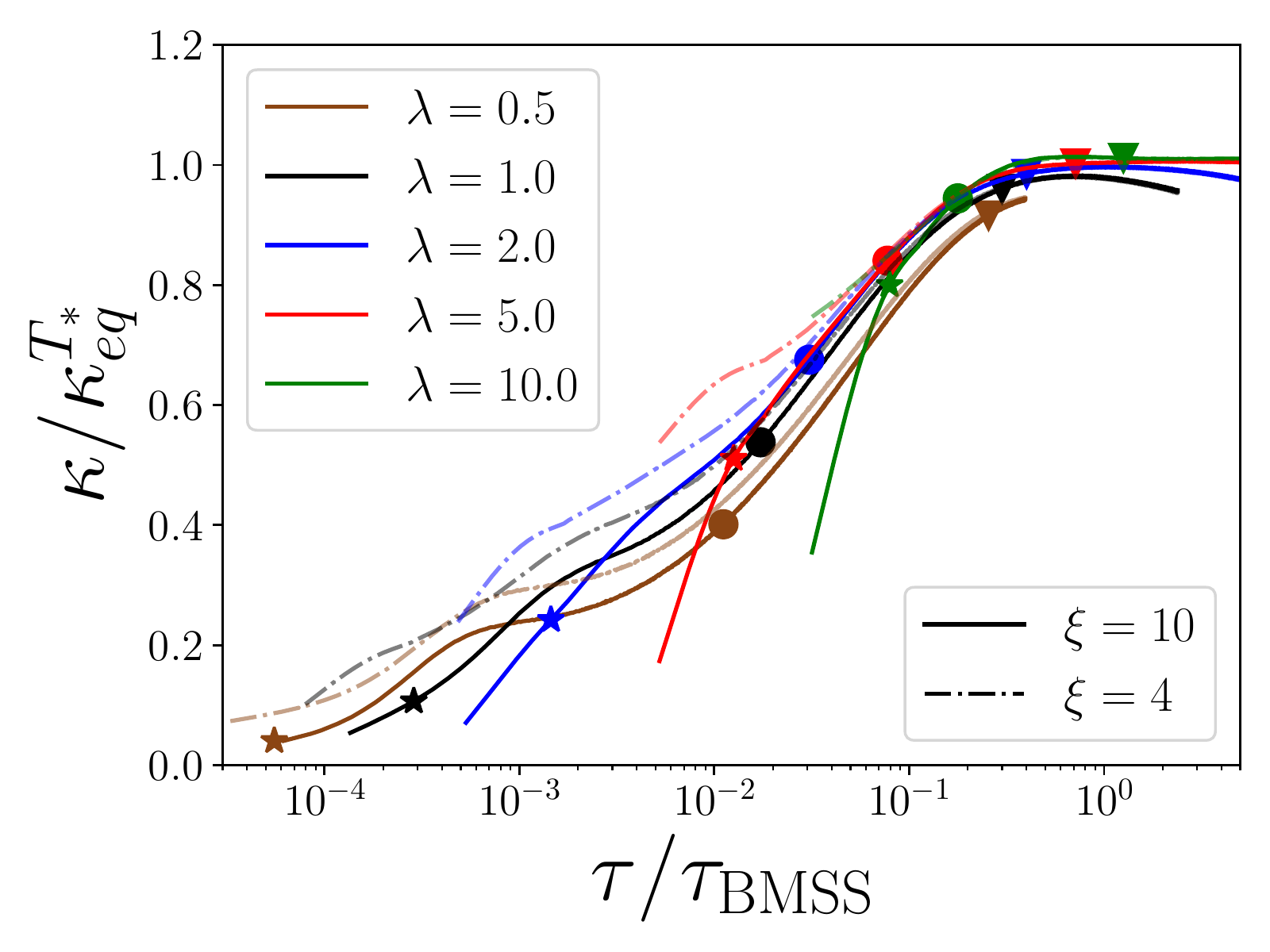}
 \caption{Ratio of $\kappa$ to the thermal value at the  same energy density $\varepsilon$ (top), the same  screening mass $\md$ (middle) and effective soft mode temperature $\tstar$ (bottom).
We have applied the Savitzky-Golay filter to smoothen the data. The filter is also applied to all the following figures involving $\kappa$.
}
\label{fig:kappa_vs_kappaeq}
\end{figure}

\begin{figure}[t]
  \includegraphics[scale=0.5]{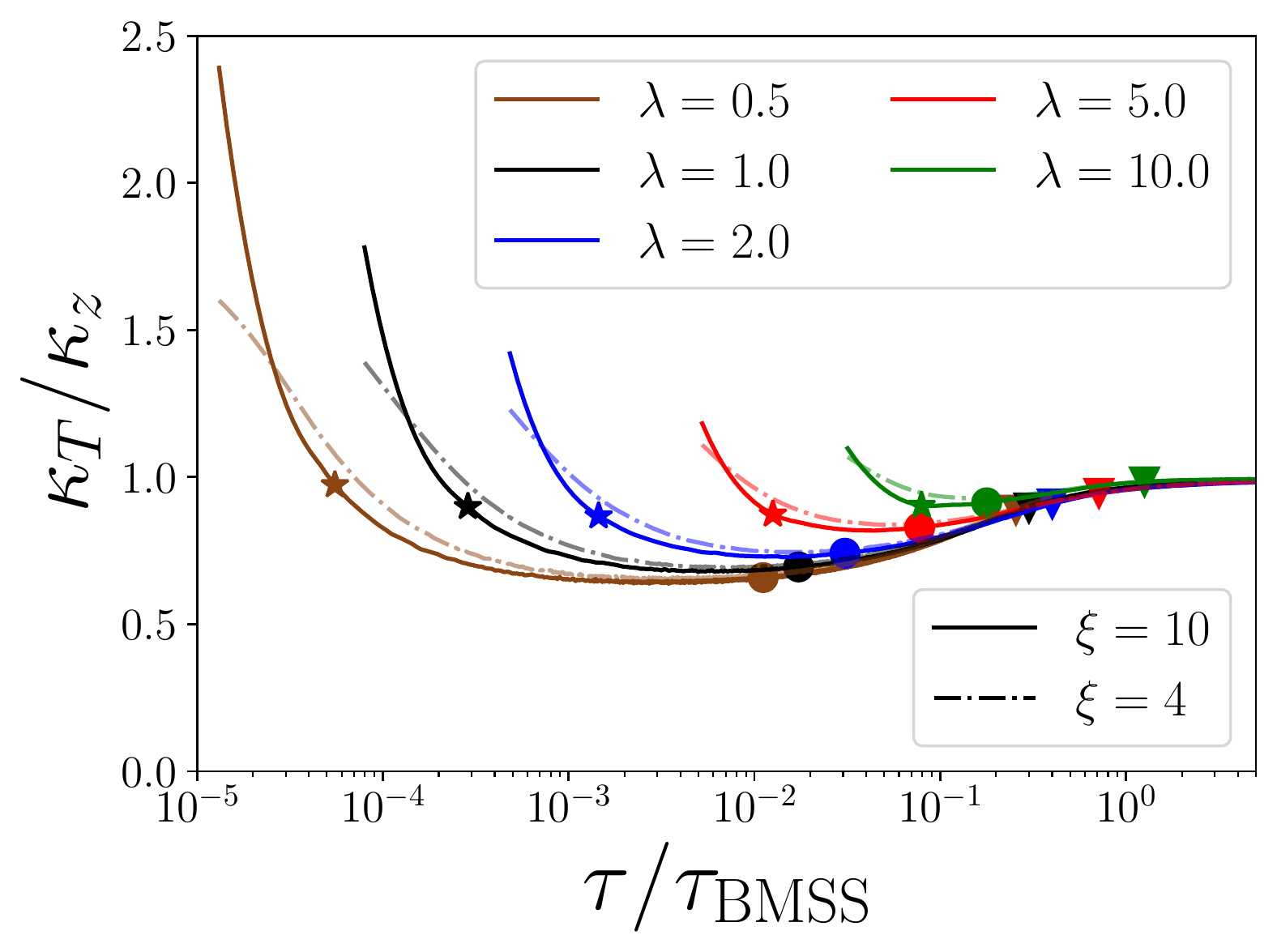}
  \caption{Ratio of the transverse and longitudinal diffusion coefficients during the bottom-up thermalization scenario. We observe that the transverse coefficient is enhanced compared to the longitudinal one during the initial evolution. When the system becomes underoccupied, this ordering reverses. 
  Finally when the system reaches approximate equilibrium the ratio approaches unity.}
\label{fig:KappaRatio}
\end{figure}

\begin{figure*}[tbh!]
\includegraphics[scale=0.47]{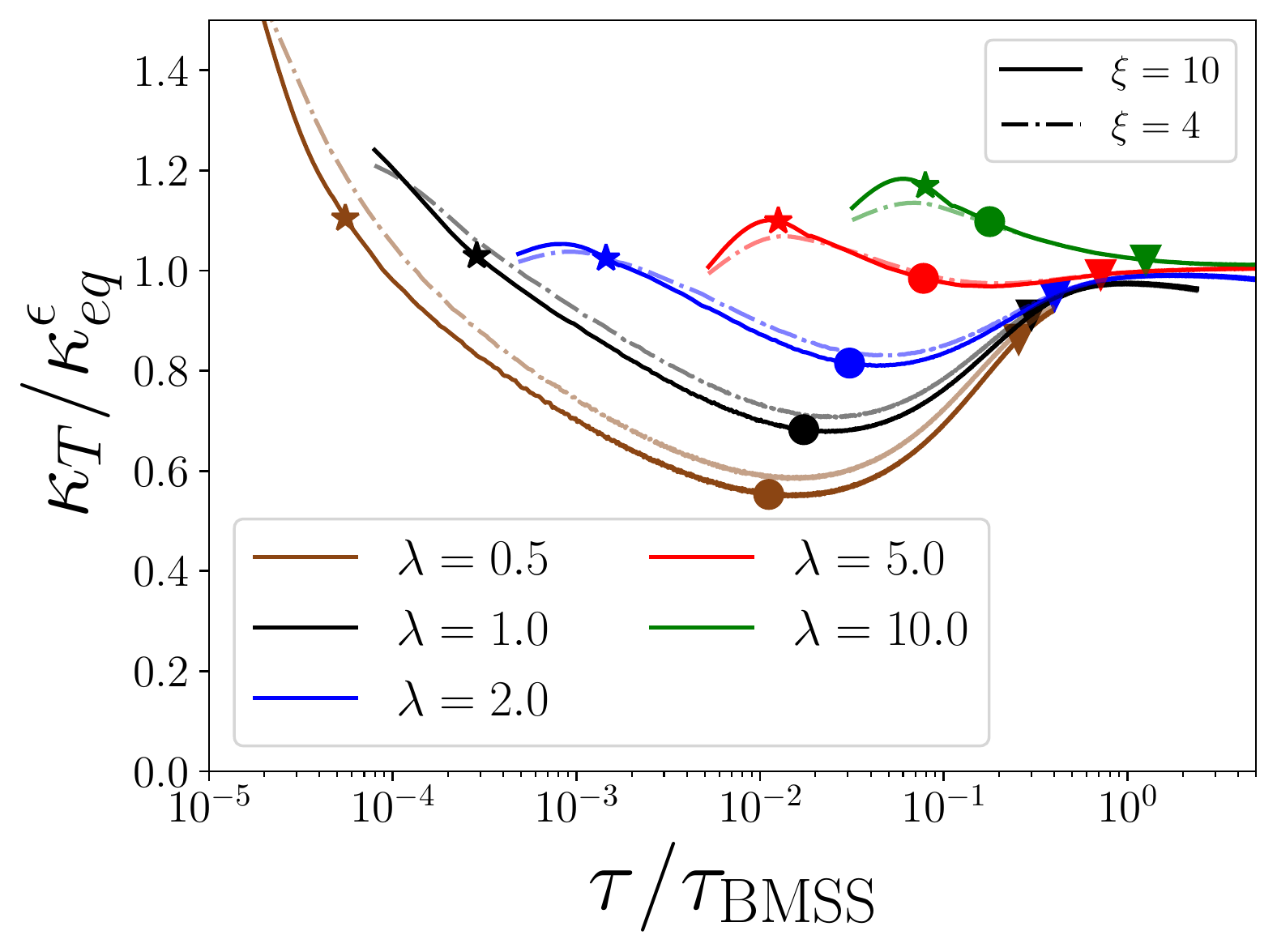}
\rule{3em}{0pt}
\includegraphics[scale=0.47]{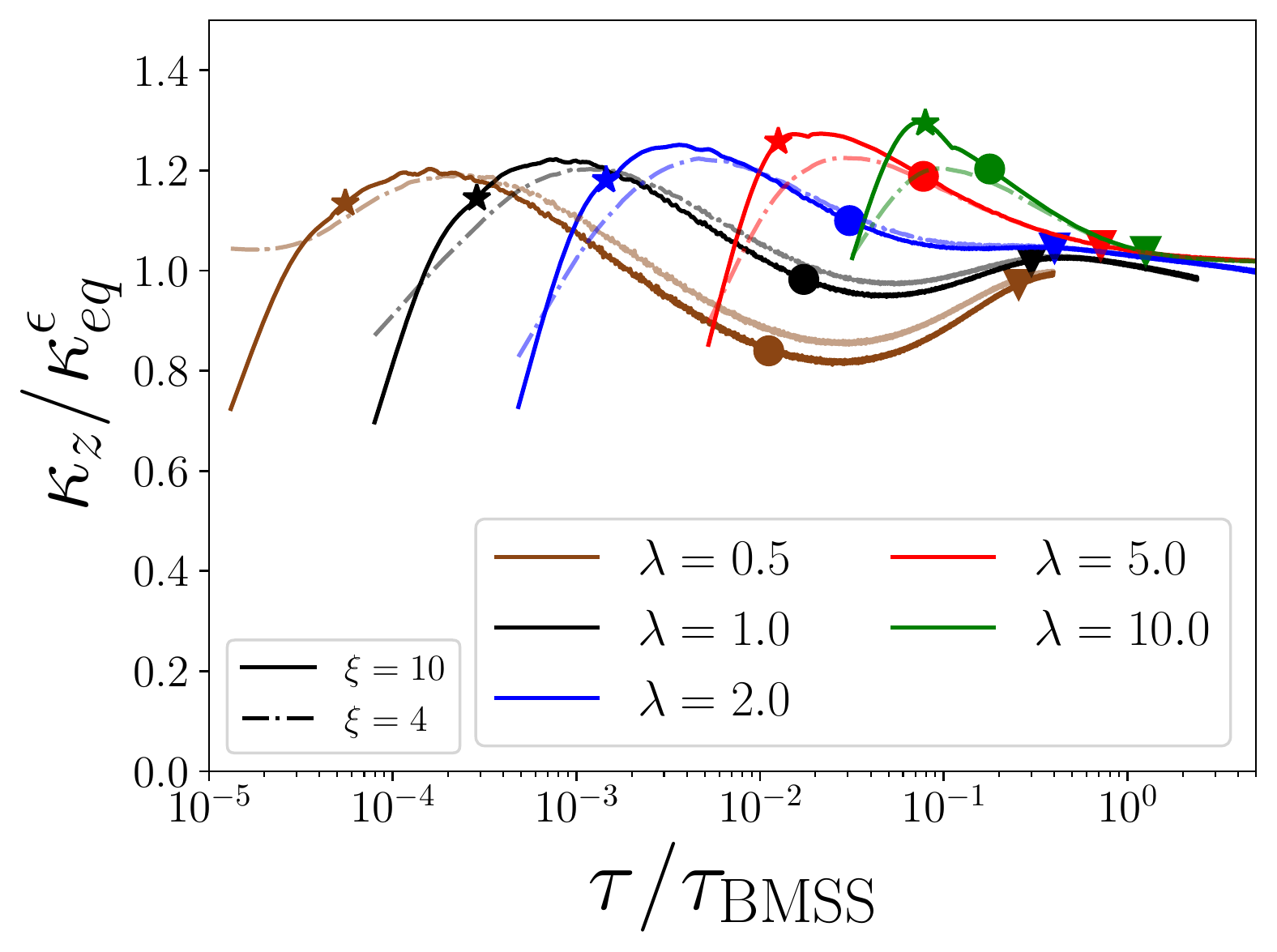}

\includegraphics[scale=0.47]{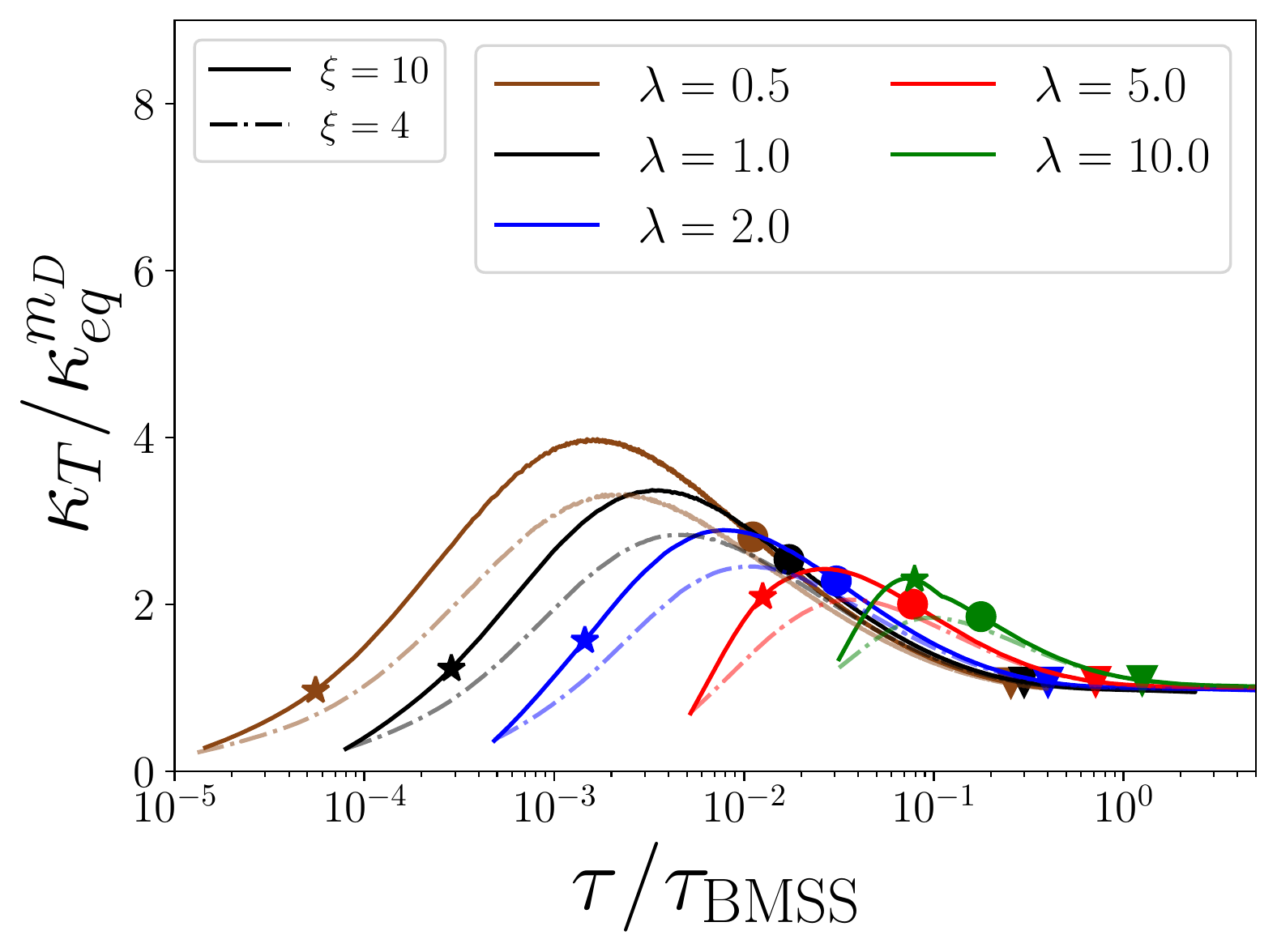}
\rule{3em}{0pt}
\includegraphics[scale=0.47]{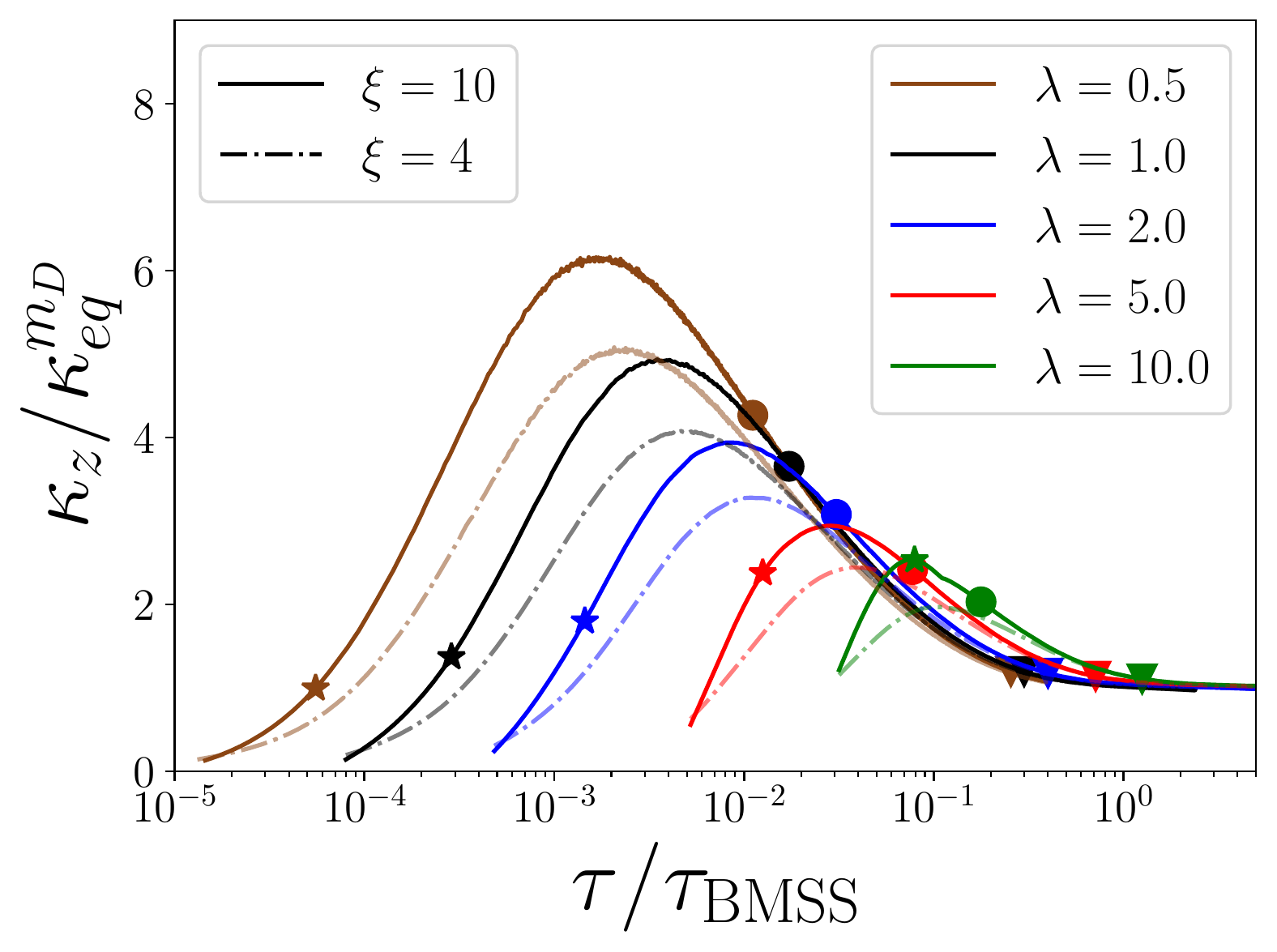}

  \includegraphics[scale=0.47]{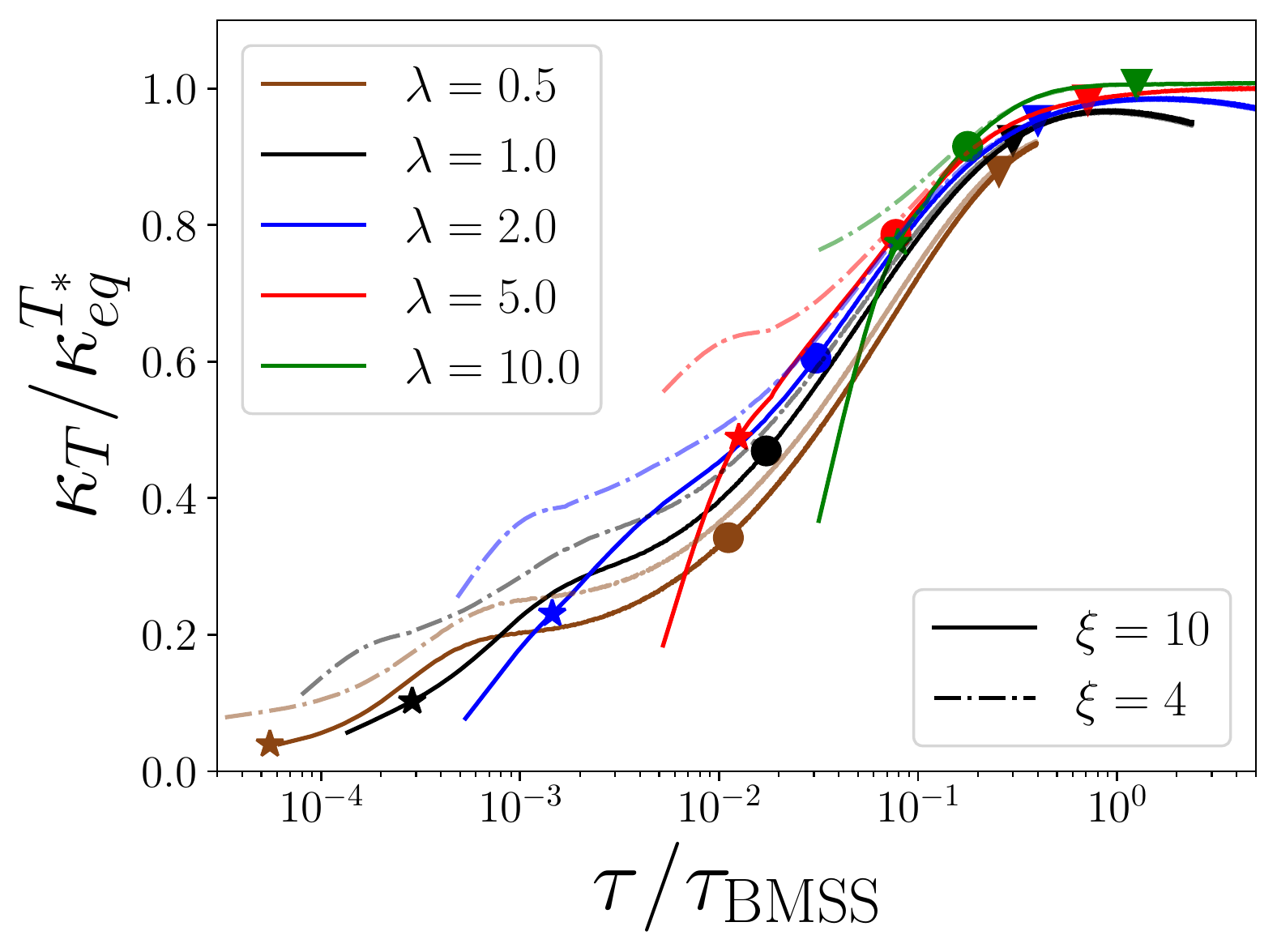}
\rule{3em}{0pt}
  \includegraphics[scale=0.47]{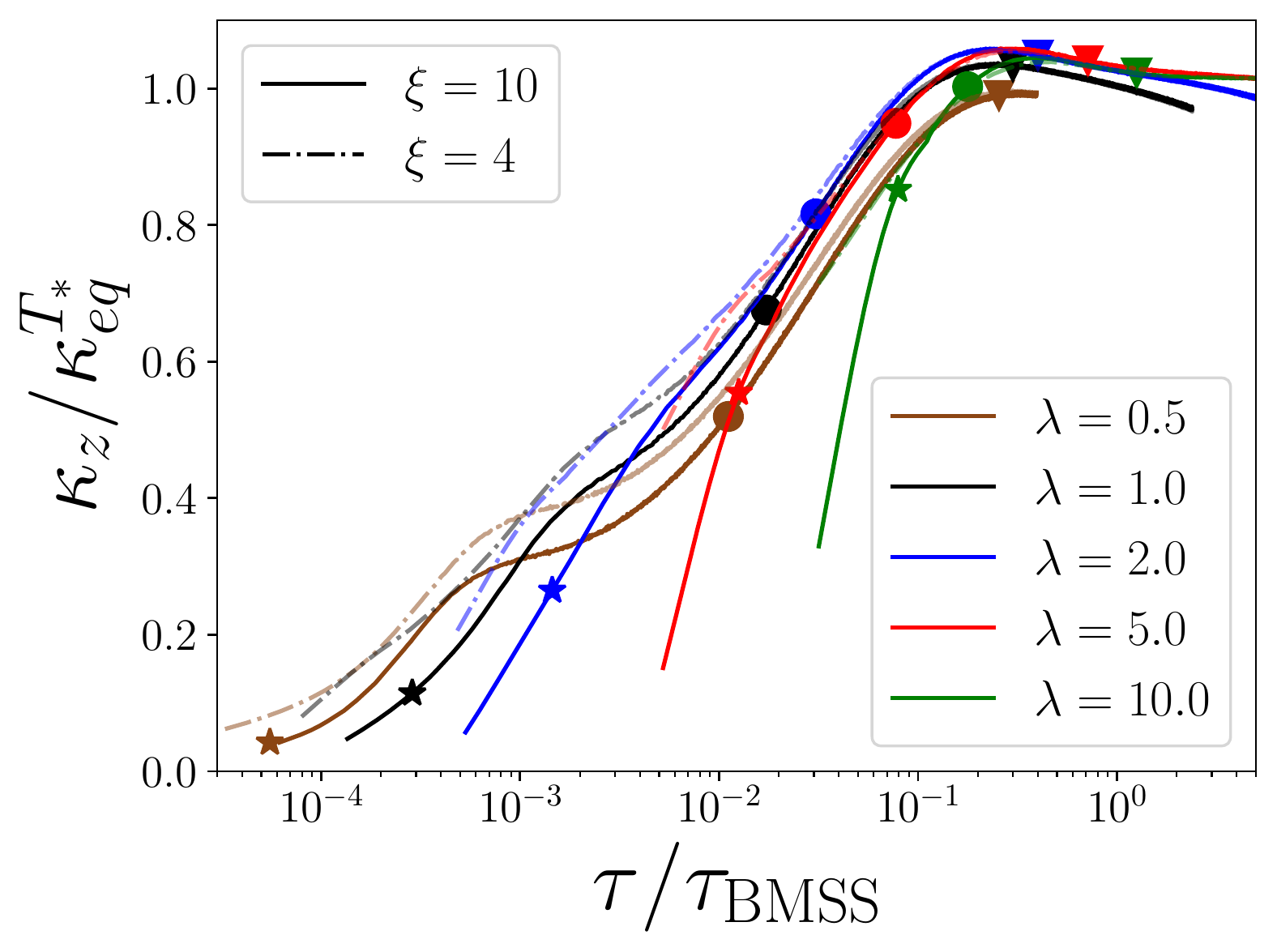}
\caption{
\label{fig:kappa_vs_kappaeq_TL}
Transverse (left panel) and longitudinal (right panel) diffusion coefficients compared to thermal values for the same $\varepsilon$ (top row), $\md$ (middle) and $\tstar$  (bottom row). 
}
\label{fig:kappa_vs_kappaeq_translong_tstarmd}
\end{figure*}

\subsection{Comparing nonequilibrium to equilibrium}
\label{sec:matching}
Our aim in this paper is to calculate 
the heavy quark momentum diffusion coefficient $\kappa$ during the hydrodynamization process 
in order to eventually assess the importance and impact of the initial nonequilibrium evolution on heavy quark observables. To facilitate the quantitative interpretation we will mostly present our results as ratios to the thermal equilibrium values. There is no unique method to compare equilibrium and nonequilibrium systems. A reasonable way to construct such a comparison is to match a dimensionful observable to its thermal counterpart. 
We will consider three different quantities - the energy density $\eps$ (which leads to the temperature defined in \eqref{eq:sensibleTemperature}), the temperature of infrared modes $\tstar$, and the screening mass $\md$. We have chosen these observables 
because they are straightforward to compute both in equilibrium and out of equilibrium  and are physical scales that play an important role in transport phenomena, as discussed in Sec.~\ref{se:MatchingEvolution}.
However, we would like to emphasize that the matching could be done with respect to any dimensionful observable that can be defined in and out of equilibrium.
The matching is performed by choosing the temperature of the thermal distribution to reproduce either the energy density  $\varepsilon$, the Debye mass $ \md$ or the effective temperature of the soft modes $\tstar$ calculated from the EKT simulation, and then calculating the diffusion coefficient $\kappa$ with this thermal distribution.

\subsection{Diffusion coefficient in and out of equilibrium}
\label{se:kappaInOutEqu}

Figure \ref{fig:kappa_vs_kappaeq} shows our result for the heavy quark diffusion coefficient during bottom-up thermalization. 
we observe an agreement between the equilibrium and thermal dynamics within $30\%$ even at early times, while at late times the system thermalizes and the ratio thus approaches unity. 
This is one of the main results in this paper.
Matching the  screening mass $\md$ (\ref{fig:kappa_vs_kappaeq} middle) or the effective infrared temperature $\tstar$ (\ref{fig:kappa_vs_kappaeq} bottom), $\kappa$ is much further away from the thermal system. The main observation is that at early times the non-equilibrium diffusion coefficient is considerably larger than the equilibrium coefficient for the same $\md$, and considerably smaller than that for the same $\tstar$.

\subsection{Transverse and longitudinal diffusion coefficient}
\label{se:kappaTransLong}

During most of the bottom up thermalization procedure the system is highly anisotropic.  This could have a significant effect on experimental observables sensitive to the initial stage of the evolution. It is therefore interesting to study also the anisotropy of the diffusion coefficient, which we parametrize in terms of the ratio of the transverse diffusion coefficient  $\kappa_T$ to the longitudinal one $\kappa_z$. Our results for this quantity are shown in \fig \ref{fig:KappaRatio}. In the initial overoccupied phase the transverse diffusion coefficient dominates.  
This can naturally be explained by the Bose enhancement, which at the early overoccupied and highly anisotropic stage benefits scatterings where only transverse momentum is exchanged.
As the system becomes underoccupied, the longitudinal diffusion coefficient becomes dominant. When the system approaches equilibrium the two become equal again, as expected due to the emerging isotropization. Thus the hierarchy of the diffusion coefficients has a natural explanation in terms of the stages of the bottom-up thermalization scenario. The ordering of the coefficients $\kappa_T < \kappa_z$ after the star marker, i.e., after the initially large occupancies,
is also in line with what is observed using squeezed thermal distributions \cite{Romatschke:2006bb} and results from the momentum anisotropy of the system.

We can compare the pre-equilibrium system to the thermal one using the same three matching procedures as in \fig \ref{fig:kappa_vs_kappaeq} to the transverse and longitudinal diffusion coefficients separately. The results are shown in 
\fig \ref{fig:kappa_vs_kappaeq_TL},
and show qualitatively similar results as for the full coefficient $\kappa$. Similarly as in \fig \ref{fig:kappa_vs_kappaeq} we find that the best way to compare our nonequilibrium results to a thermal system is by matching for the same energy density $\varepsilon$. This way the deviations from equilibrium turn out to be not larger than approximately 40 \%.  For the same screening mass $\md$ we observe very large deviations after the initially highly occupied regime, i.e., after the star time marker. 
In particular, we observe 
deviations up to a factor of 4 for the transverse and 6 for the longitudinal coefficient, respectively, for the smallest coupling.

\begin{figure}[tbh!]
\centering
\includegraphics[width=0.47\textwidth]{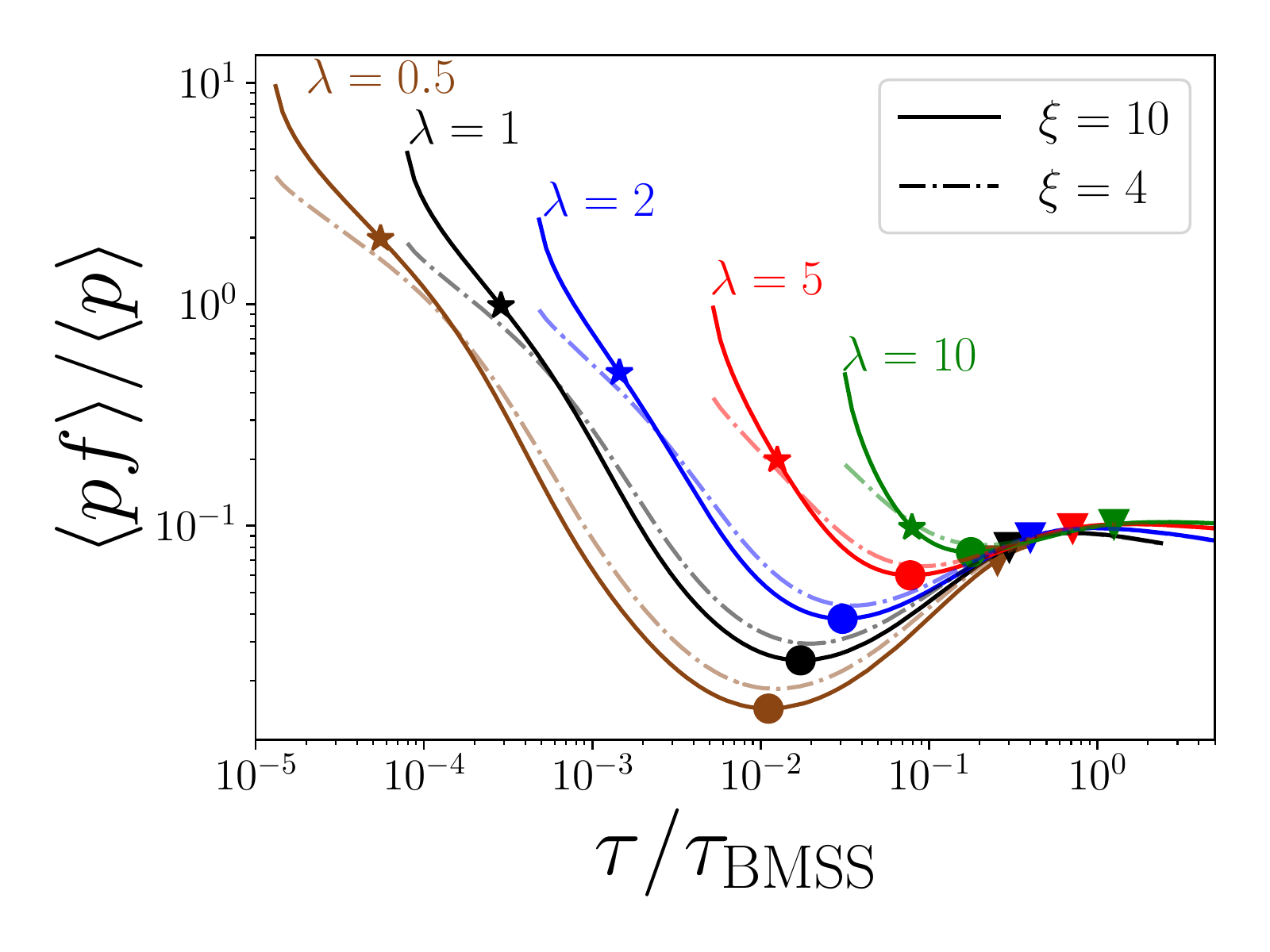}
\caption{ Time evolution of the occupation number of the hard modes. }
\label{fig:scalesVsEquil1}
\end{figure}
 
\begin{figure}[tbh!]
\centering
  \includegraphics[width=0.45\textwidth]{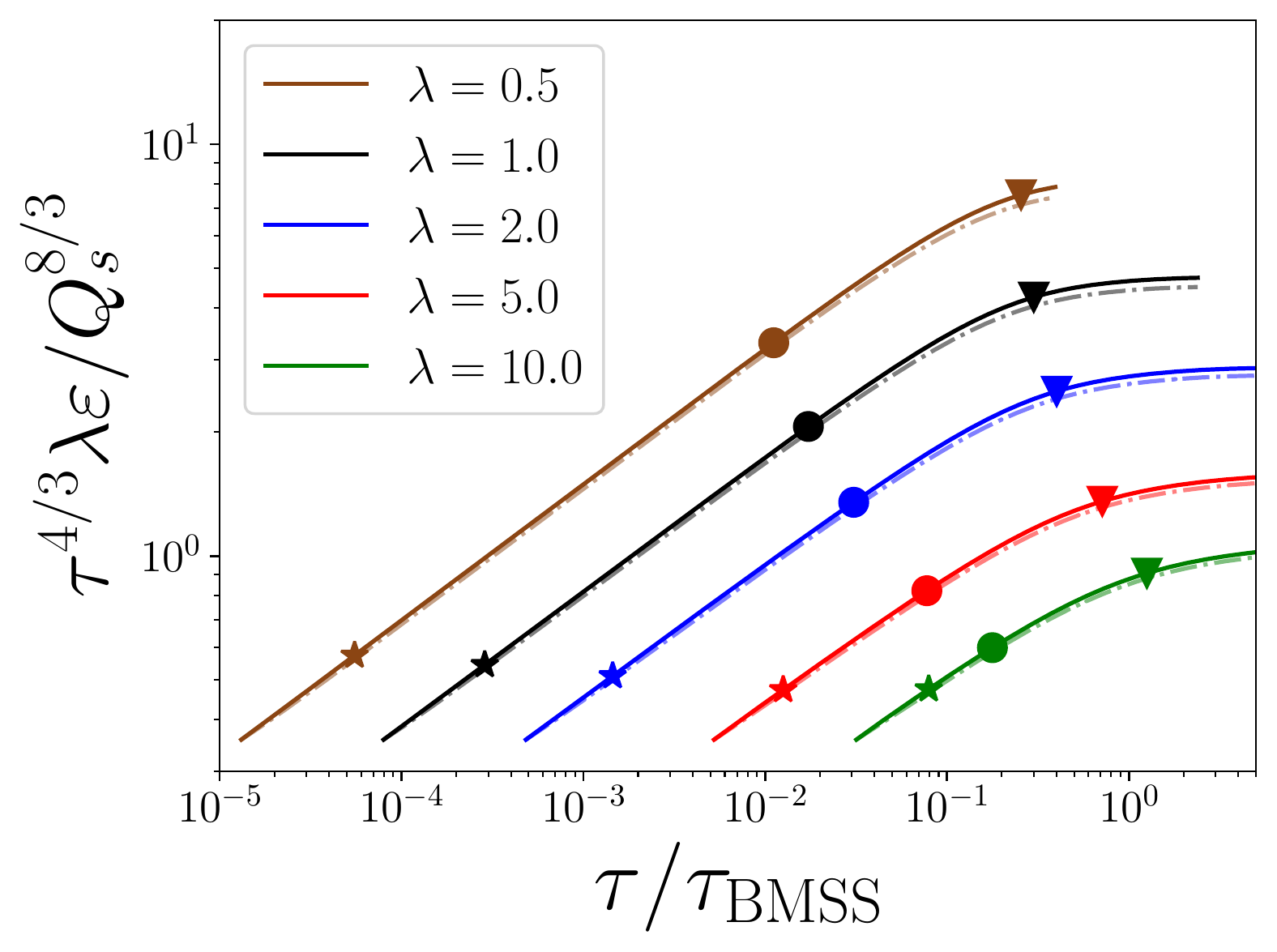}
  \includegraphics[width=0.45\textwidth]{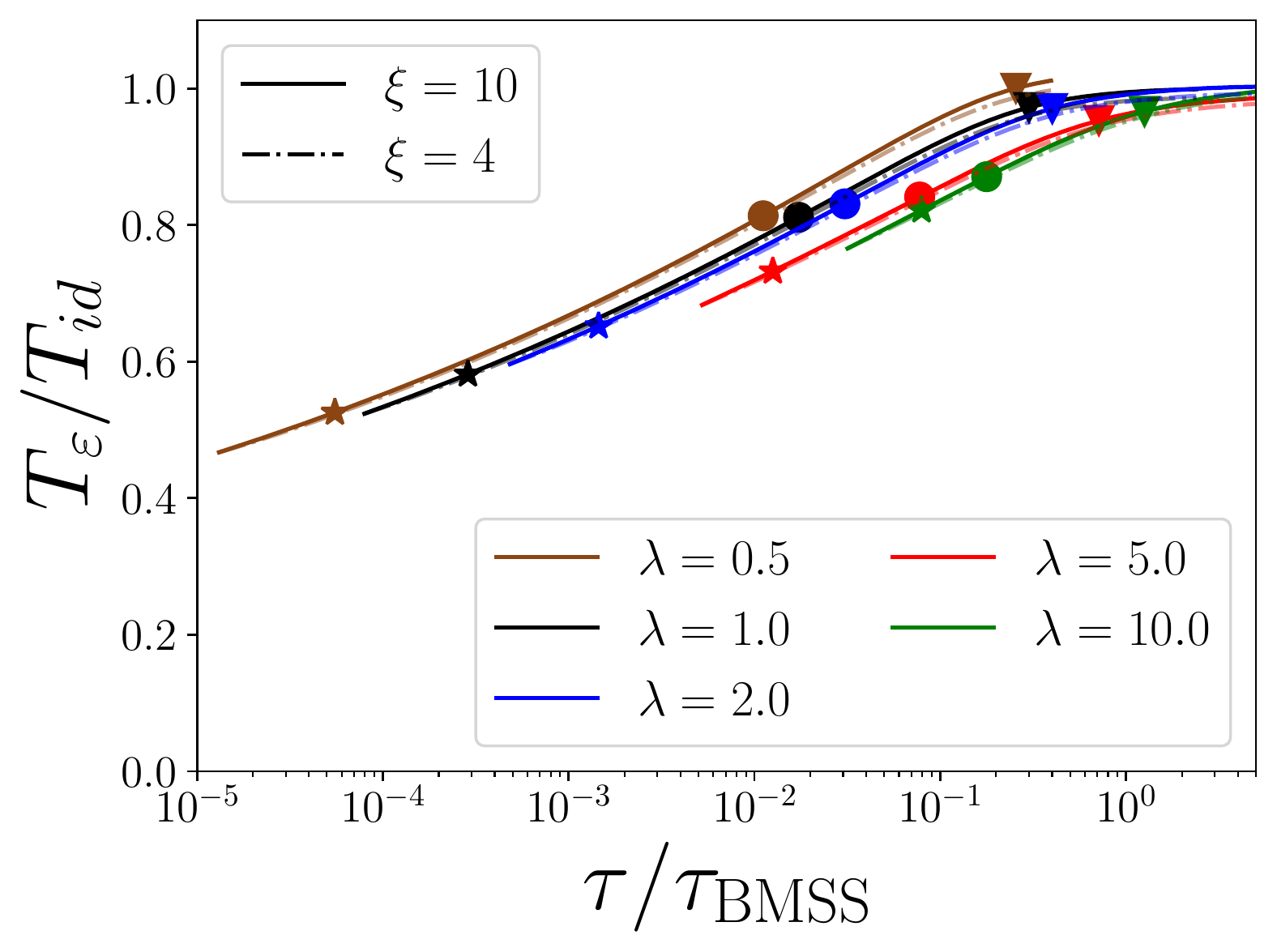}
\caption{ {\em Top:} Time evolution of the energy density $\varepsilon$ scaled by the coupling constant and the time dependence in ideal hydrodynamics $\varepsilon_{\textrm{id}} \sim \tau^{-4/3}$.  
{\em Bottom:} Time evolution of the temperature corresponding to the energy density, scaled by the time dependence of the temperature in ideal hydrodynamics $T_{\textrm{id}}\sim \tau^{-1/3}$.
}
\label{fig:scalesVsEquil2}
\end{figure}

\begin{figure}[tbh!]
\centering
  \includegraphics[width=0.46\textwidth]{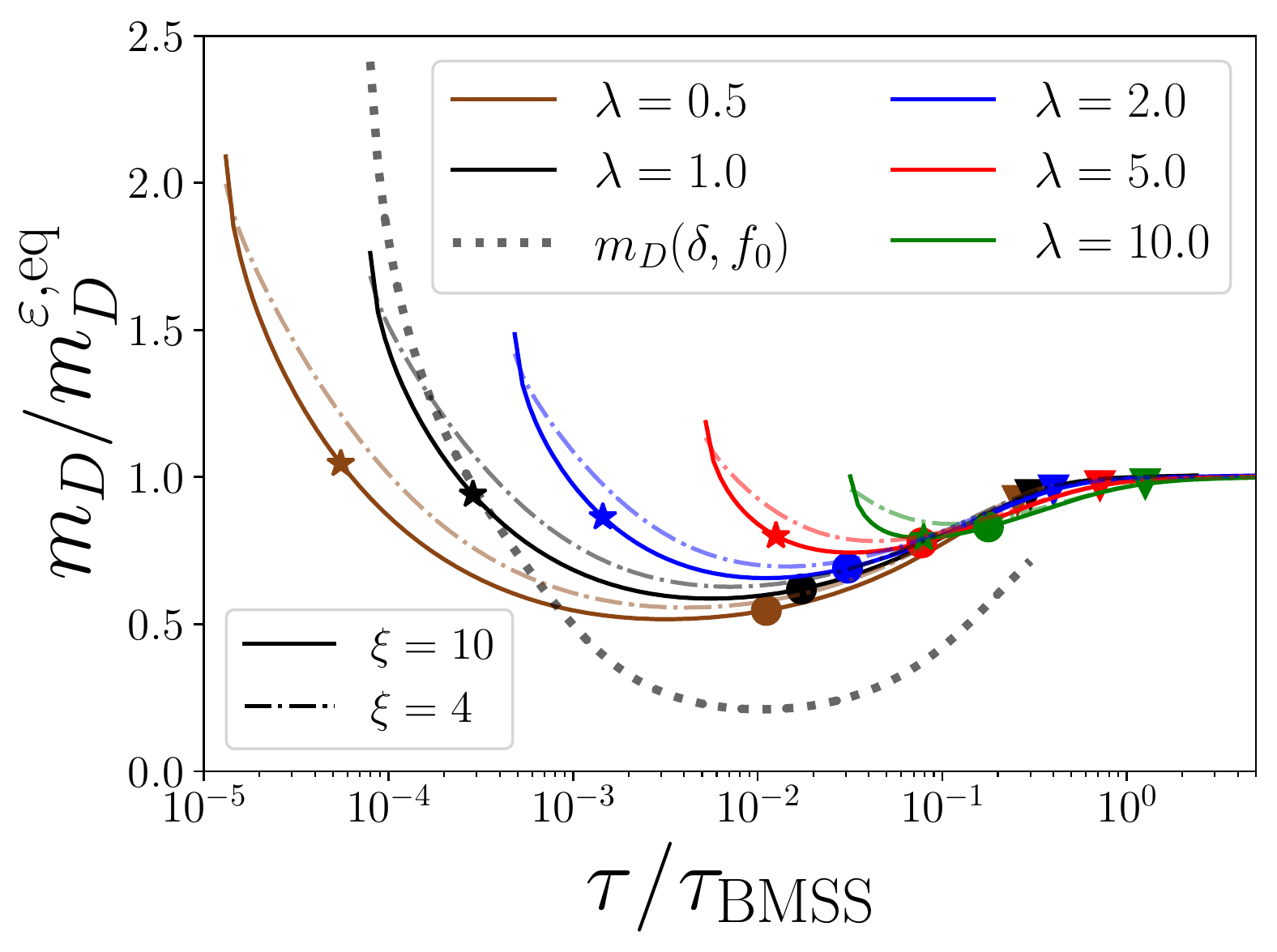}

  \includegraphics[width=0.45\textwidth]{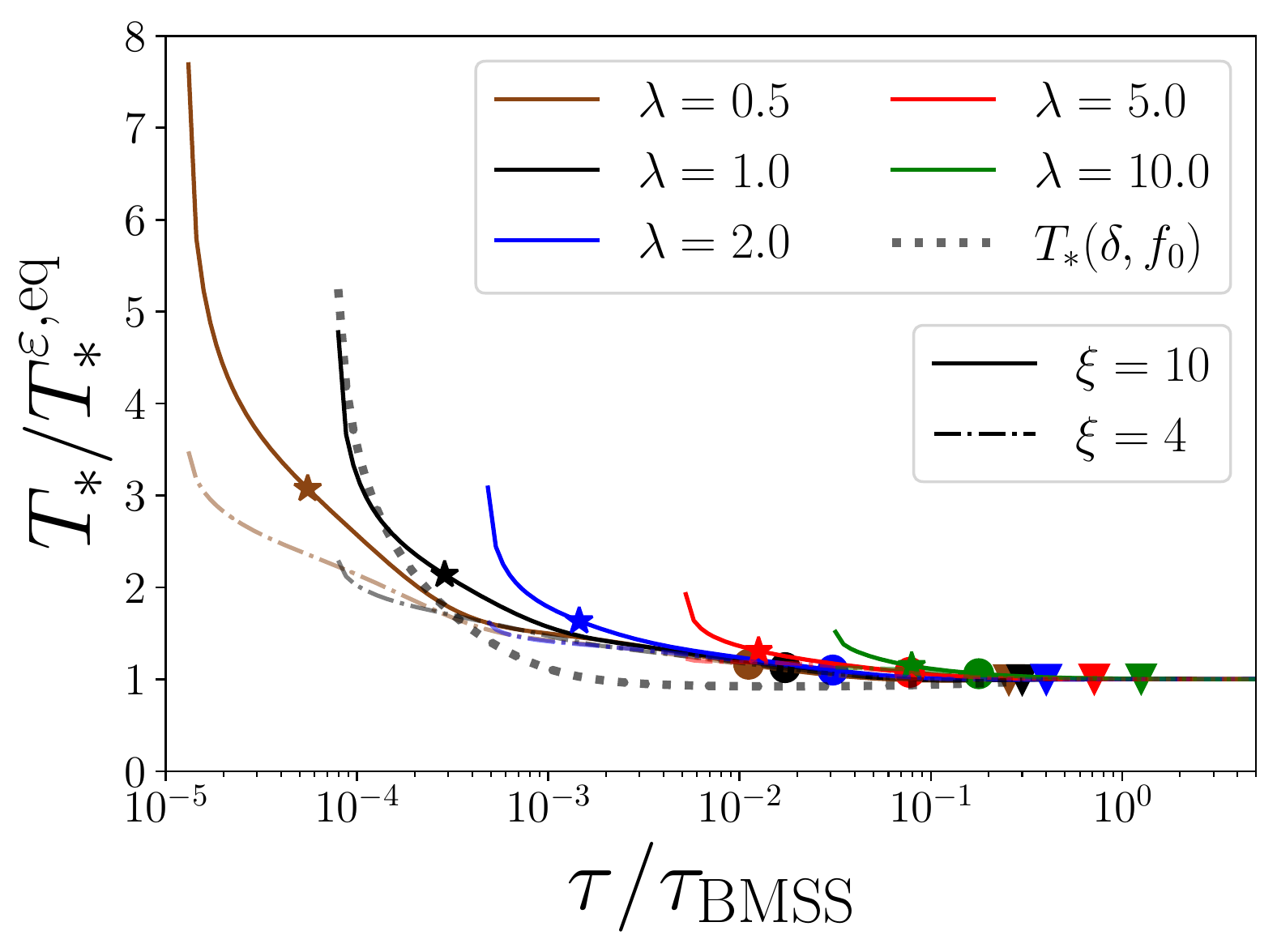}
\caption{Time evolution of the Debye mass $\md$ (top) and the effective temperature of the soft scales $\tstar$ (bottom), divided by the thermal value at the same energy density. 
The 
dotted curves correspond to the parametric estimates derived in \se \ref{se:paramest} corresponding to $\lambda=1, \xi=10$.
}
\label{fig:scalesVsEquil3}
\end{figure}

\subsection{Time-evolution of the relevant scales }
\label{se:MatchingEvolution}
To better understand the time dependence of $\kappa$, let us now discuss the evolution of relevant scales that it depends on. The different time dependences that are relevant for this discussion are summarized in \figs\ref{fig:scalesVsEquil1}, \ref{fig:scalesVsEquil2} and~\ref{fig:scalesVsEquil3}. As a first step, we start with the time dependence of the occupation number of the hard modes, shown in \fig\ref{fig:scalesVsEquil1}. It starts in an overoccupied state, then becomes underoccupied before reaching a thermal value. 

Figure \ref{fig:scalesVsEquil2} shows the time dependence of the energy density. In the upper panel, the initial value $\varepsilon\sim 1/\lambda$ is scaled out by multiplying with the coupling $\lambda$ so that all the curves start at the same value by construction. The time dependence is additionally divided by the ideal hydrodynamical behavior $\tau^{-4/3}$. We see that during most of the bottom-up thermalization, the energy density decreases as $\varepsilon\sim 1/\tau$ until it approaches the hydrodynamical behavior at the triangle marker near the end of the evolution. The bottom panel 
shows the temperature $T_\eps$ extracted from the energy density and the temperature in ideal hydrodynamics $T_{id}$, extrapolated backwards from the final state.  The latter is computed by first matching it with $T_\eps$ at the triangle time marker of near isotropy when the system exhibits an almost hydrodynamical behavior, and propagating backwards in time as in $T_{id} \sim \tau^{\nicefrac{-1}{3}}$, as discussed in \cite{Kurkela:2018oqw}. We find that the temperature deviates from the ideal hydrodynamics temperature at most by a factor of 2 at early times for the couplings we have considered. However, when the system is underoccupied at the circle time marker, the ideal temperature is in reasonably good agreement (within 20\%) with the temperature extracted from the energy density. 

Let us now return to the heavy-quark diffusion coefficient.
Parametrically, it is given by \cite{Boguslavski:2020tqz} 
\begin{align}
\kappa \sim \md^2\, g^2 \tstar \log \left(\frac{\Lambda}{\md} \right), 
\end{align}
where $\Lambda$ is the largest momentum scale of the particles in the plasma; for a thermal system we have $\Lambda\approx T$.
Ignoring the logarithmic contribution, we estimate
\begin{equation}
    \frac{\kappa}{\kappa_{eq}} \approx \frac{\md^2}{(\md^{eq})^2} \frac{\tstar}{\tstar^{eq}}.
    \label{eq:kappaRatioPocketFormula}
\end{equation}

To understand the behavior of the heavy quark diffusion coefficient, we must thus look at the time dependence of $\md$ and $\tstar$ which are shown on the top and bottom, respectively, in \fig \ref{fig:scalesVsEquil3}.
We observe that initially $\md$ is enhanced compared to the thermal system at the same energy density. When the system becomes underoccupied, especially for weak couplings, $\md$ gets relatively suppressed. The suppression is mainly driven by a decreasing occupation number, which is discussed in more detail in \se \ref{se:paramest}.
For the effective temperature $\tstar$ we observe a strong enhancement at the early stage, which is understood from the large occupancies encountered initially as also discussed in \se \ref{se:paramest}.
When the system becomes underoccupied, i.e., in the vicinity of the circle time marker, $\tstar$ is already very close to its equilibrium value for the same energy density. The combination of the curves in \fig\ref{fig:scalesVsEquil3} explains qualitatively
the behavior seen in \fig\ref{fig:kappa_vs_kappaeq}; at first an enhancement compared to the equilibrium value due to large 
$\tstar$ and $\md$ in the overoccupied phase, followed by a suppression mostly resulting 
from lower values of $\md$. Let us now construct a parametric estimate that makes this behavior explicit.

\begin{figure}[t]
\includegraphics[width=0.46\textwidth]{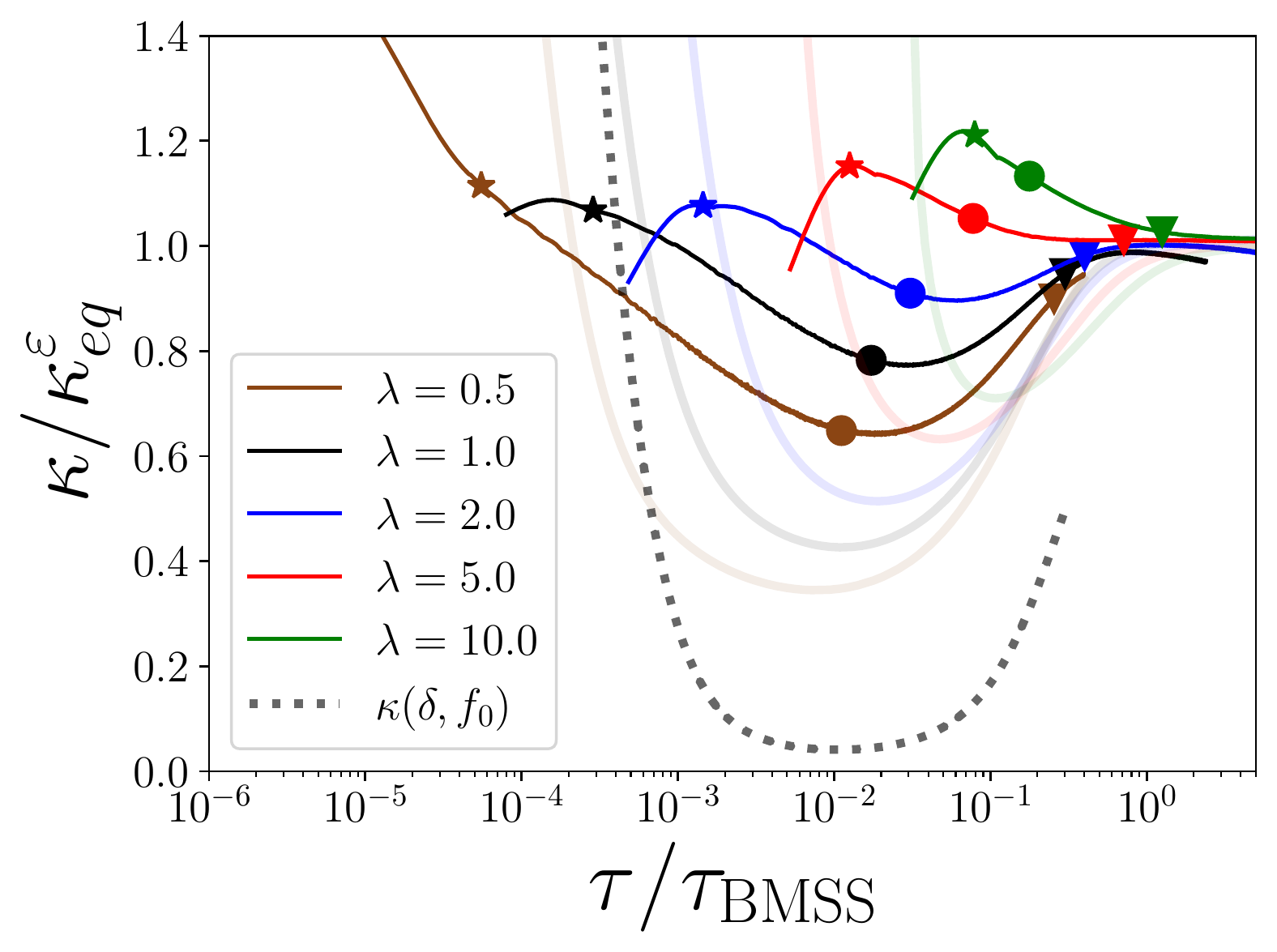}
\caption{Ratio of the diffusion coefficient to the equilibrium value for the same energy density. The wide transparent lines show the parametric estimate given by \eqref{eq:kappaRatioPocketFormula} using the extracted values for $\md$ and $\tstar$ depicted
in \fig \ref{fig:scalesVsEquil3}. The gray dotted line corresponds to the parametric estimate \eqref{eq:kappaguesstimate}.}
\label{fig:parametricGuesstimate}
\end{figure}

\subsection{Understanding the evolution of scales and $\kappa$}
\label{se:paramest}

Let us try to qualitatively understand the time evolution of the matching scales and of the heavy-quark diffusion coefficient in terms of the occupation number and anisotropy. For the purposes of our parametric estimate, we use the following squeezed and scaled distribution
\begin{equation}
\label{eq:f_params}
f(\pt, \pz) = f_0\, \theta \left(\Q^2 - (\pt^2 + \left( \nicefrac{\pz}{\delta}\right)^2 \right), 
\end{equation}
where $\theta$ is the step function. 
 We determine the typical occupation number $f_0$ during bottom-up thermalization by taking its value to be the typical occupation number of the hard modes, calculated as $f_0 = \nicefrac{\langle p f \rangle}{\langle p \rangle}$. The values of this parameter can be read off  \fig\ref{fig:scalesVsEquil1}.
The squeezing parameter $\delta$ describes the momentum anisotropy, $ \delta \sim \frac{\langle p_z \rangle}{\langle \pt \rangle }$. We estimate its value during the bottom-up evolution as $\delta = \sqrt{\nicefrac{P_L}{P_T}}$, where the values of the pressure ratio are visible in \fig\ref{fig:occupVsAnisotropy}.

For the purposes of this parametric estimate, we approximate the thermal distribution with the same simplified form \eqref{eq:f_params}. For this, we obtain the value $f_0$
by calculating the expectation values for a thermal (Bose-Einstein) distribution (corresponding to the crosses in \fig \ref{fig:occupVsAnisotropy}). The value is independent of $\lambda$ and reads
\begin{align}
    f_0^{\mathrm{BE}} = \frac{\langle p f \rangle}{\langle p \rangle} = 0.1106.
\end{align}
For a thermal system, we naturally have $\delta=1$, and we take the scale $\Q$ in \eqref{eq:f_params} to be $\Q=T$. 

The screening mass $\md$ is given by \eqref{eq:mDdefinition}. Performing the integral for the distribution \eqref{eq:f_params}, 
we get 
\begin{align}
   \md^2 \sim \delta  f_0 \lambda  \Q^2.
\end{align}
Thus, the ratio to the thermal case becomes
\begin{equation}
    \frac{\md}{\md^{eq}} = \sqrt{\delta \frac{f_0}{f_0^{\mathrm{BE}}}}\,.
    \label{eq:mdparam}
\end{equation}
This parametric estimate is shown as a gray dotted line in \fig \ref{fig:scalesVsEquil3} (top)  for $\lambda=1$ (as indicated by the color coding) and $\xi=10$. We will use these parameters throughout this section for other quantities as well. The time-evolution in \fig \ref{fig:scalesVsEquil3} can now be understood using \fig \ref{fig:occupVsAnisotropy}. The decrease in the value of $\md$ when the system evolves from overoccupation to underoccupation is mainly driven by the falling occupation number $f_0$. As the system evolves towards thermal equilibrium from the underoccupation regime (circle marker), the value of $\md$ starts to increase. This process is driven by both the growing occupation number and decreasing anisotropy. As can be seen in \fig \ref{fig:scalesVsEquil3}, this estimate does not describe the evolution quantitatively, but offers a simple qualitative description of the evolution. Possibly our parametric estimate undershoots the actual value of $\md$ in the underoccupied phase (the circle marker) because the modes contributing to the Debye mass are softer and not quite as anisotropic and underoccupied as the modes contributing to the pressure, leading to the factor $\sqrt{\delta f_0}$ in \eqref{eq:mdparam} to overestimate the effect.

Similar estimates can be applied to $\tstar$ given by \eqref{eq:Tstardef}. Using the result for $\md$, we obtain
\begin{equation}
    \tstar \sim \Q \left(f_{0} + 1\right)\,.
\end{equation}
Thus $\tstar$ is less sensitive to the anisotropy.  Comparing with the thermal estimate, we get 
\begin{equation}
    \frac{\tstar}{\tstar^{eq}} =  \frac{\left( f_0 +1\right)}{\left(  f_0^{\mathrm{BE}} + 1 \right)}.
\end{equation}
This result naturally explains the behavior of $\tstar$ observed in \fig \ref{fig:scalesVsEquil3} (bottom). The  initial enhancement of $\tstar$ is driven by the large occupation number, and the observed enhancement is larger than for $\md$ since $\tstar$ has a stronger dependence on the occupation number.
Due to the constant term, there is no suppression from underoccupation, and $\tstar$ is less sensitive to $\delta$ than $\md$. Hence, our parametric estimate is better than for $\md$, and $\tstar$ does not go below the thermal value in the underoccupied regime, but is already close to unity.

Using the estimates for $\md$ and $\tstar$, we obtain for the heavy-quark diffusion coefficient
\begin{align}
\label{eq:kappaguesstimate}
    \frac{\kappa}{\kappa_{eq}^\epsilon} =  \frac{\delta f_{0}}{f_0^{\mathrm{BE}}} \frac{\left( f_{0} + 1\right)}{\left( f_0^{\mathrm{BE}} + 1 \right)}\,.
\end{align}
Plugging in the values for $f_0$ and $\delta$ during the evolution  (calculated from $f_0$ and $\delta$ in  \figs \ref{fig:occupVsAnisotropy} and \ref{fig:scalesVsEquil1}) leads to the gray dotted curve in \fig  \ref{fig:parametricGuesstimate}. We observe that the general trend of enhancement followed by suppression and equilibration is somewhat exaggerated by this parametric estimate. 
The transparent curves in \fig  \ref{fig:parametricGuesstimate} show the estimate obtained using \eq\nr{eq:kappaRatioPocketFormula} with the actual calculated values of $\md$ and $\tstar$ from \fig\ref{fig:scalesVsEquil3}. They exhibit the same behavior in a less exaggerated way.

\FloatBarrier

\section{Comparisons with glasma and lattice calculations}
\label{sec:comparisons}
\subsection{Comparing with glasma}

For a sensible comparison between our kinetic results and $\kappa$ from the glasma stage, we first reproduce the energy density of the glasma by choosing $\Q=1.4~\mathrm{GeV}$ as in Ref.~\cite{Ipp:2020nfu} at the initial time $\Q\tau=1$.
The same value is also obtained in \cite{Keegan:2016cpi} in order to achieve consistency with the later hydrodynamic evolution. 

\begin{figure}[t]
\includegraphics[width=0.49\textwidth]{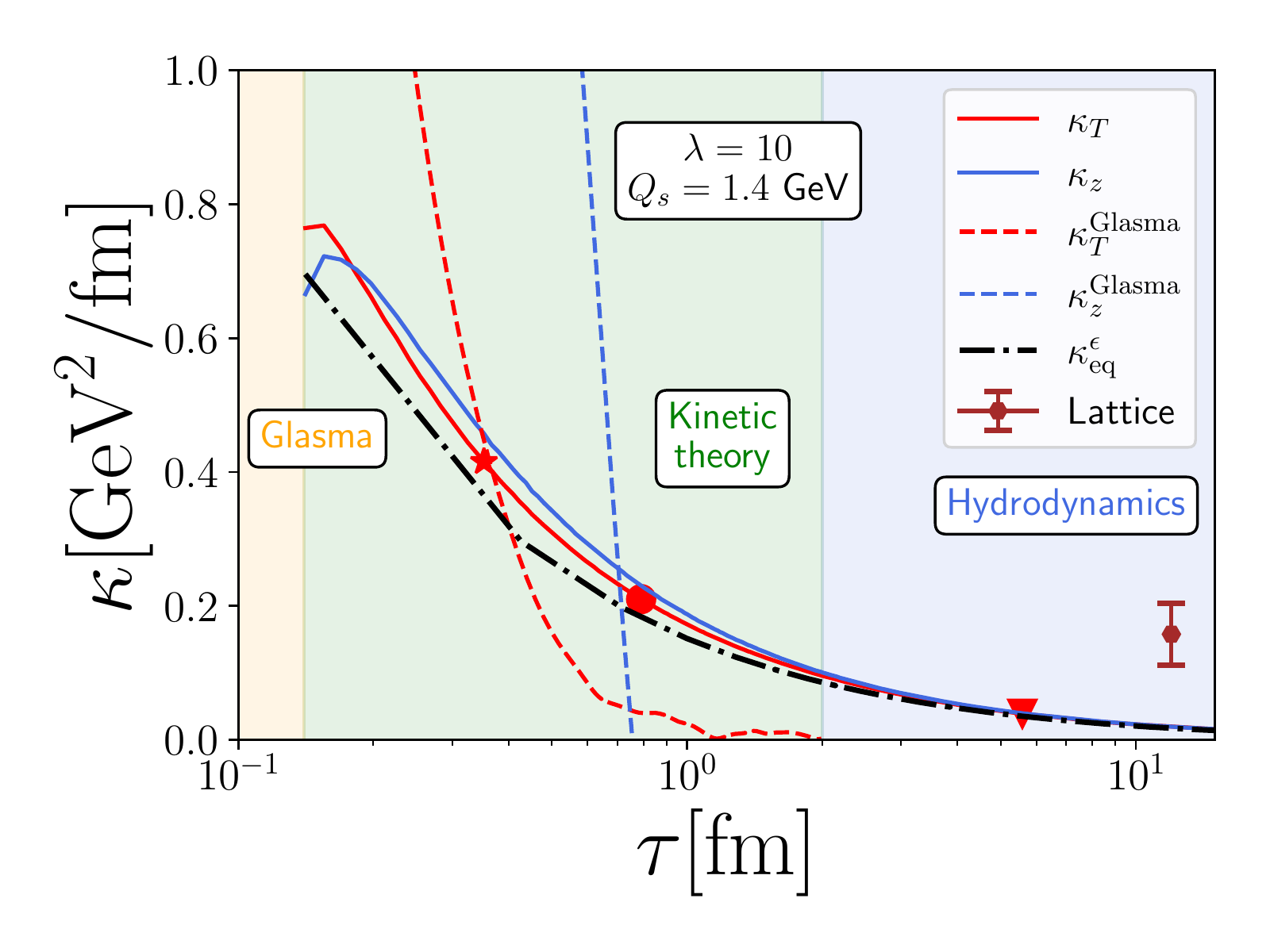}
\caption{Comparison of our results with the glasma results from \cite{Avramescu:2023vld}. Blue curves correspond to the longitudinal diffusion coefficient and red curves to the transverse diffusion coefficient. Dashed curves correspond to glasma results and solid curves to our results. The brown data point with error bars corresponds to the result of \cite{Brambilla:2022xbd} at $T=1.5~T_c$. The point is placed at such a value of $\tau$ that the nonequilibrium system has the same temperature $T_\varepsilon$ defined through energy density as the lattice system.
}
\label{fig:EKTvsGlasma}
\end{figure}

Figure \ref{fig:EKTvsGlasma} shows the transverse and longitudinal diffusion coefficients $\kappa_T^{\text{Glasma}}$ and $\kappa_z^{\text{Glasma}}$, which correspond to the situation of static quarks in the glasma \cite{Avramescu:2023vld}, 
together with our results denoted by $\kappa_T$ and $\kappa_z$. The main observation is that during the quasiparticle phase, $\kappa$ is considerably smaller than during the glasma stage at very early times (note that the glasma results peak at $\mathcal{O}(10) \frac{\mathrm{GeV}^2}{\mathrm{fm}}$). Transverse and longitudinal diffusion coefficients behave differently -- our result for $\kappa_T$ agrees with the glasma around the star marker signaling large occupation numbers, which is within the overlapping validity range of glasma and EKT. In contrast, the longitudinal coefficients intersect close to the circle marker, and the transition is not smooth. The circle marker indicates that the system is underoccupied, and hence the matching time falls outside of the validity range of classical-statistical simulations underlying the glasma description. 

The main reason for this discrepancy is the qualitatively different behavior of $\kappa$ in the glasma framework. We can always define the derivative of momentum broadening and call it ``diffusion coefficient'' $\kappa$. This, however, does not imply that the underlying behavior is that of diffusion, corresponding to Langevin-type behavior for $\langle p^2 \rangle$. In the glasma, this manifests itself in the longitudinal direction, where for large $\tau$ momentum broadening turns into momentum narrowing ($\kappa_z^{\text{Glasma}}$ becomes negative). Thus, more research is needed to understand how the transition from nondiffusive to diffusive behavior takes place between the glasma and quasiparticle pictures.

Based on the data presented in \fig \ref{fig:EKTvsGlasma}, we can also estimate the total effect of momentum broadening during the nonequilibrium evolution. It is given by the integrated diffusion coefficient
\begin{equation}
    \langle p^2 \rangle = \int \der \tau~ 3 \kappa(\tau).
\end{equation}
In the Bjorken hydro limit, we have $\varepsilon \sim \tau^{-4/3}$ and $T_\varepsilon \sim \tau^{-1/3}.$ In a rough parametric estimate $\kappa \sim T^3$, this leads to  $\kappa \sim \nicefrac{1}{\tau}$ and $\langle p^2 \rangle \sim \log(\tau)$, everything in units of $\qs$. Thus, we expect the pre-equilibrium kinetic phase of the evolution, roughly $0.1-1~\mathrm{fm}$, to have an equal contribution to heavy quark diffusion as the equilibrium phase, roughly $1-10~\mathrm{fm}$. 
Integrating over the entire EKT evolution, from $\tau = 0.14~\mathrm{fm}$ to $\tau = 1~\mathrm{fm}$ yields an estimate $\langle p^2 \rangle = 0.9~\mathrm{GeV}^2$.

\fig \ref{fig:EKTvsGlasma} also features a datapoint for the lattice result, which we will discuss in the following section \ref{sec:latticecomp}. 
 
\subsection{Comparing with lattice}
\label{sec:latticecomp}

\begin{figure}[t]
\includegraphics[width=0.46\textwidth]{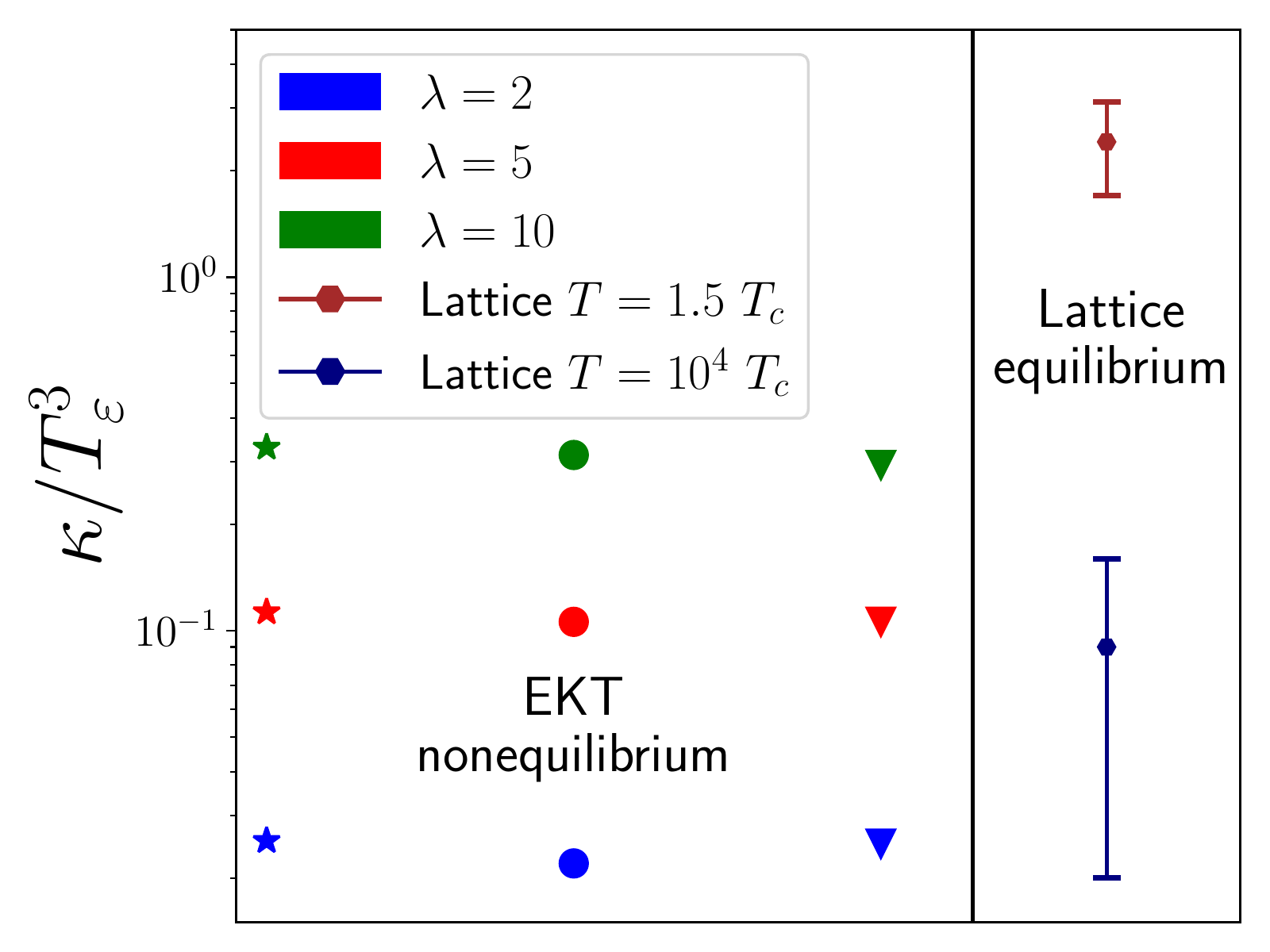}
\caption{Our results for $\lambda=2,5,10$ together with the lattice results from \cite{Brambilla:2022xbd}. The star, circle and triangle markers correspond to the values of $\nicefrac{\kappa}{T_\varepsilon^3}$ at the time when the corresponding phase of the bottom-up evolution is reached. We only consider the static correlator without higher order (magnetic) corrections here. 
}
\label{fig:latticecomp}
\end{figure}

For comparisons with lattice results, our main reference will be \cite{Brambilla:2022xbd}, where low-temperature ($T= 1.5~T_c$) and high-temperature ($T= 10^4~T_c$) estimates for $\nicefrac{\kappa}{T^3}$ are available. Other studies have also obtained comparable results at similar temperatures \cite{Banerjee:2011ra,Banerjee:2022uge,Brambilla:2020siz}. 

The obvious approach is to compare systems with the same temperature (defined by the energy density). This, however, neglects the different treatment of the QCD coupling. Since the lattice calculation is nonperturbative, it will also include effects arising from the running coupling, whereas our EKT calculation uses a fixed coupling. The second approach is to compare the systems with the same (or similar) coupling constant, which can be estimated using 
\begin{align}
    \lambda(Q)=\dfrac{4 \pi N_c}{\dfrac{33-3 N_{f}}{12 \pi} \ln \dfrac{Q^2}{\Lambda_{\mathrm{QCD}}^2}}
    \label{eq:RC}
\end{align}
at the scale $Q = 4\pi T$, 
 $N_f=0$, and  $\Lambda_{\rm QCD}/T_c \approx 2$.

 Figure \ref{fig:latticecomp} shows our EKT results for $\lambda=2,~5,~10$ as well as lattice results at $T= 1.5~T_c$ and $T= 10^4~T_c$ in terms of the ratio $\nicefrac{\kappa}{T^3}$. 
 For the higher temperature, the above expression gives $\lambda \approx 2$, whereas the lower temperature -- while being outside of the scope of the one-loop expression ---  indicates values roughly around $\lambda \sim 10$. 
 The EKT values are extracted at different stages of the non-equilibrium evolution, marked by the respective markers. 
 We observe that our results for $\lambda = 2$, 
 $\nicefrac{\kappa}{T^3} = 0.024 \pm 0.002$ 
 are in rough agreement with the lattice estimate $ 0.02 \leq \nicefrac{\kappa}{T^3} \leq 0.16$ at the larger temperature ($T= 10^4~T_c$). However, at the lower temperature corresponding to larger couplings, the lattice calculation yields a 
 larger value of $\kappa/T^3$. 
This larger value is also visible in \fig \ref{fig:EKTvsGlasma} where $\kappa$ is shown at the time of the evolution corresponding to the same energy density.

\section{Conclusions}
\label{sec:conc}

In this paper, we have extracted the heavy quark momentum diffusion coefficient using effective kinetic theory simulations during the bottom-up isotropization process. In order to better quantify  the importance of the pre-equilibrium evolution in relation to thermal equilibrium, we have displayed our result as  the ratio  $\nicefrac{\kappa}{\kappa^{eq}}$ to a thermal distribution with the same energy density $\varepsilon$, Debye mass $\md$, or effective infrared temperature $\tstar$. Our main conclusion is that during the kinetic stages of the pre-equilibrium evolution that can be described using effective kinetic theory, i.e., after the initial glasma evolution, the diffusion coefficient is within 30\% of the equilibrium value in a thermal system with the same energy density.  Thus our results suggest that modeling the pre-equilibrium diffusion coefficient with the equilibrium coefficient computed for a thermal system with the same energy density (Landau matching) is a reasonably good first approximation during the hydrodynamization process. 
In a more detailed comparison, we find that in the early overoccupied stage, $\kappa$ is 
larger than the equilibrium value and driven by the enhancement of both $\md$ and $\tstar$. In the later underoccupied phase, $\kappa$ is below the thermal comparison value due to the suppression of $\md$.
We have tested initial conditions with two different anisotropies, and we have found that our conclusions are not strongly affected by the value of the initial anisotropy parameter.

We have also studied the diffusion coefficient separately for the transverse and longitudinal (beam) directions. The transverse diffusion coefficient is larger than the longitudinal one at the very early stages when the occupation numbers are large, due to the Bose enhancement available for scatterings with a momentum exchange in the transverse direction. When the system becomes underoccupied, the longitudinal diffusion coefficient dominates. 
When the system approaches thermal equilibrium, the transverse and longitudinal diffusion coefficients become equal as expected. 
In a similar manner, we have qualitatively explained the evolution of $\md$, $\tstar$ and $\kappa$ using the behavior of typical hard occupation numbers and momentum anisotropy. Moreover, we have compared our results with those obtained in the glasma and lattice frameworks.

On the phenomenological side, our results give a rough estimate for the heavy quark momentum broadening during the kinetic stage of $\langle p^2 \rangle \approx 1~\mathrm{GeV}^2$ and display interesting angular dependence. It would be interesting to study the implications of these results in more detail. One exciting perspective is to include them into the quantum trajectories framework \cite{Brambilla:2020qwo, Brambilla:2022ynh} to study the impact of initial stages on quarkonium dynamics.
It is also possible to address the evolution of other transport coefficients during the initial non-equilibrium dynamics: in a separate paper \cite{Boguslavski:2023alu} we use the EKT setup to calculate the evolution of the jet quenching parameter $\hat{q}$. These studies open the possibility of assessing the impact of initial stages on other phenomenological observables.

\begin{acknowledgments}

The authors would like to thank D.~Avramescu, N.~Brambilla, M.~Escobedo, D.I.~M\"uller, A.~Rothkopf and M.~Strickland for valuable discussions. 
This work is supported by the European Research Council, ERC-2018-ADG-835105 YoctoLHC.  This work was also supported under the European Union’s
Horizon 2020 research and innovation by the STRONG-2020 project (grant agreement No. 824093). The content of this article does not reflect the official opinion of the European Union and responsibility for the information and views expressed therein lies entirely with the authors.  This work was funded in part by the Knut and Alice Wallenberg foundation, contract number 2017.0036. TL and JP have been supported by the Academy of Finland, by the Centre of Excellence in Quark Matter (project 346324) and project 321840.
KB and FL would like to thank the Austrian Science Fund (FWF) for support under project P 34455, and FL is additionally supported by the Doctoral Program W1252-N27 Particles and Interactions.
The authors wish to acknowledge CSC – IT Center for Science, Finland, for computational resources.  We acknowledge grants of computer capacity from the Finnish Grid and Cloud Infrastructure (persistent identifier urn:nbn:fi:research-infras-2016072533 ).
 The authors wish to acknowledge the Vienna Scientific
Cluster (VSC) project 71444 for computational resources.

\end{acknowledgments}

\appendix

\section{Numerical framework}
\label{app:numerics}

\subsection{Details of the discretization procedure}
\label{sec:discdetails}
Our numerical framework is the same as in \cite{Kurkela:2015qoa}. Instead of the distribution function, we use the 
quantity 
\begin{equation}
\nbar_{ij} = \int \derthree \lambda f(\bs{p}) w_{ij}(\bs{p}),
\end{equation}
which represents the particle number per degree of freedom and unit (spatial) volume,
as our dynamical degree of freedom.
Here the $w_{ij}$ involve the discretization in momentum $p \equiv |\bs p|$ and polar angle $\cos \theta$ as
\begin{equation}
w_{ij}\left( \bs{p} \right) = w_j(p) w_i(\cos \theta),
\end{equation}
with the wedge functions defined as
\begin{align}
w_i(z) &= \dfrac{z_{i+1}-z}{z_{i+1} - z_i} \theta(z-z_i) \theta(z_{i+1} - |z|) \nn \\
& + \dfrac{z-z_{i-1}}{z_{i} - z_{i-1}} \theta(z_i-z) \theta(z - z_i).
\end{align}
 This discretization conserves particle number, energy density and $\langle \pz \rangle$ exactly. 

Our numerical framework has in total seven discretization parameters. Two of them correspond to the number of bins used in the momentum $\in [\pmin,\pmax]$
and polar angle. There are two parameters associated to the Monte Carlo sampling procedure, one of which determines the number of samples that is needed for the time-evolution, the other one being associated to the sampling of the diffusion coefficient. For the momentum discretization we have parameters which are associated to the minimum and maximum momentum in our momentum grid $\pmin$ and $\pmax$. Finally for the time-evolution we use an adaptive step-size algorithm, which needs an initial value. 

We have verified that our numerical results are independent of the discretization parameter set by reproducing the results of \cite{Kurkela:2015qoa}. The most important dependencies on the discretization parameters are the following: Reproducing the correct behavior during the underoccupied region is very sensitive to the length of the time-step. It turns out that the minimum momentum on the grid is important to understand the approach to equilibrium -- as the system becomes thermal, it will approach a thermal distribution with an infrared cutoff $\pmin$. When we compare our simulations to thermal equilibrium, this has to be taken into account. This effect is described in detail in Appendix~\ref{app:pminDependence}. In principle we receive corrections also from the maximum momentum parameter $\pmax$ (and other discretization parameters). However, in practice the dependence on $\pmin$ appears to be by far the most important effect. We have also checked that our results do not depend on the accuracy of the discretization of the $\theta$ angle.

\subsection{Discretization of $\kappa$}
\label{se:DiscreteKappa}

Due to  cylindrical symmetry around the beam direction, the distribution function $f(\bs{p})$ out of equilibrium is assumed to only depend on the magnitude of the momentum and the polar angle $\theta$, and not on the azimuthal angle $\phi$. We start by evaluating the $\bs{p}^\prime$ integral in \eqref{eq:kappa_master_formula} by making use of the momentum conserving delta function. Furthermore we integrate over the radial component of $\bs{k}^\prime$ using the energy conserving delta function. This leads to
\begin{align}
3 \kappa &= \frac{\lambda^2 C_H}{N_c}  \int_{0}^{\infty} \dfrac{\der k}{(2 \pi)^3}\,k^2 \int_{-1}^{1}\der x_k \int_{0}^{2\pi}\der \phi_k \nn \\ 
&\times \int_{-1}^{1} \dfrac{\der x_k^\prime}{(2 \pi)^2}\int_{0}^{2\pi} \der \phi_k^\prime\,k^2 q^2\,\dfrac{1+ \cos^2 \theta_{\bs{k} \bs{k^\prime} }}{\left(q^2 + m_D^2 \right)^2} \nn \\ 
& \times f(k,x_k) \left(1+f(k,x_k^\prime)\right),
\end{align}
 where we use the notation $x_k = \cos \theta_k$, $x_k^\prime = \cos \theta_{k^\prime}$ for the polar angles. The azimuthal angles are denoted by $\phi_k$ and $\phi_k^\prime$. The cosine of the angle between $\bs{k}$ and $\bs{k}^\prime$ can be written as 
\begin{align}
 \cos \theta_{\bs{k} \bs{k^\prime} } &= 1 - \frac{q^2}{2k^2} \\
 \label{eq:azimuthdependence}
 &= \sin \theta_k \sin \theta_k^\prime \cos(\phi_k - \phi_k^\prime) + x_k x_k^\prime.
\end{align}
As can be seen from \eqref{eq:azimuthdependence}, the expression depends only on the difference of the azimuthal angles. Thus we can change the variables by $\int_{0}^{2\pi} \der \phi_k \int_{0}^{2\pi} \der \phi_k^\prime = \int_{0}^{2\pi} \der \phi_k^\prime \int_{0}^{2\pi} \der \phi_{\bs{k} \bs{k^\prime}} = 2\pi \int_{0}^{2\pi} \der \phi_{\bs{k} \bs{k^\prime}}$ and trivially carry out one of the integrals over the azimuthal angles. 
Further evaluating the integral and using the results above yields
\begin{align}
\label{eq:kappaDiscretized}
 &3 \kappa = \frac{\lambda^2}{ (2 \pi)^4}\,\,\frac{C_H}{N_c}\int_{\Pmin}^{\Pmax}\der k\,k^4 \int_{-1}^{1}\der x_k \int_{-1}^{1} \der x_k^\prime \nn \\
 &f(k,x_k) \left(1+f(k,x_k^\prime)\right) \int_{0}^{2\pi}\der\phi_{\bs{k} \bs{k^\prime}}\, q^2\, \frac{1 + (1 - q^2/(2k^2))^2}{\left(q^2 + m_D^2 \right)^2}.
\end{align}
 The momentum transfer $q^2$ can be written in terms of the integration variables as 
\begin{align}
 q^2 &= 2k^2 \left( 1 - \sqrt{1-x_k^2} \sqrt{1-(x_k^\prime)^2} \cos(\phi_{\bs{k} \bs{k^\prime}}) - x_k x_k^\prime \right).
\end{align}
We also want to study the transverse and longitudinal diffusion coefficients separately. 
These  are given by 
\begin{align}
\label{eq:kappaDiscretizedTz}
 &\kappa_i = \frac{\lambda^2}{ (2 \pi)^4}\,\,\frac{C_H}{N_c}\int_{\Pmin}^{\Pmax}\der k\,k^4 \int_{-1}^{1}\der x_k \int_{-1}^{1} \der x_k^\prime \nn \\
 &f(k,x_k) \left(1+f(k,x_k^\prime)\right) \int_{0}^{2\pi}\der\phi_{\bs{k} \bs{k^\prime}}\, q_i^2\, \frac{1 + (1 - q^2/(2k^2))^2}{\left(q^2 + m_D^2 \right)^2}.
\end{align}
 Here $q_i = q_T,q_z,$ is the momentum transfer in different directions, given by
\begin{align}
\dfrac{q_z^2}{k^2} &= \left(x_k - x_{k^\prime} \right)^2 \\
\dfrac{q_T^2}{k^2} & =  \left[ 1- \dfrac{x_k^2}{2} - \dfrac{x_{k^\prime}^2}{2} - \sqrt{\left(1- x_k^2\right) \left(1- x_{k^\prime}^2 \right)} \cos \phi_{k k^\prime}\right].
\end{align}
The normalization is such that $q^2 = 2 q_T^2 + q_z^2$, and consequently 
$ 3 \kappa = 2 \kappa_T + \kappa_z$. For a thermalized system this corresponds to $\nicefrac{\kappa_{T}}{\kappa_{z}} \to 1$ and $\nicefrac{\kappa_{T,z}}{\kappa} \to 1$.  In our numerical framework the integrals are computed using Monte-Carlo techniques. In \eqref{eq:kappaDiscretized} and \eqref{eq:kappaDiscretizedTz} we have also discretized the momentum interval with IR and UV cutoffs $[\Pmin,\Pmax]$. The dominating discretization effects are discussed in more detail in  Appendix~\ref{app:pminDependence}.

\begin{figure}
\includegraphics[width=0.46\textwidth]{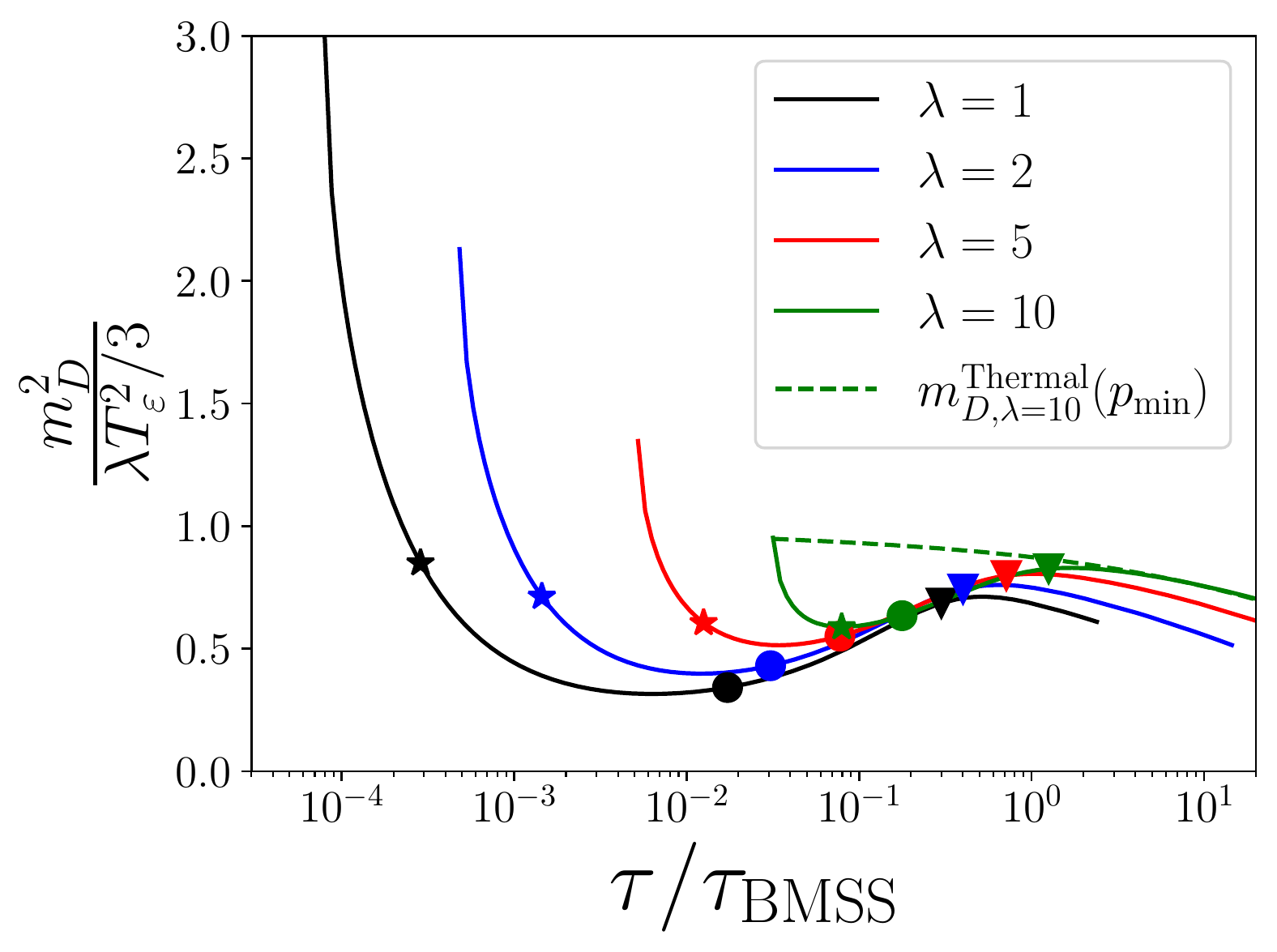}
\caption{Debye mass extracted during the bottom-up evolution, divided by the value in a \emph{continuum} thermal system with the same energy density. The dashed curve has been obtained by computing the screening mass in a thermal system with the IR regulator $\pmin$ given by \eqref{eq:mDpmin}.
The temperature $T_\varepsilon$ is extracted from the energy density using \eq\eqref{eq:sensibleTemperature}. 
The main observation is that the IR cutoff $\pmin$ decreases $\md$ by roughly 20\% from the continuum value, depending on $T/\pmin$.
}
\label{fig:mDdiscretizationEffect}
\end{figure}

\subsection{Observables and $\pmin$ dependence}
\label{app:pminDependence}

Due to the discretization effects, our system does not reach the continuum thermal equilibrium. Instead, it approaches a discretized version of thermal equilibrium. This affects our observables, as illustrated in \fig \ref{fig:mDdiscretizationEffect}. 
We observe that the numerically obtained value for $\md$ divided by its thermal expectation (obtained for a gluonic system with a Bose-Einstein distribution) deviates from unity at late times. However, when we take into account the discretization effects in terms of the momentum cutoff $\pmin$, we see a nice agreement for  $\lambda=10$. A similar agreement is observed for other curves as well, but here we show only one for clarity. In principle we could take into account also other discretization parameters such as $\pmax$ and the angular discretization, but we find that the dominant discretization contribution arises from $\pmin$.

We will now discuss the $\pmin$ dependence in $\md$, $\tstar$ and $\kappa$ in more detail.

\subsubsection{Debye mass $\md$}
\label{sec:mdpmin}

The $\pmin$ dependence in $\md$ given by \eqref{eq:mDdefinition} is obtained by inserting an IR cutoff into the integral  as follows
\begin{align}
\label{eq:mDpmin}
&m_D^2(\pmin) = \dfrac{8 \lambda}{\left(2 \pi\right)^2} \int_{\pmin}^\infty \der p p f(p) \\
& \quad = \frac{2 \lambda  T \left(T \mathrm{Li}_2\left(e^{-\frac{\pmin}{T}}\right)-\pmin \log
   \left(1-e^{-\frac{\pmin}{T}}\right)\right)}{\pi ^2}, \nn
\end{align}
where the distribution is given by the Bose-Einstein distribution. The dilogarithm is defined as 
\begin{align}
\mathrm{Li}_2(z) = \sum_{k=1}^\infty \dfrac{z^k}{k^2} = - \int_0^z \dfrac{\ln(1-u)}{u}  \der u.
\label{eq:dilogdef}
\end{align}

The curve labeled as thermal in \fig \ref{fig:mDdiscretizationEffect} is obtained using \eqref{eq:mDpmin}.

\subsubsection{Effective IR temperature $\tstar$}
\label{sec:tstarpmin}
The same procedure can be carried out for $\tstar$
\begin{align}
\lambda \tstar(t,\pmin) =  \dfrac{\lambda^2}{\pi^2  \md^2(\pmin)} \int_{\pmin}^\infty \der p p^2  f(t,p) \left(1 +f(t,p) \right).
\end{align}
For the thermal distribution the result becomes 
\begin{align}
 \label{eq:tstarpminexpression}
& \lambda \tstar(\pmin)  = \frac{\lambda }{6} \Bigg[ 6 T^2 \mathrm{Li}_2\left(e^{-\frac{\pmin}{T}}\right) \nn \\ 
&+ 3 \pmin \Bigg(\pmin
   \left(\frac{1}{e^{\pmin/T}-1}+2\right)  -2 T \log \left(e^{\pmin/T}-1\right)\Bigg)\Bigg] \nn \\ 
   & \times \Bigg[ \Big(T
   \mathrm{Li}_2\left(e^{-\frac{\pmin}{T}}\right) -\pmin \log \left(1-e^{-\frac{\pmin}{T}}\right)\Big)\Bigg]^{-1}.
 \end{align}
Expanding this for small $x = \dfrac{\pmin}{T}$ yields
 \begin{align}
 \lambda \tstar(\pmin)  =  \lambda T \dfrac{12 \pi -36x + x^3  }{12 \pi - 72 x + 18 x^2 - 2 x^3    } + \mathcal{O}(x^5).
    \end{align}
 Thus in the limit $x \to 0$ we recover the equilibrium relation $\lambda \tstar = \lambda T$.

 In the comparisons to equilibrium distributions in the main text (e.g. in \fig\ref{fig:scalesVsEquil3})
 these corrections are applied to $\md$ and $\tstar$. Consequently, the ratios approach unity when the system thermalizes, contrary to the behavior observed in \fig \ref{fig:mDdiscretizationEffect} without the corrections.

\subsubsection{Diffusion coefficient $\kappa$}
\label{sec:kappapmin}

 For $\kappa$ the corrections arising from the infrared regulator $\pmin$ are taken into account as follows. We start with the expression \eqref{eq:kappaLO} and perform the $q$ integral analytically without an infrared regulator as before. Then we replace $\md$ with the infrared regulated expression  $\md(\pmin)$ given by \eqref{eq:mDpmin}. 
 The resulting expression is 
 \begin{widetext}
\begin{align}
\kappa_{LO}^{therm} &= \int_{0}^{\infty}  \frac{\der k}{{48 \pi ^3 \text{Nc} k^2
   \left(e^{k/T}-1\right)^2}}  C_H \lambda ^2 e^{k/T} \Bigg[-\Bigg[4 k^2 \left(\frac{\lambda  T \left(T
   \text{Li}_2\left(e^{-\frac{\pmin}{T}}\right)-\pmin \log \left(1-e^{-\frac{\pmin}{T}}\right)\right)}{\pi ^2}+k^2\right) \nn \\
   & \times
   \left(\frac{6 \lambda  T \left(T \text{Li}_2\left(e^{-\frac{\pmin}{T}}\right)-\pmin \log
   \left(1-e^{-\frac{\pmin}{T}}\right)\right)}{\pi ^2}+8 k^2\right)\Bigg] \nn \\
   &\times \Big[\frac{2 \lambda  T \left(T   \text{Li}_2\left(e^{-\frac{\pmin}{T}}\right)-\pmin \log \left(1-e^{-\frac{\pmin}{T}}\right)\right)}{\pi ^2}+4
   k^2\Big]^{-1} \nn \\
   &-\frac{1}{2} \log \left(\frac{\lambda  T \left(T \text{Li}_2\left(e^{-\frac{\pmin}{T}}\right)-\pmin \log
   \left(1-e^{-\frac{\pmin}{T}}\right)\right)}{\lambda  T^2 \text{Li}_2\left(e^{-\frac{\pmin}{T}}\right)-\lambda  \pmin T
   \log \left(1-e^{-\frac{\pmin}{T}}\right)+2 \pi ^2 k^2}\right)  \nn \\
    & \times \left(\frac{16 \lambda  k^2 T \left(T
   \text{Li}_2\left(e^{-\frac{\pmin}{T}}\right)-\pmin \log \left(1-e^{-\frac{\pmin}{T}}\right)\right)}{\pi ^2}+\frac{12
   \lambda ^2 T^2 \left(\pmin \log \left(1-e^{-\frac{\pmin}{T}}\right)-T
   \text{Li}_2\left(e^{-\frac{\pmin}{T}}\right)\right){}^2}{\pi ^4}+8 k^4\right)\Bigg]
   \label{eq:EpicKappa}
\end{align}
\end{widetext}

 The remaining $k$ integral in \eqref{eq:EpicKappa} is  evaluated numerically for different values of $\lambda$ and $T$ using a small independent IR regulator for numerical stability. The results are interpolated in order to find a $\kappa$ corresponding to an arbitrary temperature within the tabulated range. We have applied these corrections to the results shown in \fig \ref{fig:kappa_vs_kappaeq}. We observe that the ratio to equilibrium is very close to unity when the system is approximately thermal.
 
 We would like to emphasize that this treatment takes only into account the discretization effects arising from $m_D$. We do not regulate the $k$ and $q$ integrals with the regulator $\pmin$. Thus this treatment takes only a subset of the corrections into account.  Hence this treatment is not expected to maintain the ratio at unity for infinitely large times. This can be seen for instance in \figs \ref{fig:kappa_vs_kappaeq} and even more prominently  \ref{fig:parametricGuesstimate}, where the ratio starts to slightly deviate from unity at very late times. 

The ratios $\nicefrac{\kappa}{\kappa_{eq}^{\md, \tstar}}$ comparing to the thermal system  at the same $\md$ and $\tstar$ are obtained using the same data. The temperature $T$ and $\pmin$ uniquely determine $\md$ given by \eqref{eq:mDpmin} and $\tstar$ given by \eqref{eq:tstarpminexpression}.

\bibliography{spires}

\begin{thebibliography}{50}%
\makeatletter
\providecommand \@ifxundefined [1]{%
 \@ifx{#1\undefined}
}%
\providecommand \@ifnum [1]{%
 \ifnum #1\expandafter \@firstoftwo
 \else \expandafter \@secondoftwo
 \fi
}%
\providecommand \@ifx [1]{%
 \ifx #1\expandafter \@firstoftwo
 \else \expandafter \@secondoftwo
 \fi
}%
\providecommand \natexlab [1]{#1}%
\providecommand \enquote  [1]{``#1''}%
\providecommand \bibnamefont  [1]{#1}%
\providecommand \bibfnamefont [1]{#1}%
\providecommand \citenamefont [1]{#1}%
\providecommand \href@noop [0]{\@secondoftwo}%
\providecommand \href [0]{\begingroup \@sanitize@url \@href}%
\providecommand \@href[1]{\@@startlink{#1}\@@href}%
\providecommand \@@href[1]{\endgroup#1\@@endlink}%
\providecommand \@sanitize@url [0]{\catcode `\\12\catcode `\$12\catcode
  `\&12\catcode `\#12\catcode `\^12\catcode `\_12\catcode `\%12\relax}%
\providecommand \@@startlink[1]{}%
\providecommand \@@endlink[0]{}%
\providecommand \url  [0]{\begingroup\@sanitize@url \@url }%
\providecommand \@url [1]{\endgroup\@href {#1}{\urlprefix }}%
\providecommand \urlprefix  [0]{URL }%
\providecommand \Eprint [0]{\href }%
\providecommand \doibase [0]{http://dx.doi.org/}%
\providecommand \selectlanguage [0]{\@gobble}%
\providecommand \bibinfo  [0]{\@secondoftwo}%
\providecommand \bibfield  [0]{\@secondoftwo}%
\providecommand \translation [1]{[#1]}%
\providecommand \BibitemOpen [0]{}%
\providecommand \bibitemStop [0]{}%
\providecommand \bibitemNoStop [0]{.\EOS\space}%
\providecommand \EOS [0]{\spacefactor3000\relax}%
\providecommand \BibitemShut  [1]{\csname bibitem#1\endcsname}%
\let\auto@bib@innerbib\@empty
\bibitem [{\citenamefont {He}\ \emph {et~al.}(2022)\citenamefont {He},
  \citenamefont {van Hees},\ and\ \citenamefont {Rapp}}]{He:2022ywp}%
  \BibitemOpen
  \bibfield  {author} {\bibinfo {author} {\bibfnamefont {Min}\ \bibnamefont
  {He}}, \bibinfo {author} {\bibfnamefont {Hendrik}\ \bibnamefont {van Hees}},
  \ and\ \bibinfo {author} {\bibfnamefont {Ralf}\ \bibnamefont {Rapp}},\
  }\bibfield  {title} {\enquote {\bibinfo {title} {{Heavy-Quark Diffusion in
  the Quark-Gluon Plasma}},}\ }\href@noop {} {\  (\bibinfo {year} {2022})},\
  \Eprint {http://arxiv.org/abs/2204.09299} {arXiv:2204.09299 [hep-ph]}
  \BibitemShut {NoStop}%
\bibitem [{\citenamefont {Rothkopf}(2020)}]{Rothkopf:2019ipj}%
  \BibitemOpen
  \bibfield  {author} {\bibinfo {author} {\bibfnamefont {Alexander}\
  \bibnamefont {Rothkopf}},\ }\bibfield  {title} {\enquote {\bibinfo {title}
  {{Heavy Quarkonium in Extreme Conditions}},}\ }\href {\doibase
  10.1016/j.physrep.2020.02.006} {\bibfield  {journal} {\bibinfo  {journal}
  {Phys. Rept.}\ }\textbf {\bibinfo {volume} {858}},\ \bibinfo {pages} {1--117}
  (\bibinfo {year} {2020})},\ \Eprint {http://arxiv.org/abs/1912.02253}
  {arXiv:1912.02253 [hep-ph]} \BibitemShut {NoStop}%
\bibitem [{\citenamefont {Brambilla}(2022)}]{Brambilla:2022fqa}%
  \BibitemOpen
  \bibfield  {author} {\bibinfo {author} {\bibfnamefont {Nora}\ \bibnamefont
  {Brambilla}},\ }\enquote {\bibinfo {title} {{Quark Nuclear Physics with Heavy
  Quarks}},}\ in\ \href {\doibase 10.1007/978-981-15-8818-1_26-1} {\emph
  {\bibinfo {booktitle} {{Handbook of Nuclear Physics}}}},\ \bibinfo {editor}
  {edited by\ \bibinfo {editor} {\bibfnamefont {Isao}\ \bibnamefont
  {Tanihata}}, \bibinfo {editor} {\bibfnamefont {Hiroshi}\ \bibnamefont
  {Toki}}, \ and\ \bibinfo {editor} {\bibfnamefont {Toshitaka}\ \bibnamefont
  {Kajino}}}\ (\bibinfo {year} {2022})\ pp.\ \bibinfo {pages} {1--43},\ \Eprint
  {http://arxiv.org/abs/2204.11295} {arXiv:2204.11295 [hep-ph]} \BibitemShut
  {NoStop}%
\bibitem [{\citenamefont {Young}\ and\ \citenamefont
  {Dusling}(2013)}]{Young:2010jq}%
  \BibitemOpen
  \bibfield  {author} {\bibinfo {author} {\bibfnamefont {Clint}\ \bibnamefont
  {Young}}\ and\ \bibinfo {author} {\bibfnamefont {Kevin}\ \bibnamefont
  {Dusling}},\ }\bibfield  {title} {\enquote {\bibinfo {title} {{Quarkonium
  above deconfinement as an open quantum system}},}\ }\href {\doibase
  10.1103/PhysRevC.87.065206} {\bibfield  {journal} {\bibinfo  {journal} {Phys.
  Rev. C}\ }\textbf {\bibinfo {volume} {87}},\ \bibinfo {pages} {065206}
  (\bibinfo {year} {2013})},\ \Eprint {http://arxiv.org/abs/1001.0935}
  {arXiv:1001.0935 [nucl-th]} \BibitemShut {NoStop}%
\bibitem [{\citenamefont {Brambilla}\ \emph {et~al.}(2017)\citenamefont
  {Brambilla}, \citenamefont {Escobedo}, \citenamefont {Soto},\ and\
  \citenamefont {Vairo}}]{Brambilla:2016wgg}%
  \BibitemOpen
  \bibfield  {author} {\bibinfo {author} {\bibfnamefont {Nora}\ \bibnamefont
  {Brambilla}}, \bibinfo {author} {\bibfnamefont {Miguel~A.}\ \bibnamefont
  {Escobedo}}, \bibinfo {author} {\bibfnamefont {Joan}\ \bibnamefont {Soto}}, \
  and\ \bibinfo {author} {\bibfnamefont {Antonio}\ \bibnamefont {Vairo}},\
  }\bibfield  {title} {\enquote {\bibinfo {title} {Quarkonium suppression in
  heavy-ion collisions: an open quantum system approach},}\ }\href {\doibase
  10.1103/PhysRevD.96.034021} {\bibfield  {journal} {\bibinfo  {journal} {Phys.
  Rev.}\ }\textbf {\bibinfo {volume} {D96}},\ \bibinfo {pages} {034021}
  (\bibinfo {year} {2017})},\ \Eprint {http://arxiv.org/abs/1612.07248}
  {arXiv:1612.07248 [hep-ph]} \BibitemShut {NoStop}%
\bibitem [{\citenamefont {Brambilla}\ \emph
  {et~al.}(2021{\natexlab{a}})\citenamefont {Brambilla}, \citenamefont
  {Escobedo}, \citenamefont {Strickland}, \citenamefont {Vairo}, \citenamefont
  {Vander~Griend},\ and\ \citenamefont {Weber}}]{Brambilla:2020qwo}%
  \BibitemOpen
  \bibfield  {author} {\bibinfo {author} {\bibfnamefont {Nora}\ \bibnamefont
  {Brambilla}}, \bibinfo {author} {\bibfnamefont {Miguel~\'Angel}\ \bibnamefont
  {Escobedo}}, \bibinfo {author} {\bibfnamefont {Michael}\ \bibnamefont
  {Strickland}}, \bibinfo {author} {\bibfnamefont {Antonio}\ \bibnamefont
  {Vairo}}, \bibinfo {author} {\bibfnamefont {Peter}\ \bibnamefont
  {Vander~Griend}}, \ and\ \bibinfo {author} {\bibfnamefont
  {Johannes~Heinrich}\ \bibnamefont {Weber}},\ }\bibfield  {title} {\enquote
  {\bibinfo {title} {{Bottomonium suppression in an open quantum system using
  the quantum trajectories method}},}\ }\href {\doibase
  10.1007/JHEP05(2021)136} {\bibfield  {journal} {\bibinfo  {journal} {JHEP}\
  }\textbf {\bibinfo {volume} {05}},\ \bibinfo {pages} {136} (\bibinfo {year}
  {2021}{\natexlab{a}})},\ \Eprint {http://arxiv.org/abs/2012.01240}
  {arXiv:2012.01240 [hep-ph]} \BibitemShut {NoStop}%
\bibitem [{\citenamefont {Brambilla}\ \emph
  {et~al.}(2021{\natexlab{b}})\citenamefont {Brambilla}, \citenamefont
  {Escobedo}, \citenamefont {Strickland}, \citenamefont {Vairo}, \citenamefont
  {Vander~Griend},\ and\ \citenamefont {Weber}}]{Brambilla:2021wkt}%
  \BibitemOpen
  \bibfield  {author} {\bibinfo {author} {\bibfnamefont {Nora}\ \bibnamefont
  {Brambilla}}, \bibinfo {author} {\bibfnamefont {Miguel~\'Angel}\ \bibnamefont
  {Escobedo}}, \bibinfo {author} {\bibfnamefont {Michael}\ \bibnamefont
  {Strickland}}, \bibinfo {author} {\bibfnamefont {Antonio}\ \bibnamefont
  {Vairo}}, \bibinfo {author} {\bibfnamefont {Peter}\ \bibnamefont
  {Vander~Griend}}, \ and\ \bibinfo {author} {\bibfnamefont
  {Johannes~Heinrich}\ \bibnamefont {Weber}},\ }\bibfield  {title} {\enquote
  {\bibinfo {title} {{Bottomonium production in heavy-ion collisions using
  quantum trajectories: Differential observables and momentum anisotropy}},}\
  }\href {\doibase 10.1103/PhysRevD.104.094049} {\bibfield  {journal} {\bibinfo
   {journal} {Phys. Rev. D}\ }\textbf {\bibinfo {volume} {104}},\ \bibinfo
  {pages} {094049} (\bibinfo {year} {2021}{\natexlab{b}})},\ \Eprint
  {http://arxiv.org/abs/2107.06222} {arXiv:2107.06222 [hep-ph]} \BibitemShut
  {NoStop}%
\bibitem [{\citenamefont {Brambilla}\ \emph
  {et~al.}(2023{\natexlab{a}})\citenamefont {Brambilla}, \citenamefont
  {Escobedo}, \citenamefont {Islam}, \citenamefont {Strickland}, \citenamefont
  {Tiwari}, \citenamefont {Vairo},\ and\ \citenamefont
  {Vander~Griend}}]{Brambilla:2023hkw}%
  \BibitemOpen
  \bibfield  {author} {\bibinfo {author} {\bibfnamefont {Nora}\ \bibnamefont
  {Brambilla}}, \bibinfo {author} {\bibfnamefont {Miguel~\'Angel}\ \bibnamefont
  {Escobedo}}, \bibinfo {author} {\bibfnamefont {Ajaharul}\ \bibnamefont
  {Islam}}, \bibinfo {author} {\bibfnamefont {Michael}\ \bibnamefont
  {Strickland}}, \bibinfo {author} {\bibfnamefont {Anurag}\ \bibnamefont
  {Tiwari}}, \bibinfo {author} {\bibfnamefont {Antonio}\ \bibnamefont {Vairo}},
  \ and\ \bibinfo {author} {\bibfnamefont {Peter}\ \bibnamefont
  {Vander~Griend}},\ }\bibfield  {title} {\enquote {\bibinfo {title}
  {{Regeneration of bottomonia in an open quantum systems approach}},}\
  }\href@noop {} {\  (\bibinfo {year} {2023}{\natexlab{a}})},\ \Eprint
  {http://arxiv.org/abs/2302.11826} {arXiv:2302.11826 [hep-ph]} \BibitemShut
  {NoStop}%
\bibitem [{\citenamefont {Moore}\ and\ \citenamefont
  {Teaney}(2005)}]{Moore:2004tg}%
  \BibitemOpen
  \bibfield  {author} {\bibinfo {author} {\bibfnamefont {Guy~D.}\ \bibnamefont
  {Moore}}\ and\ \bibinfo {author} {\bibfnamefont {Derek}\ \bibnamefont
  {Teaney}},\ }\bibfield  {title} {\enquote {\bibinfo {title} {How much do
  heavy quarks thermalize in a heavy ion collision?}}\ }\href {\doibase
  10.1103/PhysRevC.71.064904} {\bibfield  {journal} {\bibinfo  {journal} {Phys.
  Rev.}\ }\textbf {\bibinfo {volume} {C71}},\ \bibinfo {pages} {064904}
  (\bibinfo {year} {2005})},\ \Eprint {http://arxiv.org/abs/hep-ph/0412346}
  {arXiv:hep-ph/0412346 [hep-ph]} \BibitemShut {NoStop}%
\bibitem [{\citenamefont {Caron-Huot}\ and\ \citenamefont
  {Moore}(2008{\natexlab{a}})}]{Caron-Huot:2007rwy}%
  \BibitemOpen
  \bibfield  {author} {\bibinfo {author} {\bibfnamefont {Simon}\ \bibnamefont
  {Caron-Huot}}\ and\ \bibinfo {author} {\bibfnamefont {Guy~D.}\ \bibnamefont
  {Moore}},\ }\bibfield  {title} {\enquote {\bibinfo {title} {{Heavy quark
  diffusion in perturbative QCD at next-to-leading order}},}\ }\href {\doibase
  10.1103/PhysRevLett.100.052301} {\bibfield  {journal} {\bibinfo  {journal}
  {Phys. Rev. Lett.}\ }\textbf {\bibinfo {volume} {100}},\ \bibinfo {pages}
  {052301} (\bibinfo {year} {2008}{\natexlab{a}})},\ \Eprint
  {http://arxiv.org/abs/0708.4232} {arXiv:0708.4232 [hep-ph]} \BibitemShut
  {NoStop}%
\bibitem [{\citenamefont {Caron-Huot}\ and\ \citenamefont
  {Moore}(2008{\natexlab{b}})}]{Caron-Huot:2008dyw}%
  \BibitemOpen
  \bibfield  {author} {\bibinfo {author} {\bibfnamefont {Simon}\ \bibnamefont
  {Caron-Huot}}\ and\ \bibinfo {author} {\bibfnamefont {Guy~D.}\ \bibnamefont
  {Moore}},\ }\bibfield  {title} {\enquote {\bibinfo {title} {{Heavy quark
  diffusion in QCD and N=4 SYM at next-to-leading order}},}\ }\href {\doibase
  10.1088/1126-6708/2008/02/081} {\bibfield  {journal} {\bibinfo  {journal}
  {JHEP}\ }\textbf {\bibinfo {volume} {02}},\ \bibinfo {pages} {081} (\bibinfo
  {year} {2008}{\natexlab{b}})},\ \Eprint {http://arxiv.org/abs/0801.2173}
  {arXiv:0801.2173 [hep-ph]} \BibitemShut {NoStop}%
\bibitem [{\citenamefont {Brambilla}\ \emph {et~al.}(2020)\citenamefont
  {Brambilla}, \citenamefont {Leino}, \citenamefont {Petreczky},\ and\
  \citenamefont {Vairo}}]{Brambilla:2020siz}%
  \BibitemOpen
  \bibfield  {author} {\bibinfo {author} {\bibfnamefont {Nora}\ \bibnamefont
  {Brambilla}}, \bibinfo {author} {\bibfnamefont {Viljami}\ \bibnamefont
  {Leino}}, \bibinfo {author} {\bibfnamefont {Peter}\ \bibnamefont
  {Petreczky}}, \ and\ \bibinfo {author} {\bibfnamefont {Antonio}\ \bibnamefont
  {Vairo}},\ }\bibfield  {title} {\enquote {\bibinfo {title} {{Lattice QCD
  constraints on the heavy quark diffusion coefficient}},}\ }\href {\doibase
  10.1103/PhysRevD.102.074503} {\bibfield  {journal} {\bibinfo  {journal}
  {Phys. Rev. D}\ }\textbf {\bibinfo {volume} {102}},\ \bibinfo {pages}
  {074503} (\bibinfo {year} {2020})},\ \Eprint
  {http://arxiv.org/abs/2007.10078} {arXiv:2007.10078 [hep-lat]} \BibitemShut
  {NoStop}%
\bibitem [{\citenamefont {Casalderrey-Solana}\ and\ \citenamefont
  {Teaney}(2006)}]{Casalderrey-Solana:2006fio}%
  \BibitemOpen
  \bibfield  {author} {\bibinfo {author} {\bibfnamefont {Jorge}\ \bibnamefont
  {Casalderrey-Solana}}\ and\ \bibinfo {author} {\bibfnamefont {Derek}\
  \bibnamefont {Teaney}},\ }\bibfield  {title} {\enquote {\bibinfo {title}
  {{Heavy quark diffusion in strongly coupled N=4 Yang-Mills}},}\ }\href
  {\doibase 10.1103/PhysRevD.74.085012} {\bibfield  {journal} {\bibinfo
  {journal} {Phys. Rev. D}\ }\textbf {\bibinfo {volume} {74}},\ \bibinfo
  {pages} {085012} (\bibinfo {year} {2006})},\ \Eprint
  {http://arxiv.org/abs/hep-ph/0605199} {arXiv:hep-ph/0605199} \BibitemShut
  {NoStop}%
\bibitem [{\citenamefont {van Hees}\ \emph {et~al.}(2006)\citenamefont {van
  Hees}, \citenamefont {Greco},\ and\ \citenamefont {Rapp}}]{vanHees:2005wb}%
  \BibitemOpen
  \bibfield  {author} {\bibinfo {author} {\bibfnamefont {Hendrik}\ \bibnamefont
  {van Hees}}, \bibinfo {author} {\bibfnamefont {Vincenzo}\ \bibnamefont
  {Greco}}, \ and\ \bibinfo {author} {\bibfnamefont {Ralf}\ \bibnamefont
  {Rapp}},\ }\bibfield  {title} {\enquote {\bibinfo {title} {Heavy-quark probes
  of the quark-gluon plasma at {RHIC}},}\ }\href {\doibase
  10.1103/PhysRevC.73.034913} {\bibfield  {journal} {\bibinfo  {journal} {Phys.
  Rev.}\ }\textbf {\bibinfo {volume} {C73}},\ \bibinfo {pages} {034913}
  (\bibinfo {year} {2006})},\ \Eprint {http://arxiv.org/abs/nucl-th/0508055}
  {arXiv:nucl-th/0508055 [nucl-th]} \BibitemShut {NoStop}%
\bibitem [{\citenamefont {Svetitsky}(1988)}]{Svetitsky:1987gq}%
  \BibitemOpen
  \bibfield  {author} {\bibinfo {author} {\bibfnamefont {B.}~\bibnamefont
  {Svetitsky}},\ }\bibfield  {title} {\enquote {\bibinfo {title} {{Diffusion of
  charmed quarks in the quark-gluon plasma}},}\ }\href {\doibase
  10.1103/PhysRevD.37.2484} {\bibfield  {journal} {\bibinfo  {journal} {Phys.
  Rev. D}\ }\textbf {\bibinfo {volume} {37}},\ \bibinfo {pages} {2484--2491}
  (\bibinfo {year} {1988})}\BibitemShut {NoStop}%
\bibitem [{\citenamefont {van Hees}\ \emph {et~al.}(2008)\citenamefont {van
  Hees}, \citenamefont {Mannarelli}, \citenamefont {Greco},\ and\ \citenamefont
  {Rapp}}]{vanHees:2007me}%
  \BibitemOpen
  \bibfield  {author} {\bibinfo {author} {\bibfnamefont {H.}~\bibnamefont {van
  Hees}}, \bibinfo {author} {\bibfnamefont {M.}~\bibnamefont {Mannarelli}},
  \bibinfo {author} {\bibfnamefont {V.}~\bibnamefont {Greco}}, \ and\ \bibinfo
  {author} {\bibfnamefont {R.}~\bibnamefont {Rapp}},\ }\bibfield  {title}
  {\enquote {\bibinfo {title} {Nonperturbative heavy-quark diffusion in the
  quark-gluon plasma},}\ }\href {\doibase 10.1103/PhysRevLett.100.192301}
  {\bibfield  {journal} {\bibinfo  {journal} {Phys. Rev. Lett.}\ }\textbf
  {\bibinfo {volume} {100}},\ \bibinfo {pages} {192301} (\bibinfo {year}
  {2008})},\ \Eprint {http://arxiv.org/abs/0709.2884} {arXiv:0709.2884
  [hep-ph]} \BibitemShut {NoStop}%
\bibitem [{\citenamefont {Banerjee}\ \emph {et~al.}(2012)\citenamefont
  {Banerjee}, \citenamefont {Datta}, \citenamefont {Gavai},\ and\ \citenamefont
  {Majumdar}}]{Banerjee:2011ra}%
  \BibitemOpen
  \bibfield  {author} {\bibinfo {author} {\bibfnamefont {Debasish}\
  \bibnamefont {Banerjee}}, \bibinfo {author} {\bibfnamefont {Saumen}\
  \bibnamefont {Datta}}, \bibinfo {author} {\bibfnamefont {Rajiv}\ \bibnamefont
  {Gavai}}, \ and\ \bibinfo {author} {\bibfnamefont {Pushan}\ \bibnamefont
  {Majumdar}},\ }\bibfield  {title} {\enquote {\bibinfo {title} {Heavy quark
  momentum diffusion coefficient from lattice {QCD}},}\ }\href {\doibase
  10.1103/PhysRevD.85.014510} {\bibfield  {journal} {\bibinfo  {journal} {Phys.
  Rev.}\ }\textbf {\bibinfo {volume} {D85}},\ \bibinfo {pages} {014510}
  (\bibinfo {year} {2012})},\ \Eprint {http://arxiv.org/abs/1109.5738}
  {arXiv:1109.5738 [hep-lat]} \BibitemShut {NoStop}%
\bibitem [{\citenamefont {Capellino}\ \emph {et~al.}(2022)\citenamefont
  {Capellino}, \citenamefont {Beraudo}, \citenamefont {Dubla}, \citenamefont
  {Floerchinger}, \citenamefont {Masciocchi}, \citenamefont {Pawlowski},\ and\
  \citenamefont {Selyuzhenkov}}]{Capellino:2022nvf}%
  \BibitemOpen
  \bibfield  {author} {\bibinfo {author} {\bibfnamefont {F.}~\bibnamefont
  {Capellino}}, \bibinfo {author} {\bibfnamefont {A.}~\bibnamefont {Beraudo}},
  \bibinfo {author} {\bibfnamefont {A.}~\bibnamefont {Dubla}}, \bibinfo
  {author} {\bibfnamefont {S.}~\bibnamefont {Floerchinger}}, \bibinfo {author}
  {\bibfnamefont {S.}~\bibnamefont {Masciocchi}}, \bibinfo {author}
  {\bibfnamefont {J.}~\bibnamefont {Pawlowski}}, \ and\ \bibinfo {author}
  {\bibfnamefont {I.}~\bibnamefont {Selyuzhenkov}},\ }\bibfield  {title}
  {\enquote {\bibinfo {title} {{Fluid-dynamic approach to heavy-quark diffusion
  in the quark-gluon plasma}},}\ }\href {\doibase 10.1103/PhysRevD.106.034021}
  {\bibfield  {journal} {\bibinfo  {journal} {Phys. Rev. D}\ }\textbf {\bibinfo
  {volume} {106}},\ \bibinfo {pages} {034021} (\bibinfo {year} {2022})},\
  \Eprint {http://arxiv.org/abs/2205.07692} {arXiv:2205.07692 [nucl-th]}
  \BibitemShut {NoStop}%
\bibitem [{\citenamefont {Yao}\ \emph {et~al.}(2021)\citenamefont {Yao},
  \citenamefont {Ke}, \citenamefont {Xu}, \citenamefont {Bass},\ and\
  \citenamefont {M\"uller}}]{Yao:2020xzw}%
  \BibitemOpen
  \bibfield  {author} {\bibinfo {author} {\bibfnamefont {Xiaojun}\ \bibnamefont
  {Yao}}, \bibinfo {author} {\bibfnamefont {Weiyao}\ \bibnamefont {Ke}},
  \bibinfo {author} {\bibfnamefont {Yingru}\ \bibnamefont {Xu}}, \bibinfo
  {author} {\bibfnamefont {Steffen~A.}\ \bibnamefont {Bass}}, \ and\ \bibinfo
  {author} {\bibfnamefont {Berndt}\ \bibnamefont {M\"uller}},\ }\bibfield
  {title} {\enquote {\bibinfo {title} {{Coupled Boltzmann Transport Equations
  of Heavy Quarks and Quarkonia in Quark-Gluon Plasma}},}\ }\href {\doibase
  10.1007/JHEP01(2021)046} {\bibfield  {journal} {\bibinfo  {journal} {JHEP}\
  }\textbf {\bibinfo {volume} {01}},\ \bibinfo {pages} {046} (\bibinfo {year}
  {2021})},\ \Eprint {http://arxiv.org/abs/2004.06746} {arXiv:2004.06746
  [hep-ph]} \BibitemShut {NoStop}%
\bibitem [{\citenamefont {Banerjee}\ \emph
  {et~al.}(2022{\natexlab{a}})\citenamefont {Banerjee}, \citenamefont {Gavai},
  \citenamefont {Datta},\ and\ \citenamefont {Majumdar}}]{Banerjee:2022gen}%
  \BibitemOpen
  \bibfield  {author} {\bibinfo {author} {\bibfnamefont {Debasish}\
  \bibnamefont {Banerjee}}, \bibinfo {author} {\bibfnamefont {Rajiv}\
  \bibnamefont {Gavai}}, \bibinfo {author} {\bibfnamefont {Saumen}\
  \bibnamefont {Datta}}, \ and\ \bibinfo {author} {\bibfnamefont {Pushan}\
  \bibnamefont {Majumdar}},\ }\bibfield  {title} {\enquote {\bibinfo {title}
  {{Temperature dependence of the static quark diffusion coefficient}},}\
  }\href@noop {} {\  (\bibinfo {year} {2022}{\natexlab{a}})},\ \Eprint
  {http://arxiv.org/abs/2206.15471} {arXiv:2206.15471 [hep-ph]} \BibitemShut
  {NoStop}%
\bibitem [{\citenamefont {Ipp}\ \emph {et~al.}(2020{\natexlab{a}})\citenamefont
  {Ipp}, \citenamefont {M\"uller},\ and\ \citenamefont {Schuh}}]{Ipp:2020nfu}%
  \BibitemOpen
  \bibfield  {author} {\bibinfo {author} {\bibfnamefont {Andreas}\ \bibnamefont
  {Ipp}}, \bibinfo {author} {\bibfnamefont {David~I.}\ \bibnamefont
  {M\"uller}}, \ and\ \bibinfo {author} {\bibfnamefont {Daniel}\ \bibnamefont
  {Schuh}},\ }\bibfield  {title} {\enquote {\bibinfo {title} {{Jet momentum
  broadening in the pre-equilibrium Glasma}},}\ }\href {\doibase
  10.1016/j.physletb.2020.135810} {\bibfield  {journal} {\bibinfo  {journal}
  {Phys. Lett. B}\ }\textbf {\bibinfo {volume} {810}},\ \bibinfo {pages}
  {135810} (\bibinfo {year} {2020}{\natexlab{a}})},\ \Eprint
  {http://arxiv.org/abs/2009.14206} {arXiv:2009.14206 [hep-ph]} \BibitemShut
  {NoStop}%
\bibitem [{\citenamefont {Ipp}\ \emph {et~al.}(2020{\natexlab{b}})\citenamefont
  {Ipp}, \citenamefont {M\"uller},\ and\ \citenamefont {Schuh}}]{Ipp:2020mjc}%
  \BibitemOpen
  \bibfield  {author} {\bibinfo {author} {\bibfnamefont {Andreas}\ \bibnamefont
  {Ipp}}, \bibinfo {author} {\bibfnamefont {David~I.}\ \bibnamefont
  {M\"uller}}, \ and\ \bibinfo {author} {\bibfnamefont {Daniel}\ \bibnamefont
  {Schuh}},\ }\bibfield  {title} {\enquote {\bibinfo {title} {{Anisotropic
  momentum broadening in the 2+1D Glasma: analytic weak field approximation and
  lattice simulations}},}\ }\href {\doibase 10.1103/PhysRevD.102.074001}
  {\bibfield  {journal} {\bibinfo  {journal} {Phys. Rev. D}\ }\textbf {\bibinfo
  {volume} {102}},\ \bibinfo {pages} {074001} (\bibinfo {year}
  {2020}{\natexlab{b}})},\ \Eprint {http://arxiv.org/abs/2001.10001}
  {arXiv:2001.10001 [hep-ph]} \BibitemShut {NoStop}%
\bibitem [{\citenamefont {Avramescu}\ \emph
  {et~al.}(2023{\natexlab{a}})\citenamefont {Avramescu}, \citenamefont
  {B\u{a}ran}, \citenamefont {Greco}, \citenamefont {Ipp}, \citenamefont
  {M\"uller},\ and\ \citenamefont {Ruggieri}}]{Avramescu:2023qvv}%
  \BibitemOpen
  \bibfield  {author} {\bibinfo {author} {\bibfnamefont {Dana}\ \bibnamefont
  {Avramescu}}, \bibinfo {author} {\bibfnamefont {Virgil}\ \bibnamefont
  {B\u{a}ran}}, \bibinfo {author} {\bibfnamefont {Vincenzo}\ \bibnamefont
  {Greco}}, \bibinfo {author} {\bibfnamefont {Andreas}\ \bibnamefont {Ipp}},
  \bibinfo {author} {\bibfnamefont {David.~I.}\ \bibnamefont {M\"uller}}, \
  and\ \bibinfo {author} {\bibfnamefont {Marco}\ \bibnamefont {Ruggieri}},\
  }\bibfield  {title} {\enquote {\bibinfo {title} {{Simulating jets and heavy
  quarks in the glasma using the colored particle-in-cell method}},}\ }\href
  {\doibase 10.1103/PhysRevD.107.114021} {\bibfield  {journal} {\bibinfo
  {journal} {Phys. Rev. D}\ }\textbf {\bibinfo {volume} {107}},\ \bibinfo
  {pages} {114021} (\bibinfo {year} {2023}{\natexlab{a}})},\ \Eprint
  {http://arxiv.org/abs/2303.05599} {arXiv:2303.05599 [hep-ph]} \BibitemShut
  {NoStop}%
\bibitem [{\citenamefont {Carrington}\ \emph
  {et~al.}(2022{\natexlab{a}})\citenamefont {Carrington}, \citenamefont
  {Czajka},\ and\ \citenamefont {Mrowczynski}}]{Carrington:2022bnv}%
  \BibitemOpen
  \bibfield  {author} {\bibinfo {author} {\bibfnamefont {Margaret~E.}\
  \bibnamefont {Carrington}}, \bibinfo {author} {\bibfnamefont {Alina}\
  \bibnamefont {Czajka}}, \ and\ \bibinfo {author} {\bibfnamefont {Stanislaw}\
  \bibnamefont {Mrowczynski}},\ }\bibfield  {title} {\enquote {\bibinfo {title}
  {{Transport of hard probes through glasma}},}\ }\href {\doibase
  10.1103/PhysRevC.105.064910} {\bibfield  {journal} {\bibinfo  {journal}
  {Phys. Rev. C}\ }\textbf {\bibinfo {volume} {105}},\ \bibinfo {pages}
  {064910} (\bibinfo {year} {2022}{\natexlab{a}})},\ \Eprint
  {http://arxiv.org/abs/2202.00357} {arXiv:2202.00357 [nucl-th]} \BibitemShut
  {NoStop}%
\bibitem [{\citenamefont {Carrington}\ \emph
  {et~al.}(2022{\natexlab{b}})\citenamefont {Carrington}, \citenamefont
  {Czajka},\ and\ \citenamefont {Mrowczynski}}]{Carrington:2021dvw}%
  \BibitemOpen
  \bibfield  {author} {\bibinfo {author} {\bibfnamefont {Margaret~E.}\
  \bibnamefont {Carrington}}, \bibinfo {author} {\bibfnamefont {Alina}\
  \bibnamefont {Czajka}}, \ and\ \bibinfo {author} {\bibfnamefont {Stanislaw}\
  \bibnamefont {Mrowczynski}},\ }\bibfield  {title} {\enquote {\bibinfo {title}
  {{Jet quenching in glasma}},}\ }\href {\doibase
  10.1016/j.physletb.2022.137464} {\bibfield  {journal} {\bibinfo  {journal}
  {Phys. Lett. B}\ }\textbf {\bibinfo {volume} {834}},\ \bibinfo {pages}
  {137464} (\bibinfo {year} {2022}{\natexlab{b}})},\ \Eprint
  {http://arxiv.org/abs/2112.06812} {arXiv:2112.06812 [hep-ph]} \BibitemShut
  {NoStop}%
\bibitem [{\citenamefont {Carrington}\ \emph {et~al.}(2020)\citenamefont
  {Carrington}, \citenamefont {Czajka},\ and\ \citenamefont
  {Mrowczynski}}]{Carrington:2020sww}%
  \BibitemOpen
  \bibfield  {author} {\bibinfo {author} {\bibfnamefont {Margaret~E.}\
  \bibnamefont {Carrington}}, \bibinfo {author} {\bibfnamefont {Alina}\
  \bibnamefont {Czajka}}, \ and\ \bibinfo {author} {\bibfnamefont {Stanislaw}\
  \bibnamefont {Mrowczynski}},\ }\bibfield  {title} {\enquote {\bibinfo {title}
  {{Heavy Quarks Embedded in Glasma}},}\ }\href {\doibase
  10.1016/j.nuclphysa.2020.121914} {\bibfield  {journal} {\bibinfo  {journal}
  {Nucl. Phys. A}\ }\textbf {\bibinfo {volume} {1001}},\ \bibinfo {pages}
  {121914} (\bibinfo {year} {2020})},\ \Eprint
  {http://arxiv.org/abs/2001.05074} {arXiv:2001.05074 [nucl-th]} \BibitemShut
  {NoStop}%
\bibitem [{\citenamefont {Khowal}\ \emph {et~al.}(2022)\citenamefont {Khowal},
  \citenamefont {Das}, \citenamefont {Oliva},\ and\ \citenamefont
  {Ruggieri}}]{Khowal:2021zoo}%
  \BibitemOpen
  \bibfield  {author} {\bibinfo {author} {\bibfnamefont {Pooja}\ \bibnamefont
  {Khowal}}, \bibinfo {author} {\bibfnamefont {Santosh~K.}\ \bibnamefont
  {Das}}, \bibinfo {author} {\bibfnamefont {Lucia}\ \bibnamefont {Oliva}}, \
  and\ \bibinfo {author} {\bibfnamefont {Marco}\ \bibnamefont {Ruggieri}},\
  }\bibfield  {title} {\enquote {\bibinfo {title} {{Heavy quarks in the early
  stage of high energy nuclear collisions at RHIC and LHC: Brownian motion
  versus diffusion in the evolving Glasma}},}\ }\href {\doibase
  10.1140/epjp/s13360-022-02517-w} {\bibfield  {journal} {\bibinfo  {journal}
  {Eur. Phys. J. Plus}\ }\textbf {\bibinfo {volume} {137}},\ \bibinfo {pages}
  {307} (\bibinfo {year} {2022})},\ \Eprint {http://arxiv.org/abs/2110.14610}
  {arXiv:2110.14610 [hep-ph]} \BibitemShut {NoStop}%
\bibitem [{\citenamefont {Ruggieri}\ \emph {et~al.}(2022)\citenamefont
  {Ruggieri}, \citenamefont {Pooja}, \citenamefont {Prakash},\ and\
  \citenamefont {Das}}]{Ruggieri:2022kxv}%
  \BibitemOpen
  \bibfield  {author} {\bibinfo {author} {\bibfnamefont {Marco}\ \bibnamefont
  {Ruggieri}}, \bibinfo {author} {\bibnamefont {Pooja}}, \bibinfo {author}
  {\bibfnamefont {Jai}\ \bibnamefont {Prakash}}, \ and\ \bibinfo {author}
  {\bibfnamefont {Santosh~K.}\ \bibnamefont {Das}},\ }\bibfield  {title}
  {\enquote {\bibinfo {title} {{Memory effects on energy loss and diffusion of
  heavy quarks in the quark-gluon plasma}},}\ }\href {\doibase
  10.1103/PhysRevD.106.034032} {\bibfield  {journal} {\bibinfo  {journal}
  {Phys. Rev. D}\ }\textbf {\bibinfo {volume} {106}},\ \bibinfo {pages}
  {034032} (\bibinfo {year} {2022})},\ \Eprint
  {http://arxiv.org/abs/2203.06712} {arXiv:2203.06712 [hep-ph]} \BibitemShut
  {NoStop}%
\bibitem [{\citenamefont {Sun}\ \emph {et~al.}(2019)\citenamefont {Sun},
  \citenamefont {Coci}, \citenamefont {Das}, \citenamefont {Plumari},
  \citenamefont {Ruggieri},\ and\ \citenamefont {Greco}}]{Sun:2019fud}%
  \BibitemOpen
  \bibfield  {author} {\bibinfo {author} {\bibfnamefont {Yifeng}\ \bibnamefont
  {Sun}}, \bibinfo {author} {\bibfnamefont {Gabriele}\ \bibnamefont {Coci}},
  \bibinfo {author} {\bibfnamefont {Santosh~Kumar}\ \bibnamefont {Das}},
  \bibinfo {author} {\bibfnamefont {Salvatore}\ \bibnamefont {Plumari}},
  \bibinfo {author} {\bibfnamefont {Marco}\ \bibnamefont {Ruggieri}}, \ and\
  \bibinfo {author} {\bibfnamefont {Vincenzo}\ \bibnamefont {Greco}},\
  }\bibfield  {title} {\enquote {\bibinfo {title} {{Impact of Glasma on heavy
  quark observables in nucleus-nucleus collisions at LHC}},}\ }\href {\doibase
  10.1016/j.physletb.2019.134933} {\bibfield  {journal} {\bibinfo  {journal}
  {Phys. Lett.}\ }\textbf {\bibinfo {volume} {B798}},\ \bibinfo {pages}
  {134933} (\bibinfo {year} {2019})},\ \Eprint
  {http://arxiv.org/abs/1902.06254} {arXiv:1902.06254 [nucl-th]} \BibitemShut
  {NoStop}%
\bibitem [{\citenamefont {Boguslavski}\ \emph {et~al.}(2020)\citenamefont
  {Boguslavski}, \citenamefont {Kurkela}, \citenamefont {Lappi},\ and\
  \citenamefont {Peuron}}]{Boguslavski:2020tqz}%
  \BibitemOpen
  \bibfield  {author} {\bibinfo {author} {\bibfnamefont {K.}~\bibnamefont
  {Boguslavski}}, \bibinfo {author} {\bibfnamefont {A.}~\bibnamefont
  {Kurkela}}, \bibinfo {author} {\bibfnamefont {T.}~\bibnamefont {Lappi}}, \
  and\ \bibinfo {author} {\bibfnamefont {J.}~\bibnamefont {Peuron}},\
  }\bibfield  {title} {\enquote {\bibinfo {title} {{Heavy quark diffusion in an
  overoccupied gluon plasma}},}\ }\href {\doibase 10.1007/JHEP09(2020)077}
  {\bibfield  {journal} {\bibinfo  {journal} {JHEP}\ }\textbf {\bibinfo
  {volume} {09}},\ \bibinfo {pages} {077} (\bibinfo {year} {2020})},\ \Eprint
  {http://arxiv.org/abs/2005.02418} {arXiv:2005.02418 [hep-ph]} \BibitemShut
  {NoStop}%
\bibitem [{\citenamefont {Romatschke}(2007)}]{Romatschke:2006bb}%
  \BibitemOpen
  \bibfield  {author} {\bibinfo {author} {\bibfnamefont {Paul}\ \bibnamefont
  {Romatschke}},\ }\bibfield  {title} {\enquote {\bibinfo {title} {{Momentum
  broadening in an anisotropic plasma}},}\ }\href {\doibase
  10.1103/PhysRevC.75.014901} {\bibfield  {journal} {\bibinfo  {journal} {Phys.
  Rev. C}\ }\textbf {\bibinfo {volume} {75}},\ \bibinfo {pages} {014901}
  (\bibinfo {year} {2007})},\ \Eprint {http://arxiv.org/abs/hep-ph/0607327}
  {arXiv:hep-ph/0607327} \BibitemShut {NoStop}%
\bibitem [{\citenamefont {Romatschke}\ and\ \citenamefont
  {Strickland}(2004)}]{Romatschke:2004au}%
  \BibitemOpen
  \bibfield  {author} {\bibinfo {author} {\bibfnamefont {Paul}\ \bibnamefont
  {Romatschke}}\ and\ \bibinfo {author} {\bibfnamefont {Michael}\ \bibnamefont
  {Strickland}},\ }\bibfield  {title} {\enquote {\bibinfo {title} {Collisional
  energy loss of a heavy quark in an anisotropic quark-gluon plasma},}\
  }\href@noop {} {\  (\bibinfo {year} {2004})},\ \Eprint
  {http://arxiv.org/abs/hep-ph/0408275} {hep-ph/0408275} \BibitemShut {NoStop}%
\bibitem [{\citenamefont {Hauksson}\ \emph {et~al.}(2022)\citenamefont
  {Hauksson}, \citenamefont {Jeon},\ and\ \citenamefont
  {Gale}}]{Hauksson:2021okc}%
  \BibitemOpen
  \bibfield  {author} {\bibinfo {author} {\bibfnamefont {Sigtryggur}\
  \bibnamefont {Hauksson}}, \bibinfo {author} {\bibfnamefont {Sangyong}\
  \bibnamefont {Jeon}}, \ and\ \bibinfo {author} {\bibfnamefont {Charles}\
  \bibnamefont {Gale}},\ }\bibfield  {title} {\enquote {\bibinfo {title}
  {{Momentum broadening of energetic partons in an anisotropic plasma}},}\
  }\href {\doibase 10.1103/PhysRevC.105.014914} {\bibfield  {journal} {\bibinfo
   {journal} {Phys. Rev. C}\ }\textbf {\bibinfo {volume} {105}},\ \bibinfo
  {pages} {014914} (\bibinfo {year} {2022})},\ \Eprint
  {http://arxiv.org/abs/2109.04575} {arXiv:2109.04575 [hep-ph]} \BibitemShut
  {NoStop}%
\bibitem [{\citenamefont {Baier}\ \emph {et~al.}(2001)\citenamefont {Baier},
  \citenamefont {Mueller}, \citenamefont {Schiff},\ and\ \citenamefont
  {Son}}]{Baier:2000sb}%
  \BibitemOpen
  \bibfield  {author} {\bibinfo {author} {\bibfnamefont {R.}~\bibnamefont
  {Baier}}, \bibinfo {author} {\bibfnamefont {Alfred~H.}\ \bibnamefont
  {Mueller}}, \bibinfo {author} {\bibfnamefont {D.}~\bibnamefont {Schiff}}, \
  and\ \bibinfo {author} {\bibfnamefont {D.~T.}\ \bibnamefont {Son}},\
  }\bibfield  {title} {\enquote {\bibinfo {title} {'bottom-up' thermalization
  in heavy ion collisions},}\ }\href@noop {} {\bibfield  {journal} {\bibinfo
  {journal} {Phys. Lett.}\ }\textbf {\bibinfo {volume} {B502}},\ \bibinfo
  {pages} {51--58} (\bibinfo {year} {2001})},\ \Eprint
  {http://arxiv.org/abs/hep-ph/0009237} {hep-ph/0009237} \BibitemShut {NoStop}%
\bibitem [{\citenamefont {Kurkela}\ and\ \citenamefont
  {Zhu}(2015)}]{Kurkela:2015qoa}%
  \BibitemOpen
  \bibfield  {author} {\bibinfo {author} {\bibfnamefont {Aleksi}\ \bibnamefont
  {Kurkela}}\ and\ \bibinfo {author} {\bibfnamefont {Yan}\ \bibnamefont
  {Zhu}},\ }\bibfield  {title} {\enquote {\bibinfo {title} {Isotropization and
  hydrodynamization in weakly coupled heavy-ion collisions},}\ }\href {\doibase
  10.1103/PhysRevLett.115.182301} {\bibfield  {journal} {\bibinfo  {journal}
  {Phys. Rev. Lett.}\ }\textbf {\bibinfo {volume} {115}},\ \bibinfo {pages}
  {182301} (\bibinfo {year} {2015})},\ \Eprint
  {http://arxiv.org/abs/1506.06647} {arXiv:1506.06647 [hep-ph]} \BibitemShut
  {NoStop}%
\bibitem [{\citenamefont {Arnold}\ \emph {et~al.}(2003)\citenamefont {Arnold},
  \citenamefont {Moore},\ and\ \citenamefont {Yaffe}}]{Arnold:2002zm}%
  \BibitemOpen
  \bibfield  {author} {\bibinfo {author} {\bibfnamefont {Peter~Brockway}\
  \bibnamefont {Arnold}}, \bibinfo {author} {\bibfnamefont {Guy~D.}\
  \bibnamefont {Moore}}, \ and\ \bibinfo {author} {\bibfnamefont {Laurence~G.}\
  \bibnamefont {Yaffe}},\ }\bibfield  {title} {\enquote {\bibinfo {title}
  {Effective kinetic theory for high temperature gauge theories},}\ }\href
  {\doibase 10.1088/1126-6708/2003/01/030} {\bibfield  {journal} {\bibinfo
  {journal} {JHEP}\ }\textbf {\bibinfo {volume} {01}},\ \bibinfo {pages} {030}
  (\bibinfo {year} {2003})},\ \Eprint {http://arxiv.org/abs/hep-ph/0209353}
  {arXiv:hep-ph/0209353 [hep-ph]} \BibitemShut {NoStop}%
\bibitem [{\citenamefont {Mueller}(2000)}]{Mueller:1999pi}%
  \BibitemOpen
  \bibfield  {author} {\bibinfo {author} {\bibfnamefont {Alfred~H.}\
  \bibnamefont {Mueller}},\ }\bibfield  {title} {\enquote {\bibinfo {title}
  {{The Boltzmann equation for gluons at early times after a heavy ion
  collision}},}\ }\href {\doibase 10.1016/S0370-2693(00)00084-8} {\bibfield
  {journal} {\bibinfo  {journal} {Phys. Lett. B}\ }\textbf {\bibinfo {volume}
  {475}},\ \bibinfo {pages} {220--224} (\bibinfo {year} {2000})},\ \Eprint
  {http://arxiv.org/abs/hep-ph/9909388} {arXiv:hep-ph/9909388} \BibitemShut
  {NoStop}%
\bibitem [{\citenamefont {Abraao~York}\ \emph {et~al.}(2014)\citenamefont
  {Abraao~York}, \citenamefont {Kurkela}, \citenamefont {Lu},\ and\
  \citenamefont {Moore}}]{AbraaoYork:2014hbk}%
  \BibitemOpen
  \bibfield  {author} {\bibinfo {author} {\bibfnamefont {Mark~C.}\ \bibnamefont
  {Abraao~York}}, \bibinfo {author} {\bibfnamefont {Aleksi}\ \bibnamefont
  {Kurkela}}, \bibinfo {author} {\bibfnamefont {Egang}\ \bibnamefont {Lu}}, \
  and\ \bibinfo {author} {\bibfnamefont {Guy~D.}\ \bibnamefont {Moore}},\
  }\bibfield  {title} {\enquote {\bibinfo {title} {{UV cascade in classical
  Yang-Mills theory via kinetic theory}},}\ }\href {\doibase
  10.1103/PhysRevD.89.074036} {\bibfield  {journal} {\bibinfo  {journal} {Phys.
  Rev. D}\ }\textbf {\bibinfo {volume} {89}},\ \bibinfo {pages} {074036}
  (\bibinfo {year} {2014})},\ \Eprint {http://arxiv.org/abs/1401.3751}
  {arXiv:1401.3751 [hep-ph]} \BibitemShut {NoStop}%
\bibitem [{\citenamefont {Arnold}\ \emph {et~al.}(2002)\citenamefont {Arnold},
  \citenamefont {Moore},\ and\ \citenamefont {Yaffe}}]{Arnold:2002ja}%
  \BibitemOpen
  \bibfield  {author} {\bibinfo {author} {\bibfnamefont {Peter~Brockway}\
  \bibnamefont {Arnold}}, \bibinfo {author} {\bibfnamefont {Guy~D.}\
  \bibnamefont {Moore}}, \ and\ \bibinfo {author} {\bibfnamefont {Laurence~G.}\
  \bibnamefont {Yaffe}},\ }\bibfield  {title} {\enquote {\bibinfo {title}
  {{Photon and gluon emission in relativistic plasmas}},}\ }\href {\doibase
  10.1088/1126-6708/2002/06/030} {\bibfield  {journal} {\bibinfo  {journal}
  {JHEP}\ }\textbf {\bibinfo {volume} {06}},\ \bibinfo {pages} {030} (\bibinfo
  {year} {2002})},\ \Eprint {http://arxiv.org/abs/hep-ph/0204343}
  {arXiv:hep-ph/0204343} \BibitemShut {NoStop}%
\bibitem [{\citenamefont {Aurenche}\ \emph {et~al.}(2002)\citenamefont
  {Aurenche}, \citenamefont {Gelis},\ and\ \citenamefont
  {Zaraket}}]{Aurenche:2002pd}%
  \BibitemOpen
  \bibfield  {author} {\bibinfo {author} {\bibfnamefont {P.}~\bibnamefont
  {Aurenche}}, \bibinfo {author} {\bibfnamefont {F.}~\bibnamefont {Gelis}}, \
  and\ \bibinfo {author} {\bibfnamefont {H.}~\bibnamefont {Zaraket}},\
  }\bibfield  {title} {\enquote {\bibinfo {title} {{A Simple sum rule for the
  thermal gluon spectral function and applications}},}\ }\href {\doibase
  10.1088/1126-6708/2002/05/043} {\bibfield  {journal} {\bibinfo  {journal}
  {JHEP}\ }\textbf {\bibinfo {volume} {05}},\ \bibinfo {pages} {043} (\bibinfo
  {year} {2002})},\ \Eprint {http://arxiv.org/abs/hep-ph/0204146}
  {arXiv:hep-ph/0204146} \BibitemShut {NoStop}%
\bibitem [{\citenamefont {Ghiglieri}\ and\ \citenamefont
  {Moore}(2014)}]{Ghiglieri:2014kma}%
  \BibitemOpen
  \bibfield  {author} {\bibinfo {author} {\bibfnamefont {Jacopo}\ \bibnamefont
  {Ghiglieri}}\ and\ \bibinfo {author} {\bibfnamefont {Guy~D.}\ \bibnamefont
  {Moore}},\ }\bibfield  {title} {\enquote {\bibinfo {title} {Low mass thermal
  dilepton production at {NLO} in a weakly coupled quark-gluon plasma},}\
  }\href {\doibase 10.1007/JHEP12(2014)029} {\bibfield  {journal} {\bibinfo
  {journal} {JHEP}\ }\textbf {\bibinfo {volume} {12}},\ \bibinfo {pages} {029}
  (\bibinfo {year} {2014})},\ \Eprint {http://arxiv.org/abs/1410.4203}
  {arXiv:1410.4203 [hep-ph]} \BibitemShut {NoStop}%
\bibitem [{\citenamefont {Kurkela}\ and\ \citenamefont
  {Mazeliauskas}(2019)}]{Kurkela:2018oqw}%
  \BibitemOpen
  \bibfield  {author} {\bibinfo {author} {\bibfnamefont {Aleksi}\ \bibnamefont
  {Kurkela}}\ and\ \bibinfo {author} {\bibfnamefont {Aleksas}\ \bibnamefont
  {Mazeliauskas}},\ }\bibfield  {title} {\enquote {\bibinfo {title} {{Chemical
  equilibration in weakly coupled QCD}},}\ }\href {\doibase
  10.1103/PhysRevD.99.054018} {\bibfield  {journal} {\bibinfo  {journal} {Phys.
  Rev. D}\ }\textbf {\bibinfo {volume} {99}},\ \bibinfo {pages} {054018}
  (\bibinfo {year} {2019})},\ \Eprint {http://arxiv.org/abs/1811.03068}
  {arXiv:1811.03068 [hep-ph]} \BibitemShut {NoStop}%
\bibitem [{\citenamefont {Schlichting}\ and\ \citenamefont
  {Teaney}(2019)}]{Schlichting:2019abc}%
  \BibitemOpen
  \bibfield  {author} {\bibinfo {author} {\bibfnamefont {Soeren}\ \bibnamefont
  {Schlichting}}\ and\ \bibinfo {author} {\bibfnamefont {Derek}\ \bibnamefont
  {Teaney}},\ }\bibfield  {title} {\enquote {\bibinfo {title} {{The First fm/c
  of Heavy-Ion Collisions}},}\ }\href {\doibase
  10.1146/annurev-nucl-101918-023825} {\bibfield  {journal} {\bibinfo
  {journal} {Ann. Rev. Nucl. Part. Sci.}\ }\textbf {\bibinfo {volume} {69}},\
  \bibinfo {pages} {447--476} (\bibinfo {year} {2019})},\ \Eprint
  {http://arxiv.org/abs/1908.02113} {arXiv:1908.02113 [nucl-th]} \BibitemShut
  {NoStop}%
\bibitem [{\citenamefont {Berges}\ \emph {et~al.}(2021)\citenamefont {Berges},
  \citenamefont {Heller}, \citenamefont {Mazeliauskas},\ and\ \citenamefont
  {Venugopalan}}]{Berges:2020fwq}%
  \BibitemOpen
  \bibfield  {author} {\bibinfo {author} {\bibfnamefont {J\"urgen}\
  \bibnamefont {Berges}}, \bibinfo {author} {\bibfnamefont {Michal~P.}\
  \bibnamefont {Heller}}, \bibinfo {author} {\bibfnamefont {Aleksas}\
  \bibnamefont {Mazeliauskas}}, \ and\ \bibinfo {author} {\bibfnamefont {Raju}\
  \bibnamefont {Venugopalan}},\ }\bibfield  {title} {\enquote {\bibinfo {title}
  {{QCD thermalization: Ab initio approaches and interdisciplinary
  connections}},}\ }\href {\doibase 10.1103/RevModPhys.93.035003} {\bibfield
  {journal} {\bibinfo  {journal} {Rev. Mod. Phys.}\ }\textbf {\bibinfo {volume}
  {93}},\ \bibinfo {pages} {035003} (\bibinfo {year} {2021})},\ \Eprint
  {http://arxiv.org/abs/2005.12299} {arXiv:2005.12299 [hep-th]} \BibitemShut
  {NoStop}%
\bibitem [{\citenamefont {Keegan}\ \emph {et~al.}(2016)\citenamefont {Keegan},
  \citenamefont {Kurkela}, \citenamefont {Mazeliauskas},\ and\ \citenamefont
  {Teaney}}]{Keegan:2016cpi}%
  \BibitemOpen
  \bibfield  {author} {\bibinfo {author} {\bibfnamefont {Liam}\ \bibnamefont
  {Keegan}}, \bibinfo {author} {\bibfnamefont {Aleksi}\ \bibnamefont
  {Kurkela}}, \bibinfo {author} {\bibfnamefont {Aleksas}\ \bibnamefont
  {Mazeliauskas}}, \ and\ \bibinfo {author} {\bibfnamefont {Derek}\
  \bibnamefont {Teaney}},\ }\bibfield  {title} {\enquote {\bibinfo {title}
  {Initial conditions for hydrodynamics from weakly coupled pre-equilibrium
  evolution},}\ }\href {\doibase 10.1007/JHEP08(2016)171} {\bibfield  {journal}
  {\bibinfo  {journal} {JHEP}\ }\textbf {\bibinfo {volume} {08}},\ \bibinfo
  {pages} {171} (\bibinfo {year} {2016})},\ \Eprint
  {http://arxiv.org/abs/1605.04287} {arXiv:1605.04287 [hep-ph]} \BibitemShut
  {NoStop}%
\bibitem [{\citenamefont {Avramescu}\ \emph
  {et~al.}(2023{\natexlab{b}})\citenamefont {Avramescu}, \citenamefont
  {B\u{a}ran}, \citenamefont {Greco}, \citenamefont {Ipp}, \citenamefont
  {M\"uller},\ and\ \citenamefont {Ruggieri}}]{Avramescu:2023vld}%
  \BibitemOpen
  \bibfield  {author} {\bibinfo {author} {\bibfnamefont {Dana}\ \bibnamefont
  {Avramescu}}, \bibinfo {author} {\bibfnamefont {Virgil}\ \bibnamefont
  {B\u{a}ran}}, \bibinfo {author} {\bibfnamefont {Vincenzo}\ \bibnamefont
  {Greco}}, \bibinfo {author} {\bibfnamefont {Andreas}\ \bibnamefont {Ipp}},
  \bibinfo {author} {\bibfnamefont {David.~I.}\ \bibnamefont {M\"uller}}, \
  and\ \bibinfo {author} {\bibfnamefont {Marco}\ \bibnamefont {Ruggieri}},\
  }\bibfield  {title} {\enquote {\bibinfo {title} {{Heavy quark $\kappa$ and
  jet $\hat{q}$ transport coefficients in the Glasma early stage of heavy-ion
  collisions}},}\ }in\ \href@noop {} {\emph {\bibinfo {booktitle} {{11th
  International Conference on Hard and Electromagnetic Probes of High-Energy
  Nuclear Collisions}: {Hard Probes 2023}}}}\ (\bibinfo {year} {2023})\ \Eprint
  {http://arxiv.org/abs/2307.07999} {arXiv:2307.07999 [hep-ph]} \BibitemShut
  {NoStop}%
\bibitem [{\citenamefont {Brambilla}\ \emph
  {et~al.}(2023{\natexlab{b}})\citenamefont {Brambilla}, \citenamefont {Leino},
  \citenamefont {Mayer-Steudte},\ and\ \citenamefont
  {Petreczky}}]{Brambilla:2022xbd}%
  \BibitemOpen
  \bibfield  {author} {\bibinfo {author} {\bibfnamefont {Nora}\ \bibnamefont
  {Brambilla}}, \bibinfo {author} {\bibfnamefont {Viljami}\ \bibnamefont
  {Leino}}, \bibinfo {author} {\bibfnamefont {Julian}\ \bibnamefont
  {Mayer-Steudte}}, \ and\ \bibinfo {author} {\bibfnamefont {Peter}\
  \bibnamefont {Petreczky}} (\bibinfo {collaboration} {TUMQCD}),\ }\bibfield
  {title} {\enquote {\bibinfo {title} {{Heavy quark diffusion coefficient with
  gradient flow}},}\ }\href {\doibase 10.1103/PhysRevD.107.054508} {\bibfield
  {journal} {\bibinfo  {journal} {Phys. Rev. D}\ }\textbf {\bibinfo {volume}
  {107}},\ \bibinfo {pages} {054508} (\bibinfo {year} {2023}{\natexlab{b}})},\
  \Eprint {http://arxiv.org/abs/2206.02861} {arXiv:2206.02861 [hep-lat]}
  \BibitemShut {NoStop}%
\bibitem [{\citenamefont {Banerjee}\ \emph
  {et~al.}(2022{\natexlab{b}})\citenamefont {Banerjee}, \citenamefont {Datta},\
  and\ \citenamefont {Laine}}]{Banerjee:2022uge}%
  \BibitemOpen
  \bibfield  {author} {\bibinfo {author} {\bibfnamefont {D.}~\bibnamefont
  {Banerjee}}, \bibinfo {author} {\bibfnamefont {S.}~\bibnamefont {Datta}}, \
  and\ \bibinfo {author} {\bibfnamefont {M.}~\bibnamefont {Laine}},\ }\bibfield
   {title} {\enquote {\bibinfo {title} {{Lattice study of a magnetic
  contribution to heavy quark momentum diffusion}},}\ }\href {\doibase
  10.1007/JHEP08(2022)128} {\bibfield  {journal} {\bibinfo  {journal} {JHEP}\
  }\textbf {\bibinfo {volume} {08}},\ \bibinfo {pages} {128} (\bibinfo {year}
  {2022}{\natexlab{b}})},\ \Eprint {http://arxiv.org/abs/2204.14075}
  {arXiv:2204.14075 [hep-lat]} \BibitemShut {NoStop}%
\bibitem [{\citenamefont {Brambilla}\ \emph {et~al.}(2022)\citenamefont
  {Brambilla}, \citenamefont {Escobedo}, \citenamefont {Islam}, \citenamefont
  {Strickland}, \citenamefont {Tiwari}, \citenamefont {Vairo},\ and\
  \citenamefont {Vander~Griend}}]{Brambilla:2022ynh}%
  \BibitemOpen
  \bibfield  {author} {\bibinfo {author} {\bibfnamefont {Nora}\ \bibnamefont
  {Brambilla}}, \bibinfo {author} {\bibfnamefont {Miguel~\'Angel}\ \bibnamefont
  {Escobedo}}, \bibinfo {author} {\bibfnamefont {Ajaharul}\ \bibnamefont
  {Islam}}, \bibinfo {author} {\bibfnamefont {Michael}\ \bibnamefont
  {Strickland}}, \bibinfo {author} {\bibfnamefont {Anurag}\ \bibnamefont
  {Tiwari}}, \bibinfo {author} {\bibfnamefont {Antonio}\ \bibnamefont {Vairo}},
  \ and\ \bibinfo {author} {\bibfnamefont {Peter}\ \bibnamefont
  {Vander~Griend}},\ }\bibfield  {title} {\enquote {\bibinfo {title} {{Heavy
  quarkonium dynamics at next-to-leading order in the binding energy over
  temperature}},}\ }\href@noop {} {\  (\bibinfo {year} {2022})},\ \Eprint
  {http://arxiv.org/abs/2205.10289} {arXiv:2205.10289 [hep-ph]} \BibitemShut
  {NoStop}%
\bibitem [{\citenamefont {Boguslavski}\ \emph {et~al.}(2023)\citenamefont
  {Boguslavski}, \citenamefont {Kurkela}, \citenamefont {Lappi}, \citenamefont
  {Lindenbauer},\ and\ \citenamefont {Peuron}}]{Boguslavski:2023alu}%
  \BibitemOpen
  \bibfield  {author} {\bibinfo {author} {\bibfnamefont {Kirill}\ \bibnamefont
  {Boguslavski}}, \bibinfo {author} {\bibfnamefont {Aleksi}\ \bibnamefont
  {Kurkela}}, \bibinfo {author} {\bibfnamefont {Tuomas}\ \bibnamefont {Lappi}},
  \bibinfo {author} {\bibfnamefont {Florian}\ \bibnamefont {Lindenbauer}}, \
  and\ \bibinfo {author} {\bibfnamefont {Jarkko}\ \bibnamefont {Peuron}},\
  }\bibfield  {title} {\enquote {\bibinfo {title} {{Jet momentum broadening
  during initial stages in heavy-ion collisions}},}\ }\href@noop {} {\
  (\bibinfo {year} {2023})},\ \Eprint {http://arxiv.org/abs/2303.12595}
  {arXiv:2303.12595 [hep-ph]} \BibitemShut {NoStop}%
\end{thebibliography}%
\end{document}